\documentclass[%
 reprint,
 amsmath,amssymb,
 aps,
]{revtex4-2}
\usepackage{float}
\usepackage{graphicx}
\usepackage{dcolumn}
\usepackage{bm}
\usepackage{hyperref}
\usepackage{physics}
\usepackage{makecell}
\usepackage{multirow}
\usepackage{array}
\usepackage{xcolor}

\begin{document}

\preprint{APS/123-QED}

\title{Surface topological quantum criticality:\\ Conformal manifolds and Discrete Strong Coupling Fixed Points}

\author{ Saran Vijayan and Fei Zhou}
\affiliation{Department of Physics and Astronomy, University of British Columbia, 6224 Agricultural Road, Vancouver, BC, V6T 1Z1, Canada}
\date{\today}

\begin{abstract}
In this article, we study quantum critical phenomena in surfaces of symmetry-protected topological matter, i.e. surface topological quantum criticality. A generic phase boundary of gapless surfaces in a symmetry-protected state shall be a co-dimension one manifold in an interaction parameter space of dimension $D_p$ (where $p$ refers to the parameter space) where the value of $D_p$ further depends on bulk topologies. In the context of fermionic topological insulators that we focus on, $D_p$ depends on the number of half-Dirac cones $\mathcal{N}$. We construct such manifolds explicitly for a few interaction parameter spaces with various $D_p$ values. Most importantly, we further illustrate that in cases with $D_p=3$ and $6$, there are sub-manifolds of fixed points that dictate the universalities of surface topological quantum criticality. These infrared stable manifolds are associated with emergent symmetries in the renormalization-group-equation flow naturally appearing in the loop expansion. Unlike in the usual order-disorder quantum critical phenomena, typically governed by an isolated Wilson-Fisher fixed point, we find in the one-loop approximation surface topological quantum criticalities are naturally captured by conformal manifolds where the number of marginal operators uniquely determines their co-dimensions. Isolated strong coupling fixed points also appear, usually as the endpoints in the phase boundary of surface topological quantum phases. However, their extreme infrared instabilities along multiple directions suggest that they shall be related to multi-critical surface topological quantum critical phenomena rather than generic surface topological quantum criticality. We also discuss and classify higher-loop symmetry-breaking effects, which can either distort the conformal manifolds or further break the conformal manifolds down to a few distinct fixed points.
\end{abstract}

\maketitle

\section{\label{sec:intro}Introduction}
Symmetry-protected states (SPTs) \cite{kitaev(2009),wen(2009),Ryu(2010)} are known to have robust and unique gapless boundary states that can lead to various potential applications. One particularly fascinating example is the surface electronic states of 3D topological insulators\cite{fukane(2005),bernevigzhang(2006),bernevighugheszhang(2006),fukanemele(2007),moorebalents(2007),hasankanereview(2010), Qizhangreview(2011),bernevigbook(2013)} which have been attracting much attention. In the general paradigm of SPTs, these surface states usually remain gapless unless the protecting symmetry is explicitly broken by applied external fields\cite{shoucheng(2008),fukane(2008),zhang(2009)} or spontaneously broken due to strong interactions, either attractive\cite{mudry(2010),imada(2011),imada(2012),sondhi(2013),herbut(2013),sarma(2013),Ponte(2014),Grover(2014),neupert(2015),maciejko(2016),maciejko(2021)} or repulsive\cite{xu(2010),xu(2010)(2),tetsuya(2011),stern(2012),franz(2012),fritz(2013),potter(2015)}. Although the free fermion topological states and their 
gapless surfaces have been quite thoroughly understood, what happens when interactions, especially strong interactions, are present remains to be a fascinating topic.

For instance, if the protecting symmetry is unbroken in the presence of strong interactions, the surface states can become gapped only when they are topologically ordered. When the protecting symmetries are time-reversal and U(1) symmetries, there shall be 8 classes of 
3D topological distinct states when interactions are taken into account\cite{senthil(2015)}. If the 
protecting symmetry is time-reversal symmetry only, there shall be 8 classes in 1D and 16 
classes in 3D\cite{Fidkowski(2010),Fidkowski(2011),fidkowski(2013)}. Although it is unclear under which conditions or which practical Hamiltonians these exotic states can emerge in a condensed 
matter system\cite{metlitski(2015), song(2017)}, the complexity of boundary quantum dynamics always plays a paramount role in our general understanding of SPTs including topological insulators, when strong interactions are present. 

This article is focused on an interplay between topology and interactions, from a perspective of boundary quantum dynamics of topological states. We restrict ourselves to a topological insulator in a cubic crystal that is protected by the time-reversal symmetry, $U(1)$ symmetry and crystal symmetry. We will focus on surface fermions with attractive interactions. It is worth emphasizing that discussions on co-dimension one phase boundary manifolds in the space of interaction parameters are very generic and valid for other surface topological quantum criticality in SPTs. The results on the structure of conformal submanifolds are, on the other hand specific for topological insulators. 

The motivations are three-fold. Firstly, gapless surface fermions or electrons with attractive interactions can be crucial to the making of topological surface superconductivity as well as emergent Majorana fermions\cite{fukane(2008),SDShrma(2010)}. Such surface Majorana fermions can have potential applications for topological quantum computing. 

Secondly, the most commonly studied gapless states near a free fermion fixed point are very distinctly different from other gapless conformal field states near strong coupling fixed points studied below. It is essential to understand what possible universal classes of gapless states can be present as gapless boundaries of a topological state with inter-particle interactions.

Thirdly and perhaps most importantly, we are going to identify the phase boundary of free 
fermion gapless states in interaction parameter spaces. A phase boundary that separates the gapless 
surface state from other distinct states of matter usually appears as a manifold embedded in a multidimensional space of interaction parameters. For a dimension $D_p$ interaction parameter space, the phase boundary generically is a $D_p-1$ dimension manifold, i.e. a co-dimension one manifold. Each point in the manifold represents a possible quantum critical point. 

However, there only appear to be a discrete number of universality classes that determine 
dynamics on all critical states situated in the phase boundary manifold. These universality 
classes can be identified with scale-conformal symmetric fixed points naturally emerging in the studies of standard renormalization group equations. In fact, the fixed points can also form a sub-manifold but with co-dimension equal to or higher than 2 but less than $D_p-1$ itself (See details below). 

Let us further elaborate on the second and third points. To answer the question of universality classes of gapless boundaries, technically one needs to identify all possible scale symmetric fixed points in surface dynamics. While a free fermion fixed point can be naturally associated 
with a surface of weakly interacting gapless topological surfaces, strong coupling fixed points open the door to explore other possible but distinct gapless boundaries in a topological state. These scale-invariant fixed points, free or strongly interacting conformal ones, therefore can be further applied to classify universal thermodynamics and transport dynamics, either infrared or ultraviolet ones, that one can observe in a physical topological surface state. 

 It has been known that SPT surfaces can be identified via the method of gauge anomalies (or vice versa, see Ref.\cite{wen(2013)}) as the lattice symmetry can be implemented in a non-local way on surfaces. For given anomalies, one can identify phases of surfaces depending on whether spontaneous symmetry breaking(SSB) or topological orders (TO) further appear on a surface.
The surfaces of 3D bosonic SPT states and quantum criticality and dynamics have been thoroughly studied in Ref.\cite{ashvin(2013)}. We refer the readers to those discussions.

The current studies are fully consistent with this general paradigm of topological surfaces. What we have addressed here is that when interactions are attractive pairing ones, the classes of CFTs that appear on topological surfaces can be further directly related to bulk invariants, which we have used to infer the number of surface fermions. For instance, for strong TIs (with cubic crystal symmetry), the CFTs are either related to an isolated SUSY fixed point or a conformal manifold with its tangent space set by two marginal operators but without SUSY. These are the general results within a given class of interacting topological surfaces rather than within a given broader class of anomalies. Our family of gapless states are therefore a subclass with further inputs of interactions. They are related to quantum phase transitions to SSB states that break the protecting charge $U(1)$ symmetries.

There can also be other fascinating transitions, such as SSB to SSB states and TO to TO states. They are very interesting to look further into in connection to conformal manifolds. In our current studies, we have restricted ourselves to quantum phase transitions between the well-observed gapless free fermions phase and SSB states breaking the charge $U(1)$ symmetries with interaction-parameter inputs, in addition to topological data. Our main focus is to understand the phase boundary that separates the gapless phase and SSB states, breaking one of the protecting symmetries and what we can have on the phase boundary manifold.

Our main objective is then to identify all gapless dynamics in a topological surface for a wide class of short-range interactions via an analysis of possible scale symmetric fixed points. 
Unlike in the standard paradigm where it is believed there shall be a state-boundary 
correspondence, for the purpose of understanding gapless dynamics, it is more beneficial to have a correspondence between a bulk state and a family of fixed points or conformal field theories supported by a topological bulk, apart from the standard free fermion gapless surface. These emergent conformal fields also represent quantum critical points in critical manifolds that 
separate the free fermions from other phases of matter. In the following, we will pursue this goal assuming the gapless surface fermions interact via attractive interactions.

 Before presenting the outline of this article, we want to emphasize that a topological gapless surface can have one-half of a Dirac cone or $N_f=1/2$, unlike in a bulk lattice where one-half of a Dirac cone is strictly forbidden because of the Nielson-Ninomiya theorem of fermion doubling \cite{Nielson(1981),Nielson(1981)(2)}. Such anomalous surface fermions of a four-dimensional bulk are also intimately related to t'Hooft anomalies and/or gauge anomalies\cite{hooft(1976),adler(1969),bell(1969)} in three spatial dimensions.

The conformal field theories (CFT) of one-half of a Dirac cone or a quarter of a Dirac cone related to a strong coupling fixed point of surface fermions can have emergent supersymmetries(SUSY)\cite{Ponte(2014),Grover(2014),maciejko(2016),yao(2017)(2),maciejko(2021)}. The phenomena of emergent SUSY in boundaries, or surfaces or in domain-wall surfaces were previously suggested in a series of articles\cite{yao(2017),yao(2018)}. 
In all these studies, robust gapless surface fermions in a topological state play a paramount role in realizing SUSY physics. 
We refer readers to those articles for details.

In lattices, SUSY can also emerge with two half-Dirac cones in a bulk honeycomb lattice\cite{SSLee(2007)}. The kinematic constraint in the lattice effectively forbids the coupling between two half-Dirac cones resulting in two separated copies of SUSY theories.

The equivalence of a half-Dirac cone can emerge in a bulk lattice only if the charge $U(1)$ symmetry is broken spontaneously, and SUSY physics can be relevant in such cases. 
Indeed, in a more recent study on topological superconductors or superfluids, it was shown that real fermions equivalent to a quarter-Dirac cone or a half-Dirac cone can further emerge at topological quantum critical points in a superconductor where the charge $U(1)$ symmetry has been spontaneously broken. 
The degree of real fermions and bulk dynamics can then be mapped into those of $N_f=\frac{1}{4}, \frac{1}{2}$ that results in SUSY\cite{zhou(2022)}. Note that this study doesn't involve surfaces or domain walls as most others do and it is a very rare example where an equivalence of $N_f=\frac{1}{2}$ fermion or a half-Dirac cone does appear in a bulk lattice\cite{zhou(2023)}.

In this article, we will focus on more general topological surface states that can appear in a topological insulator with a crystal translation symmetry where the number of Dirac cones is defined as $N_f=\frac{\mathcal{N}}{2}$, with $\mathcal{N}=1,2,3,4$. We will exclusively focus on the cases of $\mathcal{N}=2,3$ where conformal manifolds of strong coupling fixed points naturally appear on phase boundaries of gapless free fermions. All the CFTs we discuss in this article do not exhibit SUSY, unlike in the case of $\mathcal{N}=1$; however, they represent generic strongly interacting topological surfaces of a topological cubic crystal.

We have dedicated section.\ref{sec:cfnml} exclusively for the discussions on the conformal manifolds. There, we discuss certain features of the conformal manifold that appear at the phase boundary of the gapless surface fermions. We focus on two aspects: 1) relations to generic conformal manifolds discussed in quantum field theories and 2) its distinction from conventional isolated fixed points.

Section \ref{sec:TIdfm} is dedicated to the exploration of various topological phases of a simple 3d cubic lattice model and their surface states. Being the surface of a 3d topological insulator, it could host an $\mathcal{N}$ number of surface half-Dirac cones 
(from now on we will use $\mathcal{N}=2N_f$ to count the degrees of surface fermions) where $\mathcal{N}$ can be $1$, $2$ or $3$. For the trivial phase, $\mathcal{N}$ can be either $0$ or $4$. Thus we focus only on the cases of $\mathcal{N}$ being $1,2,3$. If $\mathcal{N}$ is odd(even), then the material is a strong(weak) TI. It turns out that the topological classification does not uniquely identify the exact number of surface Dirac cones, rather it defines the even-odd parity of $\mathcal{N}$. Hence, it is possible that for the same bulk topological invariant, the surface can host different numbers of Dirac fermions, as long as the count remains odd or even. In this context, the stability of the number $\mathcal{N}$ of the surface Dirac fermions comes into question. This section addresses the issue, demonstrating that the exact number of surface Dirac cones (denoted by $\mathcal{N}$) can be stabilized by imposing additional symmetry constraints associated with a crystal (crystal translation symmetry, inversion symmetry, etc), despite identical bulk topological $Z_2$ invariants but different surfaces. This is very crucial and significant for our later discussions on phase boundaries and conformal manifolds of fixed points that define the universality of phase transitions across the phase boundaries.
As we shall show in this article, the $\mathcal{N}=1$ and $\mathcal{N}=3$ surfaces can have different phase boundary manifolds, placing them in distinctly different universality classes.

\begin{table*}
 \caption{Summary of results: List of isolated fixed points, conformal manifolds and phase boundary of an attractively interacting TI (cubic crystal) surface with $\mathcal{N}$ Dirac cones when studied upto one-loop order. Here $\hat{V}^{\mathcal{N}\times \mathcal{N}}_c$ is the critical interaction matrix of size $\mathcal{N}$ (see Eqn.\ref{RGeqn}). $\textbf{n}_{\mathcal{N}}$ is a unit vector in $\mathcal{N}$-dimensional space.}
    \label{smmry}
\renewcommand{\arraystretch}{2.0}
\begin{tabular}{|c|c|c|c|c|c|}
\hline\hline
    \makecell{Number of surface\\ Dirac cones ($\mathcal{N}$)} &
    \makecell{Dimension of the\\ parameter space\\ ($D_p = \frac{\mathcal{N}(\mathcal{N}+1)}{2}$)}  & 
    
    \makecell{Dimension of the \\ the phase boundary\\ manifold ($D_p - 1$)} & 
    
    \makecell{Isolated \\fixed points} & 
    
    \makecell{Fixed point manifold(s)\\ (Conformal manifold(s))} & 
    
    \makecell{Shape of the \\conformal \\manifold(s)} \\ 
    
    \hline
    
     \multirow{2}{*}{2}   & 
     
     \multirow{2}{*}{3} & 
     
     \multirow{2}{*}{2}  & 
     
     $\hat{V}^{2 \times 2}_c = 0 \hat{I}^{2\times2}$  & 
     
     \multirow{2}{*}{$\hat{V}_c^{2\times2} = \epsilon \textbf{n}_2\otimes \textbf{n}_2$}  & 
     
     \multirow{2}{*}{Ring} \\ 
     
     \cline{4-4} & & & 
     
     $\hat{V}^{2 \times 2}_c = \epsilon \hat{I}^{2\times2}$ & &
       \\
       
       \hline 
       
       \multirow{2}{*}{3}   &

    \multirow{2}{*}{6}  & 
    
    \multirow{2}{*}{5}  &

          $ \hat{V}^{3 \times 3}_c = 0 \hat{I}^{3\times3}$ &

  $\hat{V}^{3 \times 3}_c = \epsilon \textbf{n}_3\otimes \textbf{n}_3$ & 
  
  2-sphere \\ 
  
  \cline{4-6} & & & 
  
  $ \hat{V}^{3 \times 3}_c = \epsilon \hat{I}^{3\times3}$  &  
  
  $ \hat{V}^{3 \times 3}_c = \epsilon \hat{I}^{3\times3} - \epsilon \textbf{n}_3\otimes \textbf{n}_3$ & 2-sphere \\ \hline \hline
    \end{tabular}
    
\end{table*}

In section \ref{sec:RG}, we introduce effective attractive interactions to the 3D TI with an arbitrary $\mathcal{N}$ number of surface Dirac cones. We first develop an effective four-fermion field theory in $2+1D$ for the TI surface with $\mathcal{N}$ 2-component Dirac fermions and then use the renormalization group equation (RGE) approach in $2+\epsilon$ dimensions to study the surface quantum criticality. We find that the RG equation possesses an emergent $SO(\mathcal{N})$ symmetry at the $\mathcal{O}(\epsilon)$  order or the one-loop approximation while it is broken at the $\mathcal{O}(\epsilon^2)$ order or the two-loop approximation. In the spirit of $\epsilon$-expansion, in this section we limit our studies to the one-loop approximation, as a consequence of which the RGE possesses an emergent $SO(\mathcal{N})$ symmetry.

This emergent symmetry of the dynamical equation results in certain classes of fixed points forming a conformal manifold in the parameter space, for $\mathcal{N} > 1$. The geometry of the manifold is determined by the coset space formed by dividing the symmetry group of the RG equation ($SO(\mathcal{N})$ group in our case) with the invariant subgroup of the scale symmetric solutions.  Section \ref{sec:SSB} is dedicated to the discussion of this phenomenon in detail. 

As stated before,
the $\mathcal{N}=1$ limit is well studied and it is found that the surface possesses a UV fixed point in addition to the free-fermion fixed point. In section \ref{secN2}, we discuss the fixed points of the theory when $\mathcal{N} = 2$, that is when there are two flavors of surface fermions. We observe that in addition to the free-fermion and the unstable strong-coupling fixed points, there exists a manifold of interacting fixed points that forms a ring in the parameter space. This ring is embedded in a 2-dimensional conical phase boundary separating the gapless and superconducting phases. The infrared physics of the conical phase boundary can be described effectively by a Yukawa field theory which involves two flavors of surface fermions coupling with a single flavor of complex bosons. 

In section \ref{secN3}, we discuss the types of fixed points when there are three flavors of surface fermions on the TI surface. We observe that there are two conformal manifolds both in the shape of a 2-sphere. The 2-sphere conformal manifolds are embedded in a 5-dimensional phase boundary. Similar to the $\mathcal{N}=2$ case, the effective Yukawa field theory of the 5-dimensional phase boundary
involves $three$ flavors of surface fermions coupling with a single flavor of complex bosons.

In our current studies, we also find that the emergent symmetry in the RGEs, though a sufficient condition, is not a necessary condition for the formation of conformal manifolds. 
We show this explicitly in section \ref{sec:hghloop}, where certain three-loop effects are studied for the $\mathcal{N}=2$ TI surface. A smooth manifold persists in the presence of such $SO(\mathcal{N})$ symmetry-breaking effects, albeit in a distorted shape.
However, in general, the conformal manifolds can further break down to a few isolated fixed points when $SO(\mathcal{N})$ symmetry-breaking effects are included.

In section \ref{sec:cnclsn}, we conclude our studies and discuss future directions.

\section{Summary of main results}
The main results of our studies are (see Table.\ref{smmry}):

\paragraph*{A)}The phase boundaries of the gapless weakly interacting topological fermions strongly depend on the number of surface Dirac cones, $\mathcal{N}$. The dimension of the interaction parameter space (indicated by the subscript $p$ below) in our studies is $D_p=\mathcal{N} (\mathcal{N}+1)/2$ (see section \ref{secN2},\ref{secN3}). $D_p=3$ and $6$ respectively when $\mathcal{N}=2,3$ which we will focus on. The generic phase boundaries are given by manifolds of co-dimension $D_{co}=1$ embedded in a $D_p$ dimension parameter space.

For instance, for $\mathcal{N}=2$ and $D_p=3$, the dimension of the tangent spaces defined at any point of the phase boundary manifold, $D_T=2$ (the subscript $T$ here refers to the tangent space), is $D_T=D_p-D_{co}$. Physically, $D_T$ is identified as the number of irrelevant/marginal operators at a phase transition. By the same token, for $\mathcal{N}=3$ and a parameter space with dimension $D_p=6$, the tangent space of the phase boundary has $D_T=5$ that indicates five irrelevant and/or marginal operators at a generic critical point on the boundary.

\paragraph*{B)} The strong coupling fixed points in the phase boundary manifold can form their own manifolds with co-dimensions larger than one. The dimensions of conformal manifolds or manifolds of fixed points are always lower than the ones for the phase boundary. For $\mathcal{N}=2$ and $D_p=3$, we found a ring of strongly interacting fixed points that form a conformal manifold with co-dimension $D_{co}=2$, or a manifold of dimension $D_T=1$, in addition to an isolated fixed point with co-dimension $D_{co}=3$ (see Table.\ref{tableN2} or Fig.\ref{RGflow_N2}). For $\mathcal{N}=3$ and $D_p=6$, there are two conformal fixed point manifolds with $D_{co}=4$ or manifolds of dimension $D_T=2$. These are simple two-spheres embedded in a 6-dimension parameter space. In addition, there is an isolated fixed point with $D_{co}=6$ (see Table.\ref{n3table}). In the following section, we discuss these emergent manifolds in detail.

\paragraph*{C)} All the quantum phase transitions represented by any points in the phase boundary manifold are characterized by these conformal manifolds identified in our study. See details of the operator scaling dimensions near conformal manifolds (Appendix.\ref{apndxsclngdmnsns}).

Conformal manifolds also emerge in a recent study of topological quantum critical points (tQCPs).
It was demonstrated that in the strong coupling limit, emergent symmetries at tQCPs depend crucially on the appearance of conformal manifolds, unlike in the weak coupling limit where tQCPs are usually associated with an isolated scale symmetric fixed point\cite{zhou(2023),zhou(2024)}.

More adventurous attempts involving larger fermionic representations further imply deconfined quantum criticality 
\cite{senthil(2019)}. It is also pointed out later in Ref.\cite{zhou(2024)} that there can be no emergent gauge fields or gauge symmetries at generic tQCPs that belong to fundamental representations of protecting symmetries (with minimum number of fermions),

\paragraph*{D)} The $D_{T}=1$ conformal manifold (i.e. one-sphere) for the $\mathcal{N}=2$ and $D_T=2$ manifolds( i.e. two-spheres) for the $\mathcal{N}=3$ case are intimately related to the emergent $SO(\mathcal{N}), \mathcal{N}=2,3$ in the one-loop RGEs, respectively. 
We further illustrate the stability of these conformal manifolds in the cases of $\mathcal{N}=2,3$ against higher loop $SO(\mathcal{N})$-symmetry-breaking effects.
Certain higher loop $SO(\mathcal{N})$-symmetry breaking effects such as three-loop one-particle irreducible effects can only cause distortions (see Fig.\ref{RGflow_N23loop}).  
However, when specific two-loop $SO(\mathcal{N})$-symmetry breaking field renormalization effects are considered, we find the conformal manifolds further break down into isolated fixed points (see Table.\ref{tableN2hghloop} or Fig.\ref{RGflow_N22loop}). 

\section{\label{sec:cfnml} Conformal manifolds}

 Previously, we mentioned that the phase boundary manifold of the  $\mathcal{N} =2, 3$ flavor TI surfaces contains conformal sub-manifolds with co-dimensions $D_{co} = 1, 4$ respectively. These conformal manifolds observed at the surface quantum criticality require special attention. 
In this section, we shall describe certain generic features of the conformal manifold, especially on how it is distinct from a conventional isolated fixed point. Following this, we list down the special properties of the conformal manifold observed in our studies. 

A conformal manifold is a family of interacting Hamiltonians connected by exactly marginal operators. For instance, let $\mathcal{S}_c$ be an action defined at a point on the conformal manifold. Let $\{\mathcal{O}^{i}_M \} (i=1,..n)$ be the exactly marginal operators at the given point on the manifold. Then, after a marginal deformation,
\begin{eqnarray}
  \mathcal{S}_{c} + \delta \mathcal{S} = \mathcal{S}_c  + \sum^{n}_{i=1}\int d^{D} x \,\delta \lambda^i_M \mathcal{O}^i_M(x)    
\end{eqnarray}
the theory remains conformal invariant. Here, $\delta \lambda^{i}_M$ is the infinitesimal marginal coupling away from the fixed point corresponding to the marginal operator $\mathcal{O}^{i}_{M}$. The number of marginal operators determines the dimension of the conformal manifold. One important property of these exactly marginal operators is that their scaling dimension $\triangle_{\mathcal{O}_M} = D$, the spacetime dimensions of the theory. In contrast, the canonical scaling dimensions of $\mathcal{O}_{M}$ is $2D-2$. 

Eqns.\ref{mrgnlprtsN2} and \ref{mrgnlprtsN3}  provide the detailed expressions for the marginal operators $\mathcal{O}^i_M (x)$ of the conformal manifolds that appears on the phase boundary of $\mathcal{N}=2$ and $\mathcal{N}=3$ TI surfaces respectively.

Before proceeding further and starting our discussions on our results, we want to remark on the nature of the specific conformal manifolds discussed in this article and their relations to more generic conformal manifolds, which are well discussed in modern quantum field theories. As discussed above, typical construction of conformal manifolds usually starts with a set of exact marginal operators that form a tangent space of a manifold and can then be applied to generate other distinct conformal field theories via deformations.

Practical implementations of such procedures either require primary conformal operators with a known scale dimension matching the space-time dimension as in $(1+1)D$, or supersymmetry conformal operators to ensure such exact marginal operators as in a higher dimension. The best well-known high-dimension example is the N=4 supersymmetric Yang-Mills theory\cite{seiberg(1988),seiberg(1994)}, which can support un-compact conformal manifolds with logarithm divergent diameters measured using the standard Zamodolchikov metric\cite{zamolodchikov(1986)}.

The tangent space of our manifolds can be defined by marginal operators discussed in Appendix \ref{apndxsclngdmnsns} when they become exactly marginal in the limit of our interests. Our manifolds can then be constructed in a similar way via deformations. However, in this case, deformations simply lead to theories within the same universality class because of the higher symmetry of our manifolds. The manifold is either a one-sphere with a uniform metric or a two-sphere with a uniform curvature tensor, i.e. they form representations of $SO(2)$ and $SO(3)$ groups, respectively. The deformation of original theories induces a rotation in the interaction parameter space. Because of the emergent $SO(\mathcal{N})$ dynamic symmetry in RGEs discussed below, the operator scaling dimensions are invariant under rotations, hence the deformations.  Therefore, the deformations, although resulting in a family of distinct interaction Hamiltonians, do not result in a new universality class.

The conformal manifold in this article only hosts a single universality class, far fewer classes than the generic ones discussed before. For this reason, we name it a featureless conformal manifold to distinguish it from more generic cases, such as the ones in SCFTs, where each point on a manifold actually defines a distinct class of CFTs.

It shall be emphasized here that our result is distinctly different from an isolated Wilson-Fisher fixed point, which can be considered a zero-sphere.
An isolated fixed point forbids exact marginal operators as otherwise, it again can be deformed and induce a smooth manifold. So, from this point of view, the key distinction between a smooth manifold, as discussed in our case, and an isolated point, a more standard critical point, is the existence of a tangent space at each point of the manifold and exact marginal operators, which usually define the metric of the manifold. And the number of exact marginal operators defines the dimension of the manifold just like in all other general discussions.


\section{\label{sec:TIdfm}3D Topological insulators: Distinct surface states and deformation between them}

A 3d topological insulator is characterized by gapless boundary states protected by time-reversal symmetry (TRS). If TRS is the only symmetry of the system, then its topological phase is characterized by a single $Z_2$ invariant, namely $v_0$. Suppose we impose additional symmetries, especially crystal translation symmetry. In that case, the topological phase is characterized by four $Z_2$ invariants, namely $(v_0;v_1,v_2,v_3)$\cite{fukanemele(2007)}. Here, $v_0$ distinguishes the strong TI from the weak or trivial phase, while the other three invariants are relevant only if the system has crystal translation symmetry.
\begin{figure}[h]
 \includegraphics[width=8.5cm, height=6.5cm]{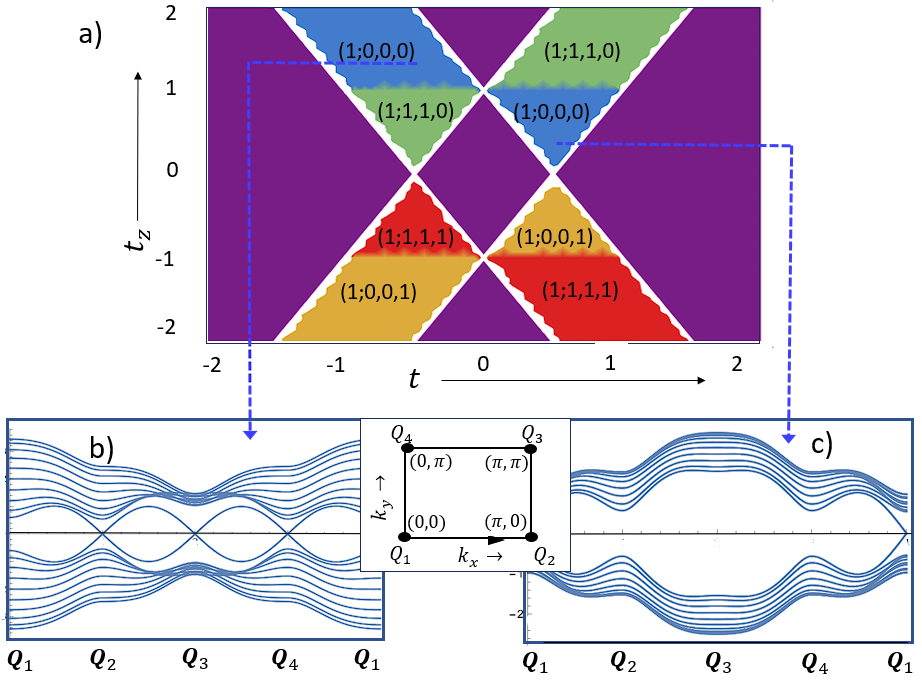} \\
 \caption{a) Phase diagram of the 3d topological insulator (TI) defined by the Hamiltonian in Eqn.\ref{anisotropicHamiltonian}, with $t_x = t_y = t$. Each topological phase is distinguished by its color, and the topological invariant is denoted by $(v_0;v_1,v_2,v_3)$. b) The surface spectrum of a bulk topological phase characterized by $(1;0,0,0)$ (represented by the blue-colored region) when $t<0$. Three surface Dirac cones are observed. c) The surface spectrum for the same bulk topological phase $(1;0,0,0)$ when $t > 0$. Only one single surface Dirac cone is observed. The inset illustrates the surface Brillouin zone (BZ) and its time-reversal invariant (TRI) points $\textbf{Q}_i (i=1,2,3,4)$.}
 \label{phsdgrm}
\end{figure}

Let us consider a 3d TI that possesses crystal translation symmetry. To make our discussions easier, we consider a 3d cubic lattice. The surface Brillouin zone (BZ) has four distinct time-reversal invariant (TRI) points, enabling a maximum of four distinct Dirac cones on the surface. We can classify the number of Dirac cones on the surface based on the topological stability. A strong TI phase($v_0 = 1$) has a single or three Dirac cones surface, while a weak TI phase ($v_0 = 0$) has two or four Dirac cones surface.

So, the topological classification does not uniquely identify the exact number of surface Dirac fermions except that it could be even or odd. For instance, for a strong TI defined in terms of the four $Z_2$ invariants $(v_0;v_1, v_2, v_3)$ with $v_0 = 1$, the surface can host a single or three Dirac cones. 


First, using a simple bulk lattice model, we show explicitly that a strong TI can host single or three Dirac cones for the same bulk topological invariant (section.\ref{sec:TIdfm}A). Then, we shall look for unitary operations that could connect these distinct surface states. We find two non-local unitary operators that are equivalent to infinite-depth ones. The first one (Eqn.\ref{U1_ltce}) does change the bulk invariant, as anticipated, therefore connecting two topologically different TIs, which is not surprising given the general paradigm about topological bulks. The second one (Eqn.\ref{U_efl_tce}), however, does not change the bulk, though of infinite depth, but changes the surface Dirac cones.

The main purpose of this part of the discussion is to understand how stable the surface Dirac cones are, given that for a given bulk, there are two different surfaces. Our analysis shows that although they are associated with the same bulk, for a given topological bulk, there appears to be no other way to deform one surface into the other surface unless it is unitary with infinite depth. Here, we have constructed the second type of infinite depth unitary that changes the surface of Dirac cones without changing the bulk.

Specifically, we find that our unitary accomplishing this necessarily partially breaks protecting symmetries and induces long-range hopping. This strongly implies that when restricted to local unitary when all protecting symmetries are enforced, the surface states remain distinct and topologically stable, reinforcing our focus on the criticality of a fixed number $\mathcal{N}$ of Dirac fermions. This is further illustrated in Fig.\ref{dfrmtn_TI}.


\subsection{\label{subsec1:level1}Distinct surface states for same bulk topological invariant: a practical example}

In this section, we begin with a minimal lattice model for a 3D topological insulator and explore its various topological phases. By numerically solving for the surface states, we show explicitly that the same topological bulk can have distinct surface states (see Fig.\ref{phsdgrm}). 

To realize a lattice model for a 3d topological insulator, we discretize a 3d Dirac Hamiltonian. We consider a cubic lattice for the purpose. The resulting minimal lattice model has the form\cite{shen(2013)},
\begin{eqnarray}
    \mathcal{H}_0 &=& \sum_{\textbf{k}} c^{\dagger}_{\textbf{k}} \mathcal{H}_{0}(\textbf{k}) c_{\textbf{k}}\label{anisotropicHamiltonian}\\ \mathcal{H}_0(\textbf{k}) &=& \sum_{i=x,y,z} \alpha_i \sin k_i + \beta \left[1 -  \sum_{i=x,y,z} t_i \cos k_i \right]\nonumber 
\end{eqnarray}
where $c^{\dagger}_{\textbf{k}}$ is a four-component spinor defined in the spin and orbital basis with the definition, 
$c^{\dagger}_{\textbf{k}} = (\begin{array}{cc}
    c^{\dagger}_{1\textbf{k},\uparrow} \,\, c^{\dagger}_{1\textbf{k},\downarrow} \,\, c^{\dagger}_{2\textbf{k},\uparrow} \,\, c^{\dagger}_{2\textbf{k},\downarrow}
\end{array})$ where 1(2) stands for the orbital index. $\alpha_i(i=x,y,z)$ and $\beta$ are the Dirac matrices that satisfy the following anti-commutation relations:
$$ \{\alpha_i, \alpha_j \} = 2\delta_{ij},\,\, \{\alpha_i, \beta \} = 0$$
and they have the following definition: $\alpha_i = \tau_x \otimes s_i$ and $\beta = \tau_z \otimes s_0$. 
Here $\tau_i$ and $s_i$ ($i=x,y,z$) are Pauli matrices defined in atomic orbital space and spin space, respectively. 

By tuning the parameters $t_i (i=x,y,z)$,  we can study all 16 distinct topological phases (including weak and strong TI phases). However, our prime objective here is to look for the possibility of distinct surface states for the same bulk topological invariant. Therefore, we reduce the number of tuning parameters by setting $t_x = t_y = t$. This simplification leaves us with two tuning parameters: $t$ and $t_z$.

Now, we explore the various topological phases of the system as a function of the tuning parameters $t$ and $t_z$. Since the Hamiltonian has inversion symmetry, calculating the bulk topological invariants $(v_0;v_1, v_2, v_3)$ is straightforward. The $Z_2$ invariants can be calculated directly from the parity eigenvalues of the ground state at the eight TR invariant points in the bulk Brillouin zone(BZ)\cite{fukane(2007)}. Fig.\ref{phsdgrm}(a), shows the phase diagram in the $t-t_z$ space.

To study the surface states explicitly, we derive the surface spectrum by solving a slab geometry of the lattice with open boundary conditions at the edges. We show the surface spectrum for the surface normal to the $z-$axis (Figs.\ref{phsdgrm}(b), \ref{phsdgrm}(c)).

Now consider the blue-colored topological phase identified by the topological invariant $(1;0,0,0)$ in Fig.\ref{phsdgrm}. A bulk gap closing separates two regions. The region with $t > 0$ has a single Dirac cone phase on the surface normal to the $z-$axis (see Fig.\ref{phsdgrm})(c)). On the other hand, the blue region with $t < 0$ has three Dirac cones surface, as evident from the surface spectrum shown in Fig.\ref{phsdgrm}(b). The bulk gap closes at the boundary separating these topological phases (the white line in Fig.\ref{phsdgrm}), pointing towards a topological phase transition.

Examining Fig.\ref{phsdgrm}, we observe that the regions with different surface states are not continuously connected despite belonging to the same bulk topological phase. We must cross a critical point to transition from a surface with a single Dirac cone to one with three Dirac cones.

This observation highlights that the same topological bulk can exhibit either a single or three Dirac cones on the surface, as shown in Fig. \ref{phsdgrm}. A critical point separates the two distinct surface states, where the bulk gap closes. This is due to the higher symmetry of the lattice model, which, in addition to time-reversal symmetry (TRS), also possesses crystal symmetries. If the model only has TRS as the protecting symmetry, an adiabatic path connecting the two surface states should exist without closing the bulk gap. This section aims to look for a unitary deformation that connects these distinct surface states. 

Even though realizing such a unitary deformation may seem straightforward, at least theoretically, it is tricky. We have to preserve the time-reversal symmetry throughout the unitary transformation. In addition, the Hamiltonian and the unitary operator should be smoothly defined on the surface of the toroid defined by the Brillouin zone. Starting from an effective field theory, we shall derive the explicit form of the unitary operator that could connect the two distinct surfaces.

\subsection{Distinct surface states: A consequence of flipping of mass signs}

\begin{figure}[b]
 \includegraphics[width=8.5cm, height=6.8cm]{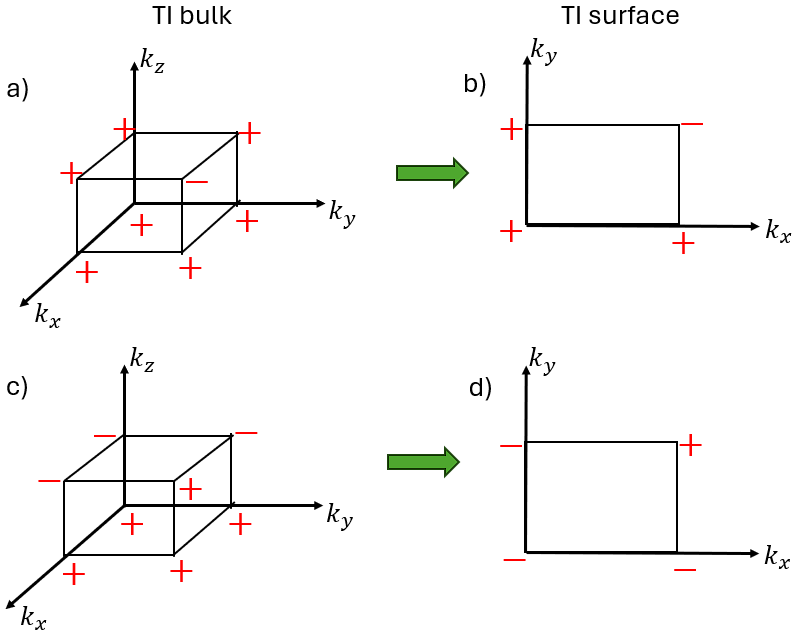}
 \caption{The figure illustrates how flipping the mass signs on one of the planes in the bulk Brillouin zone (BZ) leads to a change in the number of surface Dirac cones without altering the bulk topological invariant. a) Mass signs ($\delta_{\Gamma_i}$) at the eight TRI points in the bulk BZ. The TI phase is $(1;1,1,1)$.  b) Mass signs ($\pi_{\Lambda_i}$) at the four TRI momenta in the surface Brillouin zone in the xy-plane. c) Flipping of mass signs at the $k_z = \pi$ plane for the same bulk topological phase $(1;1,1,1)$. d) Resultant mass signs ($\pi_{\Lambda_i}$) in the xy-plane after flipping the mass signs. (See Eqns.\ref{TI_inv1},\ref{TI_inv2},\ref{srfce_ms} for details).}
 \label{mass_signs}
\end{figure}

Here, we study the bulk topological invariants and examine the underlying physics that leads to distinct surface states for the same topological bulk. We demonstrate that the transition between the distinct surface states is made possible by flipping the fermion mass signs $\delta_{\Gamma_{i}}$'s (defined in Eqns.\ref{TI_inv1},\ref{TI_inv2}) at the TRI points on any one of the three $k_a = \pi (a=x,y,z)$ planes in bulk BZ (see Fig.\ref{mass_signs}).

For a 3d lattice, there are eight TRI points in the Brillouin zone. If the lattice is cubic, then we could represent the TRI momenta simply as $\Gamma_{i =  (n_1,n_2,n_3)} = \pi(n_1 \hat{x} + n_2 \hat{y} + n_3 \hat{z})$ where $n_i$ can be either 0 or 1. From ref.\cite{fukanemele(2007)} the four $Z_2$ invariants $(v_0;v_1,v_2,v_3)$ have the following definition,
\begin{subequations}
\begin{eqnarray}
    (-1)^{v_0} &=& \prod_{n_i = 0,1} \delta_{n_1,n_2,n_3}\label{TI_inv1} \\ (-1)^{v_j} &=& \prod_{n_{(i\neq j)} = 0,1; n_j = 1} \delta_{n_1,n_2,n_3}\label{TI_inv2}
\end{eqnarray}    
\end{subequations}
where $\delta_{n_1, n_2, n_3}$ can be $\pm 1$. For an inversion symmetric system, $\delta_{\Gamma_i}$ is simply the parity eigenvalue of the ground state at the TR-invariant point $\Gamma_i$\cite{fukane(2007)}. 

It is clear from Eqn.\ref{TI_inv1} that the system is topologically non-trivial for an odd number of negative fermion mass signs $\delta_{\Gamma_{i}}$'s; hence, a gapless surface state exists. The other three invariants ($v_1,v_2,v_3$) correspond to the product of $\delta_{\Gamma_i}$'s at the four TR-invariant points on the three distinct $k_a=\pi$ planes (here $a=x,y,z$ respectively for $v_1, v_2, v_3$). 

Now consider the TI surface normal to the $z$-axis. The surface BZ is 2-dimensional with four TR-invariant points defined by $\Lambda_{i=(n_1,n_2)} = \pi (n_1 \hat{x} + n_2 \hat{y})$. To determine whether a Dirac cone exists at the surface TRI point $\Lambda_i$, we should calculate the time-reversal (TR) polarization $\pi_{\Lambda_i}$ defined at the TRI point. It is given by,
 \begin{eqnarray}
     \pi_{\Lambda_i} = \delta_{\Gamma_{i1}}\delta_{\Gamma_{i2}} \label{srfce_ms}
 \end{eqnarray}
where $\Gamma_{i1}$ and $\Gamma_{i2}$ have the same $k_x$ and $k_y$ components as $\Lambda_i$ but differ from each other in their $k_z$ component, which is $0$ for $\Gamma_{i1}$ and $\pi$ for $\Gamma_{i2}$. If $\pi_{\Lambda_i} = -1$, a surface Dirac cone exists at the $\Lambda_i$ point. So the bulk mass must change sign as it travels from $\Gamma_{i1}$ to $\Gamma_{i2}$ for a surface Dirac cone to appear at the $\Lambda_i$ point.

We use this procedure to locate the gapless points on the surface BZ in Fig. \ref{mass_signs}. In Fig. \ref{mass_signs}(a), there is a single negative sign at the $(\pi, \pi, \pi)$ point in the BZ. The bulk topological invariant is $(1;1,1,1)$. Using Eqn.\ref{srfce_ms}, we find that the TR polarization value $\pi_{\Lambda_i} = -1$ at the surface TRI point $\Lambda_i = (\pi, \pi)$ as shown in Fig.\ref{mass_signs}(b).
As a result, there must be a Dirac cone at the $(\pi, \pi)$ point in the surface BZ.

In Fig. \ref{mass_signs}(c), we flipped the bulk mass signs at the four TR points in the bulk BZ's $k_z = \pi$ plane. The four topological invariants did not change under this transformation and remained $(1;1,1,1)$. However, in Fig. \ref{mass_signs}(d), we see that the projected surface BZ has its TR polarization value $\pi_{\Lambda_i}$ negative at three of the four TR invariant points. The TI surface now has three Dirac cones at these three TR invariant points. Note again that the bulk topology has not changed under this transformation.

Thus, deforming the TI surface from a single Dirac cone state to a three Dirac cone state or vice versa is fundamentally connected to the flipping of the sign of $\delta_{\Gamma_i}$'s at the four TRI points on any one of the three distinct $k_a = \pi$($a=x,y,z$) planes in bulk BZ. In the following, we develop a unitary operation that could flip the fermion mass signs at one of the three planes, thereby deforming the surface states.  

\subsection{Unitary deformation between distinct surface states}
Here, we shall develop an explicit unitary operator that can smoothly connect the two distinct surface states of a TI with the same bulk topological invariant. We consider the 3D TI to have inversion symmetry so that $\delta_{\Gamma_i}$ (see Eqn.\ref{TI_inv1}) is just the sign of the mass gap (parity eigenvalue) at the TRI point $\Gamma_{i=(n_1,n_2,n_3)}$.

\subsubsection{Effective field theory approach}

\begin{figure}[b]
 \includegraphics[width=4cm, height=3.5cm]{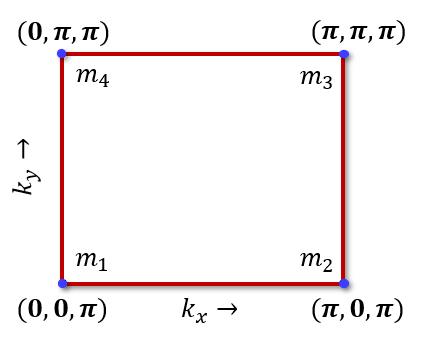}
 \caption{The four TRI points at the $k_z = \pi$ plane in the bulk Brillouin zone (BZ). Also shown are the corresponding mass gaps $m_i(i=1,2,3,4)$ at each of the four TRI points. }
 \label{bulkbz}
\end{figure} 

We use the effective field theory(EFT) approach to derive the unitary operator. Here, we look at the effective low-energy model of the Hamiltonian given in Eqn.\ref{anisotropicHamiltonian}. By expanding to first order in $\textbf{k}$ around all the time-reversal (TR) invariant points in the bulk Brillouin zone (BZ), only the mass terms in the Hamiltonian persist. Since the effective model only includes TR invariant points from the bulk BZ, it becomes simpler to construct a subspace of the BZ spanned solely by the TR invariant momenta. Consequently, the effective Hamiltonian in momentum space reduces to an $8\times8$ matrix with the diagonal entries corresponding to the respective mass gaps. 

To deform the surface state, we require the mass to be flipped only on the $k_z = \pi$ plane as discussed in section.\ref{sec:TIdfm}B. So, to look for possible ways to flip the masses, we construct an effective $4\times4$ Hamiltonian matrix in the subspace spanned by the four TRI points in the $k_z = \pi$ plane (see fig.\ref{bulkbz}). Then, the effective $4\times4$ Hamiltonian has a diagonal representation as follows:
\begin{eqnarray}
    \mathcal{H}_{\text{eff}} = \beta \otimes \begin{bmatrix} m_1 & 0 & 0 & 0 \\ 0 & m_2 & 0 & 0 \\  0 & 0 & m_3 & 0 \\ 0 & 0 & 0 & m_4
    \end{bmatrix} \label{Heff1}
\end{eqnarray}
here $m_1, m_2, m_3, m_4$ are the mass gaps at the TRI points $(0,0,\pi)$, $(\pi,0,\pi)$, $(\pi,\pi,\pi)$ and $(0,\pi,\pi)$ respectively. In this representation, an off-diagonal element, if present, would imply a coupling between the TRI momentum states in the TI bulk.

Now, we shall develop a compact representation for the Hamiltonian to make further computations easier. Let us introduce the Pauli matrices $\Sigma_i (i=x,y,z)$ acting on this subspace. Using this, we shall make the following definitions for the mass matrices $\hat{M}_1$ and $\hat{M}_2$ as,
\begin{subequations}
\begin{eqnarray}
    \hat{M}_1 &=&  \frac{1 + \Sigma_z}{2} m_1 + \frac{1 -\Sigma_z}{2} m_2 \\ \hat{M}_2 &=&  \frac{1 + \Sigma_z}{2} m_3 + \frac{1 -\Sigma_z}{2} m_4
\end{eqnarray} 
\end{subequations}
We define another set of Pauli matrices $\sigma_i (i=x,y,z)$ in this subspace. We can recast the Hamiltonian defined in Eqn.\ref{Heff1} in terms of $\Sigma_i$'s and $\sigma_i$'s as,
\begin{eqnarray}
\mathcal{H}_{\text{eff}} &=& \beta \otimes \left[\frac{ (1 + \sigma_z)}{2} \otimes\hat{M}_1 + \frac{(1 - \sigma_z)}{2}\otimes\hat{M}_2 \right]\label{Heff2}
\end{eqnarray}

We want to construct a unitary transformation that could flip the signs of the four mass gaps without breaking the time-reversal symmetry. To facilitate this, we define the chirality matrix as,
\begin{eqnarray}
    \gamma = -i \alpha_x \alpha_y \alpha_z
\end{eqnarray}
which anti-commutes with $\beta$ matrix but commute with the $\alpha_i$ matrices. Applying $\gamma$ alone will flip the sign of mass gaps, but the unitary operator breaks the time-reversal symmetry(TRS).

One good strategy is to devise a unitary transformation that realizes coupling between different mass points in the bulk Brillouin zone. For instance, consider the following unitary operation,
\begin{equation}
    U_{1}(\lambda) = \text{Exp}\left[i\frac{\lambda \pi}{2} \sigma_y \otimes \gamma  \right]\label{U1}
\end{equation}
Here, $\sigma_y$ signifies a coupling between the mass points separated by the momentum $\textbf{K} = (\pi, \pi, 0)$. Applying the unitary transformation to the Hamiltonian in Eqn.\ref{Heff2}, we see that the signs of the four mass gaps flip when $\lambda$ approaches one.

However, it simultaneously changes their position too. Since the unitary transformation is equivalent to introducing coupling between the points $\textbf{k}$ and $\textbf{k} + \textbf{K}$ points in the BZ(where $\textbf{K} = (\pi, \pi, 0)$), the mass points separated in the BZ by the vector $\textbf{K}$ will get interchanged when $\lambda$ equals one. Explicitly, the Hamiltonian matrix reads,
\begin{eqnarray}
    U^{-1}_{1}(\lambda = 1)\mathcal{H}_{\text{eff}} &U_{1}&(\lambda = 1)\nonumber \\ &=& \beta \otimes \begin{bmatrix} -m_3 & 0 & 0 & 0 \\ 0 & -m_4 & 0 & 0 \\  0 & 0 & -m_1 & 0 \\ 0 & 0 & 0 & -m_2
    \end{bmatrix} \label{U1blkmtrx}
\end{eqnarray}
The interchange of the mass points is evident in the above Hamiltonian matrix. Consequently, this operation results in a change of the bulk topology. This is no surprise as the three weak topological invariants $v_i$($i=1,2,3$) are robust only if the crystal translation symmetry is respected. Here, we broke the translation symmetry by introducing the momentum coupling. So the three invariants can change while keeping the strong topological invariant $v_0$ unchanged.

We fix this by constructing another unitary operator that rotates the mass signs back to their original locations. The unitary operation that can implement this step is,
\begin{eqnarray}
    U_2(\lambda) = \text{Exp}\left[i\frac{\lambda \pi}{2} \sigma_y \otimes I \right]\label{U2}
\end{eqnarray}

Together, the full unitary operator has the following structure,
\begin{eqnarray}
    U(\lambda) = U^{-1}_2(\lambda)U_{1}(\lambda)\label{U_eft}
\end{eqnarray}
The effective Hamiltonian after the unitary rotation to $\lambda = 1$ attains the form,
\begin{eqnarray}
    \mathcal{H}_{\text{eff}}(\lambda = 1) \nonumber &=& U^{-1}(\lambda = 1)\mathcal{H}_{\text{eff}}U(\lambda = 1)\\ &=& \beta \otimes \begin{bmatrix} -m_1 & 0 & 0 & 0 \\ 0 & -m_2 & 0 & 0 \\  0 & 0 & -m_3 & 0 \\ 0 & 0 & 0 & -m_4
    \end{bmatrix}
\end{eqnarray}
So, one can see that the unitary transformation changes the sign of masses at the four Dirac points.

Therefore, to achieve the inversion of mass signs at the four top corners in the Brillouin zone, we initially rotate the effective Hamiltonian with the unitary operator $U_1 (\lambda)$ until $\lambda$ equals one and subsequently rotate it back using $U_2(\lambda)$ to restore the mass points to their initial positions.

Notably, for any intermediate value of $\lambda$, the lattice translation and the inversion symmetries are broken. The breaking of translation symmetry is because we introduced the coupling between the momentum states. Furthermore, since $U_1(\lambda)$ contains the $\gamma$ operator (see Eqn.\ref{U1}), which anti-commutes with the mass term $\beta$, it also breaks the inversion symmetry. However, when $\lambda = 1$, these symmetries are restored. Restoring inversion symmetry is crucial since inversion symmetry is essential for expressing the topological invariants as functions of the parity eigenvalues at each time-reversal invariant (TRI) point.

\subsubsection{Microscopic model}
So far, we have obtained the unitary transformation that could deform the TI surface using the effective field theory.
Here, we devise a unitary operator that realizes the deformation of the TI surface on a microscopic bulk lattice Hamiltonian. Since the unitary evolution involves coupling between momentum states, we shall use the following representation for the bulk Hamiltonian,
\begin{eqnarray}
\mathcal{\hat{H}}_{\textbf{K}}(\textbf{k})  &=& \begin{bmatrix} \mathcal{H}(\textbf{k}) & 0 \\ 0 & \mathcal{H}(\textbf{k} + \textbf{K})
\end{bmatrix} \label{Hmcrscpc} \\ &=& \left(\frac{1 + \sigma_z}{2} \right)\mathcal{H}(\textbf{k}) + \left(\frac{1 - \sigma_z}{2} \right) \mathcal{H}(\textbf{k} + \textbf{K})\nonumber
\end{eqnarray}
where we fix $\textbf{K} = (\pi, \pi, 0)$. We used the Pauli matrices $\sigma_i (i=x,y,z)$  for the representation. The off-diagonal terms, if present, represent a coupling between momentum states $\textbf{k}$ and $\textbf{k} + \textbf{K}$.  

We consider the Hamiltonian given in Eqn.\ref{anisotropicHamiltonian}. We shall take an isotropic model by setting $t_x=t_y=t_z = t$. In the above representation, the bulk Hamiltonian takes the form,
\begin{eqnarray}
    \mathcal{\hat{H}}_{\textbf{K}}(\textbf{k}) &=& \sigma_z \left( \alpha_x \sin k_x + \alpha_y \sin k_y \right) + \alpha_z \sin k_z\nonumber \\ &+& \beta \left(1 - t \cos k_z \right) - \sigma_z \beta \left(t\cos k_x + t\cos k_y \right)\nonumber\\ \label{H0mcrscpc}
\end{eqnarray}

Recall that while deriving the unitary operator using the effective field theory, we reduced the full $8\times8$ mass matrix to a simple $4 \times 4$ matrix by assuming that the unitary evolution affects only the four TRI points in the $k_z = \pi$ plane(see Eqn.\ref{Heff1}). We implicitly assumed this while deriving the unitary operator. However, when we construct a microscopic lattice model for the unitary operator, we must introduce a momentum-dependent function in the unitary operator such that the action of the unitary operator is restricted to the top four mass points. To realize this, we introduce the following Kronecker delta function into the unitary operator,
\begin{eqnarray}
    \phi_{\textbf{k}} = \delta_{k_z, \pi}\label{krnckr}
\end{eqnarray}
This ensures that the unitary operation acts solely on the Hamiltonian's $k_z = \pi$ plane. 

\begin{figure}[b]
 \includegraphics[width=8.5cm, height=5.2cm]{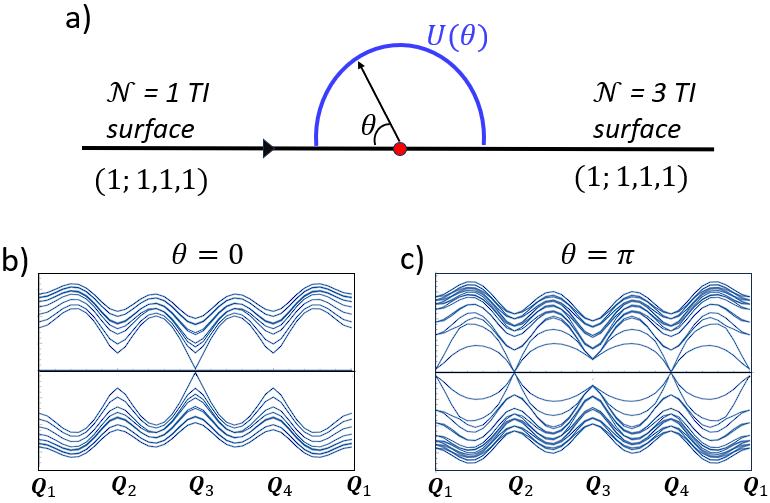}
 \caption{a) Schematic of two different pathways connecting the $\mathcal{N}=1$ and $\mathcal{N}=3$ TI surfaces for the same bulk topological invariant. Along the black line, all symmetries are preserved, but the $\mathcal{N}=1$ surface can be connected to $\mathcal{N}=3$ surface only by a gap closing or via a critical point. On the other hand, along the semi-circle, the $\mathcal{N}=1$ surface can be smoothly connected to the $\mathcal{N}=3$ surface without closing the bulk gap. However, several crystal symmetries are broken. Also shown is the surface spectrum when $\theta = 0$ (b) and $\theta = \pi$ (c) by solving the Hamiltonian in Eqn.\ref{Hklambda} for the bulk topological phase $(1;1,1,1)$. $\textbf{Q}_i$'s ($i=1,2,3,4$) are the TRI points in the surface BZ. (See Fig.\ref{phsdgrm} for the exact definition). } 
\label{dfrmtn_TI}
\end{figure}

However, the resulting Hamiltonian will lead to an infinite-range hopping effect in the bulk lattice. For numerical simulations, it is beneficial to replace the Kronecker delta function in Eqn.\ref{krnckr} with a smooth function, such as a Gaussian function as,
\begin{eqnarray}
    \phi_{\textbf{k}} = \text{Exp}\left[ -\frac{\left(f_{\textbf{k}}-1\right)^2}{2\epsilon}\right]\,\, , \,  f_{\textbf{k}} = \frac{1 - \cos k_z}{2}\label{gaussian}
\end{eqnarray}
with $\epsilon > 0$ being a localization parameter. In Fig.\ref{dfrmtn_TI}, where the surface spectrum before and after the unitary transformation is shown, we set $\epsilon = 0.025$.

We now insert $\phi_{\textbf{k}}$ defined in Eqn.\ref{gaussian} into the unitary operator in Eqn.\ref{U_eft} to restrict the action of unitary evolution to the $k_z = \pi$ plane of the bulk BZ. The modified unitary operator now reads,
\begin{subequations}\label{unitaryoprtr}
\begin{eqnarray}
    U_{1}(k_z, \lambda) &=& \text{Exp}\left[i\frac{\lambda \pi \phi_{\textbf{k}}}{2} \sigma_y \otimes \gamma  \right]\label{U1_ltce}\\  U_2(k_z, \lambda) &=& \text{Exp}\left[i\frac{\lambda \pi \phi_{\textbf{k}}}{2} \sigma_y \otimes I \right]\\ 
    U(k_z, \lambda) &=& U^{-1}_2(k_z, \lambda)U_{1}(k_z, \lambda) \label{U_efl_tce}
\end{eqnarray}    
\end{subequations}

The Hamiltonian in Eqn.\ref{Hmcrscpc} under the unitary operation $U(k_z, \lambda)$ takes the form,
\begin{widetext}
    \begin{eqnarray}
        \hat{\mathcal{H}}_{\textbf{K}}(\textbf{k}, \lambda) &=& U^{-1}(k_z, \lambda) \hat{\mathcal{H}}_{\textbf{K}}(\textbf{k}) U(k_z, \lambda)\nonumber \\ &=& \left[\sigma_z \cos^2 \lambda \pi \phi_{\textbf{k}} + \sigma_z \otimes \gamma \sin^2 \lambda \pi \phi_{\textbf{k}} + \sigma_x \otimes (\gamma - 1)\frac{1}{2} \sin 2\pi\lambda \phi_{\textbf{k}}\right]\left(\alpha_x \sin k_x + \alpha_y \sin k_y \right)\nonumber \\ &+& \alpha_z \sin k_z + \left[\sigma_z \otimes \beta \cos \lambda \pi \phi_{\textbf{k}} + \sigma_x \otimes \beta\gamma \sin \lambda \pi \phi_{\textbf{k}} \right](1 - t \cos k_z)\nonumber \\ &-& \beta\left[\cos \lambda \pi \phi_{\textbf{k}} - \sigma_x \otimes I \sin \lambda \pi \phi_{\textbf{k}} \right] (t\cos k_x + t\cos k_y) \label{Hklambda}
    \end{eqnarray}
\end{widetext}

This is the explicit form of the general Hamiltonian after the unitary transformation for arbitrary $\lambda$. The coefficients of $\sigma_x$ matrix give the amplitude of coupling between the momentum states $\textbf{k}$ and $\textbf{k} + \textbf{K}$. These terms break the crystal translation symmetry. Since $[\beta,\gamma]\neq 0$, the Hamiltonian does not have the inversion symmetry. Hence, the symmetry is lowered for $\lambda \neq 0,1$. In addition, $\phi_{\textbf{k}}$ is a Kronecker delta function. Therefore, the Hamiltonian at any $\lambda \neq 0, 1$ would have infinite-ranged hopping terms.

In short, the unitary operation,
\begin{itemize}
    \item breaks translation symmetry (momentum coupling)
    \item breaks inversion symmetry (due to the $\gamma$ dependent term)
    \item introduces infinite-ranged hopping in the bulk lattice (since the unitary operation is restricted to $k_z = \pi$ plane in the bulk BZ).
\end{itemize}
for any intermediate value of $\lambda$. 

When $\lambda = 1$ and on taking the limit of $\epsilon \rightarrow 0$, the Hamiltonian in Eqn.\ref{Hklambda} becomes,
\begin{eqnarray}
    \mathcal{H}_{\textbf{K}}(\textbf{k}, \lambda = 1) &=& \sigma_z \otimes \left(\alpha_x \sin k_x + \alpha_y \sin k_y \right) + \alpha_z \sin k_z \nonumber\\ &+& \sigma_z \otimes \beta \cos \pi \phi_{\textbf{k}}\left(1 - t \cos k_z \right)\nonumber\\  &-& \beta \cos \pi \phi_{\textbf{k}} \left(t\cos k_x + t\cos k_y \right). \label{Hfnl}
\end{eqnarray}
The terms containing $\sigma_x$ and $\gamma$ vanished in this limit, thus restoring both the inversion and translation symmetry. On the other hand, the infinite range hopping persists since the term $\phi_{\textbf{k}} = \delta_{k_z,\pi}$ is localized in the momentum space.

 Since $\phi_{\textbf{k}} = \delta_{k_z, \pi}$, the mass signs at the four TRI points in the $k_z = \pi$ plane are flipped when $\lambda$ is tuned to unity. Therefore, as evident from the fig.\ref{mass_signs}, if the Hamiltonian before the unitary transformation is in a strong TI phase with a single Dirac cone, then after the unitary transformation, we arrive at the strong TI phase of the same set of topological invariants but with three Dirac cones without closing the gap.

Fig.\ref{dfrmtn_TI} shows a schematic picture of this smooth connection between the two surface states and the corresponding surface spectrum before and after the deformation is applied. 

Recall that the unitary operation $U_{1}(k_z,\lambda)$ flips the mass signs in the $k_z=\pi$ plane; however, it also interchanges the mass points separated by momentum $\textbf{K}$ (see Eqn.\ref{U1blkmtrx}). As a result, the unitary operation changes the bulk topological invariants, even though it deforms a single Dirac cone surface state to three Dirac cone surface states (see Fig.\ref{dfrmtn_TIprt1}). Details are moved to the Appendix.\ref{appndxuntry1}. Just like $U(k_z,\lambda)$,  $U_1$ is also a non-local unitary operation and breaks the translation and inversion symmetries of the bulk for intermediate values of $\lambda$. 

Here, we proposed two non-local unitary operations that could connect distinct surface states but without breaking the protecting symmetry. While the first one $U_{1}(k_z,\lambda)$(Eqn.\ref{U1_ltce}) changes the bulk topological invariants, the final unitary $U (k_z,\lambda)$(Eqn.\ref{U_efl_tce}) connects distinct surface states without changing the topological bulk. Thus, we showed that an adiabatic path exists that could connect distinct surface states without closing the bulk gap and without breaking the protecting symmetry. However, the cost of introducing such a unitary deformation is that it lowers the symmetry of the system. The translation and the inversion symmetry are broken for intermediate values of the parameter $\lambda$ between $0$ and $1$. In addition, the unitary operation is non-local in the sense that it introduces infinite-range hopping in the bulk lattice.


A practical realization of such a non-local unitary path would be challenging. Therefore, we argue that the number of surface fermion flavors, $\mathcal{N}$, can be stabilized by imposing crystal symmetry constraints. This gives us the confidence to study the effect of strong interactions on the TI surface with $\mathcal{N}$ flavors of surface fermions. Unless the interactions break the symmetries mentioned above or introduce infinite-range hopping, the number of flavors $\mathcal{N}$ should remain topologically stable.

Although we have demonstrated this point in the case of a strong topological insulator (TI), the same principles also apply to a weak TI. Consider the Hamiltonian in Eqn.\ref{anisotropicHamiltonian} representing a weak TI phase with two Dirac cones ($\mathcal{N}=2$). Upon applying the unitary operation $U(\lambda)$, we observe that at $\lambda = 1$, the TI surface retains two Dirac cones. Yet, they are located at complementary points in the surface Brillouin zone (BZ). For instance, if the Dirac cones were at $(0,0)$ and $(\pi,0)$ for $\lambda = 0$, then after applying the unitary deformation at $\lambda = 1$, the Dirac cones would be found at $(\pi,\pi)$ and $(0,\pi)$. For this reason, the locations of the Dirac cones before and after the transformation are complementary.


\section{\label{sec:RG} Interacting dynamics of the TI surface with \texorpdfstring{$\mathcal{N}$}{}-flavors of surface Dirac fermions}

In the previous section, we have seen that a surface with $\mathcal{N}$ (where $\mathcal{N} \leq 4$) Dirac cones can be stabilized if we impose crystal symmetry constraints on the system. 
This is despite the fact that the surface states with distinct values of $\mathcal{N}$ could have an identical bulk topological invariant (see Fig.\ref{phsdgrm}). In this context, we introduce strong attractive interactions to the TI surface with $\mathcal{N}$ flavors of surface fermions. Our ultimate goal is to identify various conformal-field-theory states and classify universality families of surface topological quantum criticality in a strongly interacting TI surface.

\subsection{Uniqueness of the gapless surface states}
Before introducing interactions, we emphasize a property of the gapless surface fermions that distinguishes them from the gapless Dirac fermions in a 2D lattice. To begin with, let us define a 'pseudo-chirality' operator as
\begin{eqnarray}
\Gamma_s &=&\beta\alpha_z, \,\,\,   \Gamma^2_s = 1; \nonumber \\
\left[\Gamma_{s},\alpha_x\right]&=&[\Gamma_s, \alpha_y]=0.
\end{eqnarray}
In the $\alpha-\beta$  representation introduced in this article, $\Gamma_s = \tau_y s_z$,
where $\tau_y$ is the Pauli matrix for the orbital part and $s_z$ is the Pauli matrix for spins (see Eqn.\ref{anisotropicHamiltonian}). 

The structure of the $\Gamma_s$ operator suggests that the eigenvalue can only be $\pm 1$. We call this term 'pseudo-chirality' because the conventional chirality operator is usually not defined for even space dimensions. The important message here is that the surface fermions of a 3d TI must be $eigenstates$ of this 'pseudo-chirality' operator. Since a strong TI has an odd number of Dirac cones, there would be an odd number of eigenstates for a specific eigenvalue of $\Gamma_s$ operator at a given surface. The majority of the eigenstates would have a specific pseudo-chirality so that the TI surface can be associated with that specific pseudo-chirality value. The location of the surface TRI point at which the Dirac cone is defined is irrelevant here. 

This is rather an intrinsic property of the topological insulator, disregarding the detailed structure of the surface Brillouin zone. Now, let us compare it with a gapless Dirac fermion in a bulk 2d lattice. Each Dirac point will always have both positive and negative pseudo-chiralities. Therefore, a 2d lattice would always have an equal number of eigenstates for each eigenvalue of the $\Gamma_s$ operator. So a bulk 2d lattice cannot be associated with a specific chirality. This is an intrinsic property of lattices due to the well-known fermion-doubling theory\cite{Nielson(1981),Nielson(1981)(2)}. 

For a weak TI, there is an even number of surface Dirac cones ($\mathcal{N}=2$). In general, there could be an equal number of eigenstates of positive and negative pseudo-chiralities. However, each Dirac cone is uniquely associated with a specific pseudo-chirality. From this perspective, the TI surface fundamentally differs from a bulk 2D lattice when considering individual Dirac cones. However, it is also possible that all (or a majority of) the surface Dirac cones share the same pseudo-chirality eigenvalue, giving the surface a definite pseudo-chirality. Therefore, one can assert that the TI surface is always $distinct$ from the bulk 2d lattice, even if the state is a weak TI.

 Given this unique property of the topological surface fermions, it would be illuminating to study the effect of interactions between them which would be fundamentally different from the case of a 2d lattice.
 Here, we shall look at the dynamics of the surface subject to strong, attractive interactions. We are mainly interested in the singlet-pairing interactions that could result in a phase transition to a superconductor. We shall construct an effective 2-dimensional interaction action that could capture the essential low-energy dynamics of the system. We then apply the renormalization group analysis to study the effect of these four-fermion interactions on the TI surface. Since the four-fermion interaction is marginal in $1+1D$, we perform a $(2+\epsilon)$ expansion to study the RGE flow. In the coming sections, we explore the fixed point manifolds in the case of $\mathcal{N}=2$ followed by $\mathcal{N}=3$.

\subsection{Effective 2d interacting model}

In section \ref{sec:TIdfm}, we started with a 3d Dirac lattice model suitable for a cubic lattice and solved it numerically to determine the surface states. 
However, since we want to study the effects of interactions on the TI surface, we need to better understand the analytical form of the surface states, at least near the corresponding TRI points. 
As discussed at the beginning of this section, we cannot simply start with a 2d lattice model to discuss surface fermions. 

Here, we begin with a bulk 3d Hamiltonian, expand around those TRI points to first order in crystal momentum where the sign of mass gap is negative, and solve for the surface fermions analytically. We then end up with a 2-component spinor state localized in the direction normal to the surface. This is equivalent to the domain-wall construction typically employed in field theory to avoid the fermion-doubling problem\cite{fradkin(1987),boyanovsky(1987),kaplan(1992),jansen(1996),Nielson(1981),Nielson(1981)(2)}. The details of the derivation are moved to the appendix \ref{EFT_deriv}.

\begin{figure}[b]
 \includegraphics[width=8.5cm, height=6.8cm]{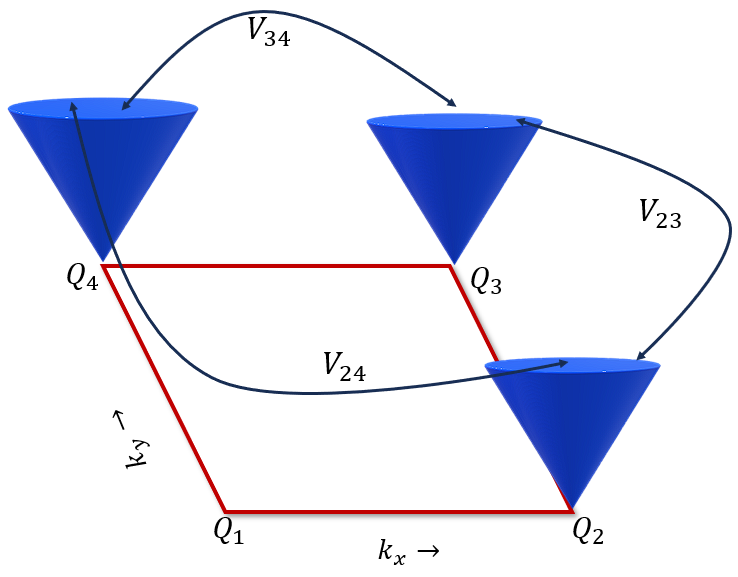} \\
 \caption{Schematic picture of a $\mathcal{N}=3$ flavor TI surface with Dirac cones present at the TRI points $\textbf{Q}_2$, $\textbf{Q}_3$ and $\textbf{Q}_4$ (See Fig.\ref{phsdgrm} for their exact definition). The diagram further depicts several scatterings of Cooper pairs between Dirac cones, resulting from the attractive pairing interactions among the surface fermions. Here $V_{ij}$ denotes the matrix element of interaction between the $i$th and the $j$th Dirac cone (see Eqn.\ref{Lfrfmn}). }
 \label{frfrmnintrc}
\end{figure}

Now we are ready to formulate the effective 2D low-energy theory for the TI surface. As discussed in the previous section, the surface BZ contains four TRI points, allowing the surface to host up to four ($\mathcal{N} \leq 4$) Dirac cones at these TRI points, depending on the bulk topology. We assign a flavor index $'i'$ ($i=1,2,3,4$) to differentiate them. A surface fermion of flavor $'i'$ belongs to the Dirac cone at the TRI point $\textbf{Q}_i$ (refer to Fig.\ref{frfrmnintrc} for $\textbf{Q}_i$'s definition).

Utilizing this framework, we present the Lagrangian governing the free-fermion dynamics on the TI surface.
\begin{eqnarray}
    \mathcal{L}_{f} &=& \sum_{i=1}^{4}\varepsilon_i \psi^{\dagger}_{i} \left[ \partial_\tau + \hat{h}_{i} \right]\psi_{i}\label{Lfree}\\ \text{where} \,\, \hat{h}_{1} &=& -i s_x \partial_{y} + i s_y \partial_{x}\, ,\,\,\, \hat{h}_{2} = -i s_x \partial_{y} - i s_y \partial_{x}\nonumber \\ \hat{h}_{3} &=& i s_x \partial_{y} - i s_y \partial_{x}\,\, ,\,\,\,\, \hat{h}_{4} = i s_x \partial_{y} + i s_y \partial_{x} \nonumber \\ 
    \text{and}\,\, \varepsilon_i &=& \begin{cases}
        1, \,\,\, \text{Dirac cone present at } Q_i \\ 0 \,\,\, \text{No Dirac cone at } Q_i \nonumber
    \end{cases}
\end{eqnarray}
where $\hat{h}_i$ could effectively describe the free-fermion dynamics near the TRI point $\textbf{Q}_i$. Here $\psi_{i}$ is a 2D 2-component fermion field operator with the definition $\psi^{\dagger}_{i} = (\psi^{\dagger}_{i,\uparrow} \,\,\, \psi^{\dagger}_{i,\downarrow})$ that represents the low-energy excitations near the TRI point $\textbf{Q}_i$. 
Here we inserted a term $\varepsilon_i$ artificially to take into account the presence or absence of the Dirac cone at the TRI point $\textbf{Q}_i$. 

Whether a Dirac cone is present or absent at the given TRI point is solely determined by the TI bulk, as we saw in detail in section \ref{sec:TIdfm}.
Therefore, one must start with the bulk topological phase and find out the precise value of $\varepsilon_i, (i=1,2,3,4)$, and the locations of Dirac cones in the surface BZ to construct the effective surface theory.

Let's consider a scenario where a net attractive interaction exists between electrons in a topological insulator. 
We'll assume that the ultraviolet (UV) cut-off of this interaction is smaller than the bulk mass gaps at all eight time-reversal invariant (TRI) points, allowing us to treat the bulk fermions as non-interacting. 
We further assume that the pairing is of singlet order, intra-orbital, and is between time-reversal Kramer doublets. We start with a 3d interacting theory for the domain-wall fermions. We then integrate out the degrees of freedom normal to the surface to arrive at an effective 2d interacting theory\cite{saran(2022)}. The details of the derivation are presented in the appendix \ref{EFT_deriv} for brevity.

Finally, we arrive at the following interaction Lagrangian,
 \begin{eqnarray}
\mathcal{L}_{\text{I}} &=& -\sum^{4}_{i,j = 1} \varepsilon_i\varepsilon_j V_{ij}\,\,\psi^{\dagger}_{i}(\textbf{r},\tau) s_y \psi^{\dagger T}_{i}(\textbf{r},\tau) \psi^{T}_{j}(\textbf{r},\tau)s_y \psi_{j} (\textbf{r},\tau) \nonumber \\ \label{Lfrfmn}
\end{eqnarray}
where,
\begin{eqnarray}
    V_{ij} = V \frac{m_i m_j}{m_i + m_j}. \nonumber
\end{eqnarray}

Here, $V$ is the attractive interaction strength in the 3D bulk. It is non-zero within the energy scale $[-\Lambda, \Lambda]$, where $\Lambda$ is the UV cut-off scale of the interaction. Here $i,j$ runs over the surface fermion flavors, subject to their presence. $m_{i}$ and $m_j$ are the bulk mass gaps of the Dirac cones between which the scattering of the singlet pair of fermions occurs due to effective attractive interaction. In the effective theory, the singlet pair of fermions from a surface Dirac cone can scatter to either the same or to another Dirac cone (if present), the strength of which is determined by the respective bulk mass gaps assigned to each of the Dirac cones. See Fig.\ref{frfrmnintrc} for a schematic picture of the scattering.

\subsection{Symmetries of the free surface fermions}

Let us now study the symmetries of the non-interacting part of the theory defined in Eqn.\ref{Lfrfmn}. A surface with $\mathcal{N}$ Dirac cones can be pictured as a theory of $\mathcal{N}$ distinct flavors of 2-component Dirac fermions. We could define an internal flavor space of $\mathcal{N}$-dimensions. In this flavor space, we define a $\mathcal{N}$ component spinor as $\Psi^{\dagger} = \left(\begin{array}{cccc}
\psi^{\dagger}_1 & \psi^{\dagger}_2 & \psi^{\dagger}_3 & ... 
\end{array}\right)$. Here $\psi^{\dagger}_i (i=1,2,..\mathcal{N})$ is a 2-component spinor (see Eqn.\ref{Lfree}). In this section, we shall show that the free fermions can be $invariant$ under the $SU(\mathcal{N})$ rotations of the Dirac spinor $\Psi$ in the flavor space if a proper spinor-base is chosen.

To begin with, we notice here that the free electron dynamics at different Dirac cones in a TI are usually not identical. The helicity of the Dirac cone at each of the TRI points differs. Here we define helicity as the orientation of the fermion spin relative to its linear momentum. For instance, the expectation value of the spin operator for the surface fermions at the TRI point $\textbf{Q}_1$ reads as $ \bra{\psi_1}  \hat{\textbf{s}} \ket{\psi_1} = \frac{1}{|\textbf{k}|} \left(k_y \hat{x} - k_x \hat{y}\right)$, while for the Dirac cone at $\textbf{Q}_2$, we see that $\bra{\psi_2} \hat{\textbf{s}} \ket{\psi_2} = \frac{1}{|\textbf{k}|} \left(k_y \hat{x} + k_x \hat{y}\right)$ in the momentum space.

However, they are all related by a mirror reflection.
Therefore, the helicity of all the Dirac cones could be made equivalent to each other by applying a simple $SU(2)$ $\pi$-rotation (clockwise or anticlockwise) in the spinor space at each Dirac cone. After such a helicity transformation, 
$SU(\mathcal{N})$ flavor symmetry naturally arises in free fermions.

Consider a Dirac cone at the surface TRI point $\textbf{Q}_i$ (see Fig.\ref{frfrmnintrc} for its definition). We look for a possible $SU(2)$ rotation in the spinor space such that the helicity of the Dirac cone at $\textbf{Q}_i, i=1,2,3,4$ is made equal to $\textbf{Q}_1$. Let $U_i$ be the unitary matrix defined for the Dirac cone at $\textbf{Q}_i$, which realizes this transformation. $U_i$'s would have the following definitions,
   \begin{eqnarray}
       U_1 &=& \hat{I}, \,\,\, U_2 = i s_x, \,\,\,  
       U_3 = i s_z, \,\,\, U_4 = i s_y
   \end{eqnarray}

 In general, a unitary matrix $U_i$, $i=2,3,4$ or $i s_{\alpha} (\alpha = x,y,z)$ realizes an $SU(2)$ spinor rotation about the $\alpha$-direction by an angle $\pi$  for a surface fermion at the TRI point $\textbf{Q}_i$. Rotating each flavor of the surface fermions independently as $\psi_i \rightarrow U_i \psi_i$, the free-fermion dynamics of all the Dirac cones appear to be identical.
   
   Now we shall examine the interaction term defined in Eqn.\ref{Lfrfmn} under this helicity transformation. For this purpose, it would suffice to study how the singlet pair annihilation operator $\psi^{T}_i is_y \psi_i$ transforms under helicity transformation. 
   \begin{eqnarray}
       \psi^{T}_i is_y \psi_i &\rightarrow& 
       \psi^{T}_i U^{T}_{i} is_y U_i \psi_i \nonumber\\ &=& \psi^{T}_i is_y  \psi_i \nonumber
   \end{eqnarray}
We find that the pair annihilation operator is always {\em invariant} under the helicity transformation for all $U_i$. Hence the effective action defined by Eqns.\ref{Lfrfmn} is invariant under the helicity transformation. 
To study the effect of these interactions, we find it most convenient to work with a basis where different Dirac cones have the same helicity and appear to be identical.

   As a result,  we express the free theory of surface fermions in Eqn.\ref{Lfree} in terms of the $\mathcal{N}$-component spinor field $\Psi(\textbf{x})$ as,
   \begin{eqnarray}
       \mathcal{L}_{f} &=& \Psi^{\dagger} \left[ \left(\partial_{\tau} + \hat{h}_1 \right) \otimes \hat{I}_{\mathcal{N}} \right] \Psi \label{LfreeNflvr}
   \end{eqnarray}
   Here $\hat{I}_{\mathcal{N}}$ is an identity matrix in the flavor space.  $\hat{h}_{1}$ is defined in Eqn.\ref{Lfree}. Also, notice that we haven't used the term $\varepsilon_i$ here. This is because the helicity invariance implies the location of the surface Dirac cones is not relevant to the theory. Only the number of surface Dirac cones $\mathcal{N}$ matters.
   
   One can see that the free theory of surface fermions in Eqn.\ref{LfreeNflvr} is invariant under $SU(\mathcal{N})$ flavor rotations of the Dirac spinor. The single-particle Green's function in the flavor space has the simple form of 
   \begin{eqnarray}
       G(\textbf{k}, i\omega_n) &=& \frac{1}{i\omega_n - \left(s_x k_y - s_y k_x \right)} \otimes \hat{I}_{\mathcal{N}}.
       \label{eq:GF}
   \end{eqnarray}
The Green's function here is proportional to the identity operator in the flavor space, i.e. it is an $SU(\mathcal{N})$ singlet. 
However, the interaction part in Eqns.\ref{Lfrfmn} does not have such an $SU(\mathcal{N})$ flavor symmetry.

\subsection{Symmetries of the interaction matrix $V_{ij}$}

As noticed before in Eq.\ref{eq:GF} , the non-interacting fermion Green's functions are $SU(\mathcal{N})$ invariant. The same can be said about products of multiple Green's functions that don't involve flavor symmetry-breaking terms. These properties of Green's functions play an
important role in the discussions of the renormalization group equations (RGEs) and their solutions.
Following the free theory symmetry above, it turns out that the RGE (see next section) can have an emergent $SO(\mathcal{N})$ symmetry, i.e. RGEs are invariant under an $SO(\mathcal{N})$ flavor rotation of the interaction matrix $V_{ij}, i,j=1,...,\mathcal{N}$
 which is real and Hermitian. That is $V_{ij} =V_{ji}=V^*_{ij}$.


 
 We emphasize here that the generic structure of $V_{ij}$ for surface fermions of the topological bulk always breaks the $SO(\mathcal{N})$ flavor symmetry. That can be explicitly seen in the Eqn.\ref{Lfrfmn} specifically suitable for phonon-mediated interactions(\cite{saran(2022)}). At least two mechanisms lead to the breaking of $SO(\mathcal{N})$ symmetry. One is the surface Dirac cone of different flavors that have flavor-dependent localization lengths or are induced by different bulk mass gaps $m_i$. That leads to differences among the diagonal components of $V_{ij}$. The second mechanism is that the bulk interactions with gaped bulk fermions can induce interactions between different flavors of surface fermions. This results in non-zero off-diagonal components of $V_{ij}$. These will effectively break the symmetry of interactions in the flavor space.

However, we shall see in the next subsection that $SO(\mathcal{N})$ flavor rotation symmetry appears as an emergent symmetry in the renormalization group equations (RGEs) when the renormalization effect of interactions is studied up to the one-loop level. This emergent $SO(\mathcal{N})$ symmetry in the RGEs plays a crucial role in forming conformal manifolds in the parameter space, which we shall explore in subsequent sections.

\subsection{Renormalization group study of the effective interacting model for a general \texorpdfstring{$\mathcal{N}$}{}: \texorpdfstring{$2+\epsilon$}{} expansion}

\begin{figure}[b]
 \includegraphics[width=7.5cm, height=9.35cm]{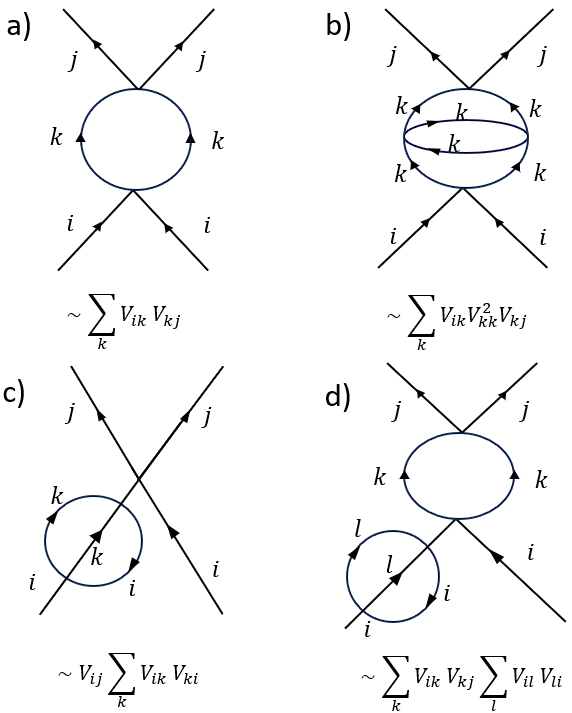}
 \caption{\label{fnmndgrms} One-loop (a) and three-loop (b)  one-particle irreducible diagrams that contribute to the RGE of the interaction matrix element $V_{ij}$(see Eqn.\ref{Lfrfmn}). (c) The two- and (d) three-loop contributions involve the fermion field renormalization effect. The indices $i,j,k$ are the flavor indices.}
\end{figure}

So far, we have developed an effective interacting model for the TI surface. We are now ready to study the surface quantum criticality using the renormalization-group-equation (RGE) technique. In Eqn.\ref{Lfrfmn}, we have an interacting theory of $\mathcal{N}$ flavors of 2-component Dirac fermions in $(2+1)D$. Dimensional analysis suggests that the interaction term in Eqn.\ref{Lfrfmn} is marginal in $1+1$D dimensions. So in $(2+1)D$, the interaction term is irrelevant in the infrared (IR) limit if the interaction is weak. However, in the case of $\mathcal{N} = 1$, it had been demonstrated that the interacting theory has an interacting UV fixed point \cite{kopnin(2008),castro(2009),mudry(2010),metzner(2010),imada(2011),imada(2012),sondhi(2013)}. Here, we study surface quantum criticality phenomena in a more general case with $\mathcal{N} > 1$ surface Dirac cones and search for strong coupling conformal-field-theory fixed points that shall define possible universality classes of surface topological quantum criticality for a given $\mathcal{N}$.

Fig.\ref{fnmndgrms} illustrates the diagrams contributing to the renormalization of the interaction matrix element $V_{ij}$. As pointed out in the previous subsection, the effective field theory is not invariant under the $SO(\mathcal{N})$ flavor rotation. Nevertheless, we find here that there can be an emergent dynamic symmetry of $SO(\mathcal{N})$ in RGEs. For instance,
the contribution from the one-loop diagram (Fig.\ref{fnmndgrms}(a)), which is proportional to $\sum_{k} V_{ik} V_{kj}$, transforms as a rank-2 tensor when $V_{ij}$ is rotated in the flavor space. 

The two-loop contributions are due to the fermion line renormalization, shown in Fig.\ref{fnmndgrms}(c). Its contribution to the RGE flow of the matrix element $V_{ij}$ does not behave as a rank-2 tensor
as it has the form of $V_{ij} \sum V_{ik} V_{ki}$. In our model, the fermion field renormalization has a flavor dependence. Hence, the resulting RGE with these two-loop contributions does not exhibit an $emergent$ $SO(\mathcal{N})$ symmetry.

Furthermore, the contribution from the three-loop diagram (Fig.\ref{fnmndgrms}(b)) which is proportional to $\sum_{k} V_{ik} V^2_{kk}V_{kj}$ also does not transform as a tensor of any rank. So the emergent $SO(\mathcal{N})$ symmetry is broken only if we include the two-loop and other higher-loop contributions.
One shall also notice that the contributions from Fig.\ref{fnmndgrms} (a) and (b) are the typical one-particle irreducible diagrams while (c) and (d) both are related to the standard fermion-field renormalization effect.
The ones that are due to the field renormalization shown here do break the emergent $SO(\mathcal{N})$ symmetry.

The renormalization group study is performed in the $2+\epsilon$ space-time dimensions. Restricting ourselves to the one-loop order in the spirit of $\epsilon$ expansion, we can anticipate an emergent $SO(\mathcal{N})$ symmetry in the RG equations. 
Then we obtain the following RG equation\cite{Shankar(2017),justinbook(2002),fradkinbook(2019)}, 
\begin{eqnarray}
 \frac{d V_{ij}}{d l} = - \epsilon V_{ij} + \frac{1}{c^2}\sum_{k} V_{ik} V_{kj}\label{RGE1}
\end{eqnarray}
where $\epsilon = D - 2$ and $D$ is the spacetime dimensions.  Note that by increasing $l$, we are moving towards the IR limit of the theory. The numerical constant $c^2 = \frac{\pi}{8}$. $V_{ij}$ is the renormalized dimensionless interaction matrix element. The derivation details of Eqn.\ref{RGE1} are moved to the appendix.\ref{appndxRGE} for brevity.

As discussed before, limiting our calculation to one loop allows us to observe an emergent $SO(\mathcal{N})$ symmetry in the RG equation. Hence, it is possible to write the RG equation as a matrix differential equation. We could express $V_{ij}$ as an $\mathcal{N}\times \mathcal{N}$ dimensional symmetric matrix $\hat{V}$. The second term in the RHS is simply the square of this matrix. Let us also absorb the coefficient $c^2$ into the interaction strength by rescaling $\hat{V} \rightarrow \frac{\hat{V}}{c^2}$. So we could re-write this RG equation as,
\begin{eqnarray}
    \frac{d \hat{V}}{d l} = - \epsilon\hat{V} +  \hat{V}^2 \label{RGeqn} 
\end{eqnarray}
Here, we have expressed the interaction strength as an $\mathcal{N}\times \mathcal{N}$ symmetric matrix. 

One can easily verify that Eq.\ref{RGeqn} has an emergent $SO(\mathcal{N})$ symmetry. That is, the RGEs are invariant under the following transformation.

\begin{eqnarray}
\hat{V}(l) \rightarrow \hat{R}^{-1} \hat{V}(l) \hat{R}
\label{ST}
\end{eqnarray}  
if $\hat{R}$ is in an $SO(\mathcal{N})$ rotation group.

This dynamic symmetry of RGEs further implies that for any solution $\hat{V}_c(l)$ that is not an $SO(\mathcal{N})$ singlet, one can generate a smooth manifold of solutions by simply applying an $SO(\mathcal{N})$ rotation.
This plays an important role in understanding the conformal manifolds below.

Note that the $SO(\mathcal{N})$ symmetry of the RG equation in Eqn.\ref{RGeqn} is $emergent$ and not fundamental. The theory does not have flavor rotation symmetry. The RG equation has this symmetry because we limited our calculations to one-loop contributions. A higher-loop contribution explicitly breaks the flavor rotation symmetry, as evident from the two- or three-loop diagrams shown in Fig.\ref{fnmndgrms} (b), (c), and (d).

\begin{table*}
\caption{\label{tableN2}Table showing the list of fixed points, the shape of the manifold in the parameter space, and their stabilities for the $\mathcal{N} = 2$ flavor surface. Here $\phi \in [0,2\pi]$ and $\epsilon = D - 2$, where $D$ stands for spacetime dimensions.}
\begin{ruledtabular}
\begin{tabular}{cccc}
 Isolated fixed points/Manifolds  & Stability & Invariant subgroup($H$) & Manifold\\
 \hline
  $\hat{V}_c = 0 \hat{I}$   & Three irrelevant operators. Stable fixed point &$SO(2)$ & Point  \\
 $\hat{V}_c = \frac{\epsilon}{2} \hat{I} + \frac{\epsilon}{2} \cos \phi \, \sigma_z + \frac{\epsilon}{2} \sin \phi \, \sigma_x$  & One relevant, one irrelevant and one marginal operator.
 &$\hat{I}$ & ring  \\
 $\hat{V}_c =  \epsilon\hat{I} $ & Three relevant directions. Unstable fixed point. &$SO(2)$ & Point
 \end{tabular}
\end{ruledtabular}
\end{table*}

\section{\label{sec:SSB} Symmetries of the RGE and its dynamical solutions}
Before we begin discussing the fixed points and their properties for different $\mathcal{N}$, here we present a fascinating observation from our study on the fixed points and the formation of conformal manifolds. Whether a set of fixed points forms a smooth manifold is conveniently and intuitively related to the dynamical symmetry of the RG equation and the invariant subgroup of the fixed point coupling matrix. In this section, we shall put forward the general argument. In the upcoming sections, where we solve for the fixed points of the RG Eqn.\ref{RGeqn} for $\mathcal{N}=2$ and $\mathcal{N}=3$ flavor surfaces, we shall show explicitly that this general argument is followed in all cases. 

The argument follows: "Let $G$ be the symmetry group of the RG equation. Let the interaction matrix at the fixed point be invariant under a subgroup $H$ of $G$. Then, we find that the space of critical points formed in the parameter space has a group theoretic interpretation as the coset space,
\begin{eqnarray}
    C = \frac{G}{H}
\end{eqnarray}
The dimension of the coset manifold formed is given by,
\begin{eqnarray}
     \text{dimension of manifold} = \text{dim}(G) - \text{dim}(H) " 
\end{eqnarray}

If the dimension of the coset space is greater than zero, a smooth conformal manifold is formed in the parameter space. This symmetry argument proves to be very useful in understanding the shape and dimensions of the conformal manifolds if it exists. We shall make explicit use of this theory to solve the fixed points of the RG equation (Eqn.\ref{RGeqn}) in the case of $\mathcal{N}=3$ TI surface, where the parameter space is six-dimensional and a direct brute-force calculation is challenging. 

Interestingly, we shall find in section.\ref{sec:hghloop} that the symmetry of the dynamical equation is $not$ a necessary condition for the existence of conformal manifolds. When considering the 3-loop effects in the RG equation, we find that the $SO(\mathcal{N})$ symmetry is broken (see Fig.\ref{fnmndgrms}). However, the conformal manifold persists, although in a distorted shape.

\section{\label{secN2}
Conformal manifolds and phase boundary in the $\mathcal{N}=2$ flavor TI surface}
Here, we shall look at the fixed points of the theory for the $\mathcal{N} = 2$ flavor TI surface subject to attractive interactions. The two Dirac cones could be at any two TRI points out of the four available ones on the surface BZ. However, the exact location does not matter since the effective interacting theory of Eqn.\ref{frfrmnintrc} is invariant under helicity transformations. The $\mathcal{N} = 2$ flavor surface is a weak TI phase, which means that even TR invariant perturbations can gap the surface. However, we will consider the TI to be in the clean limit, allowing us to examine the dynamics of the gapless surface state. 

The main objective here is to study the isolated fixed points and conformal manifolds of the interacting theory in the three-dimensional parameter space ($D_p = 3$).  Once the fixed point manifolds are found, we intend to demonstrate how the conformal manifolds or isolated fixed points can be represented as a coset space $G/H$, as outlined in section.\ref{sec:SSB}. Following this, we shall look for the phase boundary that separates the gapless surface state from the superconducting state. Finally, we show that the infrared physics at the phase boundary can be conveniently studied via an effective Yukawa field theory that involves a single complex boson strongly interacting with two flavors of gapless fermions.



The interaction matrix $\hat{V}$ defined in Eqn.\ref{RGeqn} is a $2\times2$ symmetric matrix. Thus, the RG equation possesses emergent $SO(2)$ symmetry. Given that $\hat{V}$ is a real symmetric matrix, we could re-parametrize the $\hat{V}$-matrix in the following simple way:
\begin{eqnarray}
    \hat{V} &=& V_0 \hat{I} + V_z \sigma_z + V_x \sigma_x \label{N2dcmp1}
\end{eqnarray}
where $\sigma_x$ and $\sigma_z$ are Pauli matrices defined in the parameter space. This decomposition makes sense because a real symmetric matrix of size $2$ has $3$ independent parameters, which is the case here. Physically, $V_x$ signifies the scattering of Cooper pairs between the surface fermions of different flavors. $V_z$ represents the difference in the intra-flavor interactions of surface fermions. Plugging this back to the RG equation in Eqn.\ref{RGeqn}, we get, 
\begin{eqnarray}
    \frac{dV_0}{dl} &=& - \epsilon V_0 + V^2_0 + V^2_x + V^2_z\nonumber \\ \frac{dV_x}{dl} &=& V_x \left(-\epsilon + 2 V_0 \right)\nonumber \\ \frac{dV_z}{dl} &=& V_z \left(-\epsilon + 2 V_0 \right)
    \label{RGE3}
\end{eqnarray}

Following the general discussions about the $SO(\mathcal{N})$ symmetry of Eq.\ref{RGeqn}, \ref{ST}, one can easily show that the RG flow lines exhibit an axial $SO(2)$ rotation symmetry.
This is explicitly evident from the invariance of the RGE(given in Eqn.\ref{RGE3}) under $\hat{R}(\chi)$,  $SO(2)$ rotations about the $V_0$ axis in the $V_0-V_x-V_y$ representation introduced in Eq.\ref{N2dcmp1},

\begin{eqnarray}
\hat{R}(\chi)=\exp(i  \frac{\chi}{2} \sigma_y), \chi \in [0,2\pi].
\end{eqnarray}
Another important consequence of the $SO(2)$ invariance is that the scaling dimensions of the operators are the same for all points on the ring manifold (see Appendix.\ref{apndxsclngdmnsns}).

\subsection{ Isolated fixed points Vs Conformal manifolds}

We find the fixed points by setting the beta function in Eq.\ref{RGE3} to zero. We have listed all the possible fixed points and their properties in the table \ref{tableN2}. The non-interacting or the trivial fixed point is the stable one. It is expected because the four-fermion interaction is marginal in $ 1+1D$ dimensions. 
On the other hand, the strong-coupling fixed point $\hat{V}_c = \epsilon \hat{I}$ is unstable in the infrared limit. Both the Dirac cones are strongly interacting but independently. Inter-cone scattering is not there, as evident from $V^c_x = 0$. If one needs to study this fixed point dynamics, all three interacting parameters need to be fine-tuned to this point.

\begin{figure}[b]
\includegraphics[width=5.5cm, height=7.10cm]{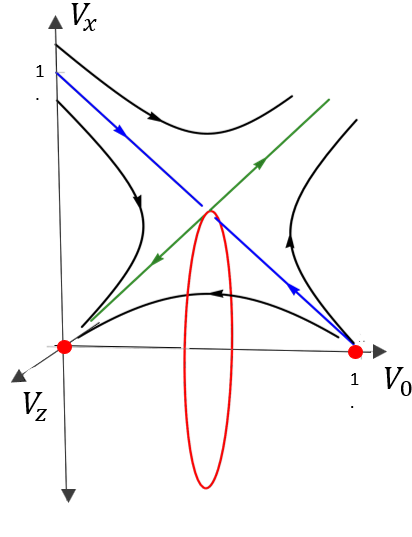}
\caption{RG flows of the parameters $V_0$, $V_z$ and $V_x$ defined in Eqn.\ref{N2dcmp1}. The red points denote the isolated fixed points, while the red-colored ring denotes the fixed point manifold. The blue line indicates the direction of an irrelevant operator corresponding to the $V_{x,c} = 0, V_{z,c} = \frac{\epsilon}{2}, V_{0,c} = \frac{\epsilon}{2}$ point on the ring manifold while the green line represents a relevant operator.}
\label{RGflow_N2}
\end{figure}

While the above two are expected for a four-fermion interaction, we have an interesting set of fixed points defined by 
\begin{eqnarray}
   \hat{V}_c = \frac{\epsilon}{2} \hat{I} + \frac{\epsilon}{2} \cos \phi \, \sigma_z + \frac{\epsilon}{2} \sin \phi \, \sigma_x,\, \phi \in [0,2\pi]
\end{eqnarray}
Rather than an isolated point, these fixed points constitute a smooth manifold within the parameter space. The manifold is a ring with radius $\epsilon/2$ centered at $(\epsilon/2, 0,0)$ in the $V_0 - V_z - V_x$ parameter space. In terms of $V_x$, $V_0$ and $V_z$, the equation of the ring manifold reads,
\begin{equation}
    V^{2}_{x, c} + V^{2}_{z, c} = \frac{\epsilon^2}{4}\,,\,\, V_{0,c} = \frac{\epsilon}{2}
\end{equation}

 As discussed in section.\ref{sec:cfnml}, a conformal manifold is associated with the presence of exactly marginal operators. We find one exact marginal operator at any point on the ring manifold, as expected since the manifold is one-dimensional. The marginal operator at a point identified by the metric $\phi$ on the manifold is given by,
\begin{eqnarray}
    \mathcal{O}^{\phi}_{M}(x) &=& - \sin \phi \left(\Phi^{\dagger}_1 \Phi_1 (x) -  \Phi^{\dagger}_2\Phi_2 (x) \right)\nonumber \\ &+& \cos \phi \left(\Phi^{\dagger}_1 \Phi_2 (x) + \Phi^{\dagger}_2 \Phi_1 (x) \right)\label{mrgnlprtsN2}
\end{eqnarray}
where $\Phi_i = \psi^T_i s_y \psi_i$ is the Cooper pair annihilation operator associated with the flavor '$i$'($i=1,2$). Notice that the marginal operator is a function of the angle $\phi$, meaning that it is different at different locations on the manifold. It is directed tangential to the ring. Also  the scaling dimension $\triangle_{\mathcal{O}_{M}} = D$, the spacetime dimensions.

We also find one relevant and one irrelevant operator on the manifold. Both the operators are functions of $\phi$. However, their scaling dimensions are independent of the location on the manifold, implying that the entire ring manifold belongs to a single universality class. See appendix.\ref{apndxsclngdmnsns} for the exact form of the operators and their respective scaling dimensions.  

The RG flow diagram is shown in fig.\ref{RGflow_N2}. This ring manifold can be described as a conformal manifold of co-dimension $D_{co} = 2$ in the three-dimensional parameter space.   

\subsection{Conformal manifolds as a coset space \texorpdfstring{$G/H$}{}}
Having found out all the fixed points from the RG equation, let us now see how the argument made in section \ref{sec:SSB} on the connection between the dynamical symmetries and the geometry of the coset manifolds can be applied to each of the cases. We show here explicitly that the geometry of coset space for each fixed point matches what we found above using brute calculations.

\paragraph{The weak and the strong coupling fixed points.}
The weak and the strong coupling fixed points have the following matrix forms,
\begin{eqnarray}
    \text{weak}\,\rightarrow\,\, V_c = \left(\begin{array}{cc}
        0 & 0 \\
        0 & 0
    \end{array} \right), \,\,\,\, \text{strong}\,\,\rightarrow \,\, V_c = \epsilon\left(\begin{array}{cc}
        1 & 0 \\
        0 & 1
    \end{array} \right) \nonumber \\
\end{eqnarray}
One can see clearly that both the matrices are invariant under the $SO(2)$ transformations. So, the subgroup $H$ here is the same as the group $G$. So, the coset space is trivial,
\begin{eqnarray}
    C = \frac{SO(2)}{SO(2)} = S^0
\end{eqnarray}
$S^{0}$ means the coset space is a point in the parameter space. These are the isolated fixed points in the parameter space. 

\paragraph{The ring manifold}
 Let us pick up the diagonal form of the intermediate fixed point by setting $\phi = 0$ in table \ref{tableN2},
\begin{eqnarray}
    V^c = \epsilon \left(\begin{array}{cc}
        1 & 0 \\
        0 & 0
    \end{array} \right)
\end{eqnarray}
The matrix is not invariant under any group transformations. Therefore, the subgroup is $H = \hat{I}$, the identity operation. Hence, the coset space is,
\begin{eqnarray}
    C = \frac{SO(2)}{\hat{I}} = S^1
\end{eqnarray}
$S^1$ implies the manifold is a ring. Consequently, the manifold derived in the parameter space can be regarded as the coset space of $G/H$.

Here we explicitly showed that the geometry of the fixed points/manifolds in the parameter space can be expressed as the coset space $G/H$, where $G$ is the symmetry group of the RG equation and $H$ is the invariant subgroup of the fixed point. 

\subsection{Phase boundary}
Having studied the renormalization group flows and the fixed point manifolds of an interacting $\mathcal{N}=2$ TI surface, we now identify the phase boundary separating the gapless and superconducting states. We determine the phase boundary by studying the RG flow diagram in fig.\ref{RGflow_N2}. Any point on the ring manifold has an irrelevant operator. On either side of this line representing the irrelevant operator (blue line), the RGE solutions flow towards the gapless or gapped superconducting phase in the infrared limit. This is evident from the RG flow of the sole relevant operator (green line) in Fig.\ref{RGflow_N2}. Hence, the ring manifold, together with the lines of irrelevant operators, naturally defines the phase boundary. 

\begin{figure}[b]
\includegraphics[width=7cm, height=6.5cm]{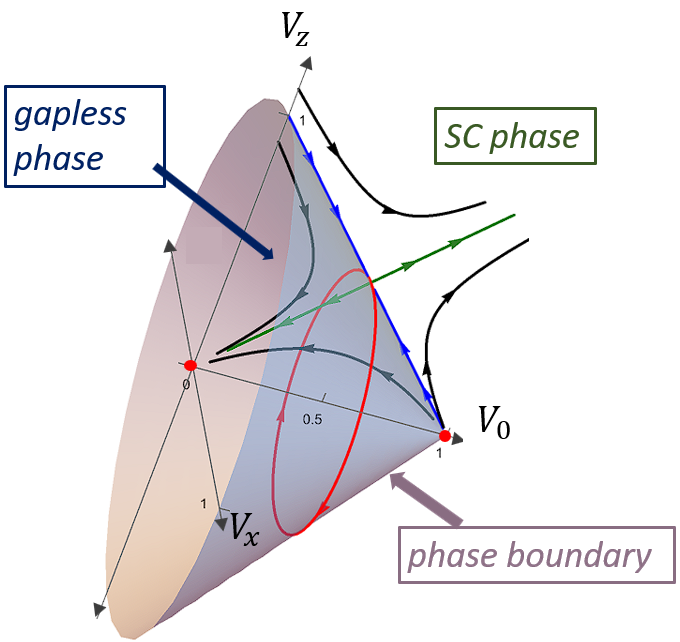}
\caption{Phase diagram showing a conical phase boundary separating a strongly interacting gapped phase from a weakly interacting gapless phase in the three-dimensional parameter space for the TI surface with two Dirac cones. Also shown are the RG flow lines same as in the previous fig.\ref{RGflow_N2}. The ring manifold (red circle) and its irrelevant operators (blue lines) lie on the surface of the conical phase boundary.}
\label{phsdgrmN2}
\end{figure}

Fig.\ref{phsdgrmN2} shows the phase boundary in the parameter space. It forms a 2-dimensional surface in the shape of a cone. It has its apex on the strong coupling fixed point $V_{0,c} = 1$ and intersects the conformal ring manifold. Quantitatively, the following equation describes the phase boundary,
\begin{eqnarray}
 V^2_z + V^2_x &=& \left(\epsilon - V_0\right)^2, \,\,\, 0 \leq V_0 < \epsilon 
\end{eqnarray}  

Here is an interesting point regarding the physics at the phase boundary. Notice that the renormalization group flows of all points on the phase boundary towards the conformal ring manifold. Hence, the universality class of the topological surface at the 2-dimensional conical phase boundary is effectively described by the one-dimensional conformal fixed point manifold in the shape of a ring. This unique feature proves invaluable in formulating an emergent single boson field theory for the phase boundary, a topic we will explore in the following section.

\subsection{Effective Yukawa field theory for the phase boundary}
 
 Here, we present another exciting physics of the ring conformal manifold. We aim to show that a single boson Yukawa field theory can effectively capture the quantum critical dynamics of this manifold. Only one relevant operator for the conformal manifold supports this single boson theory, as the boson mass parameter can be effectively attributed to the relevant operator in the parameter space.

 Let us revisit the expression of the interaction matrix $\hat{V}_c$ at the conformal manifold once again. From the table.\ref{tableN2}, it has the form, $\hat{V}_c = \frac{\epsilon}{2} \hat{I} + \frac{\epsilon}{2} \cos \phi \, \sigma_z + \frac{\epsilon}{2} \sin \phi \, \sigma_x$. After simple rearrangements, $\hat{V}_c$ can be written as a tensor product of a vector in 2-dimensional space with itself. Let $\textbf{n} = \sqrt{\epsilon}\left(\cos \theta, \,\,\, \sin \theta \right)$ where $\theta = \phi/2$ be the unit vector. Then, $\hat{V}_c$ can be expressed in terms of $\textbf{n}$ as,
 \begin{eqnarray}
     \hat{V}_c &=& \textbf{n} \otimes \textbf{n}
 \end{eqnarray}

  In other words, any point on this conformal manifold is factorizable. Plugging this back into the interaction term defined in Eqn.\ref{Lfrfmn},
  \begin{eqnarray}
      \mathcal{L}_{I} = -\sum^{2}_{i,j} \left(n_i \psi^{T}_{i}(\textbf{r}, \tau) s_y \psi_i (\textbf{r}, \tau) \right)^{\dagger} \left(n_j \psi^{T}_{j}(\textbf{r}, \tau) s_y \psi_j (\textbf{r}, \tau) \right)\nonumber \\\label{frfrmn-HS}
  \end{eqnarray}
  where $n_i$ and $n_j$ are components of $\textbf{n}$. Essentially, the interaction term can be expressed in an $A^{\dagger}_i A_j$-type form. 
  
  Let us introduce a complex Boson field $\phi (\textbf{r}, \tau)$ into the theory. Physically, this boson can be regarded as the Cooper pair of a spin-singlet pair of surface fermions.  After performing the standard Hubbard-Stratonovich transformation and introducing the boson dynamics into the picture, we arrive at the following Yukawa field theory for the conformal manifold,
\begin{eqnarray}
    \mathcal{Z} &=& \int \prod^{2}_{i=1} D\psi_{i}D\psi^{\dagger}_{i}D\phi^{*}D\phi \,\, e^{-\mathcal{S}[\phi,\psi_i]}\nonumber \\ \mathcal{S}[\phi,\psi_i] &=& \int d^{d}\textbf{r}d\tau \left[\mathcal{L}_{f} + \mathcal{L}_{b} + \mathcal{L}_{fb} \right]\nonumber \\ \mathcal{L}_{f} &=& \sum_{i=1}^{2}\psi^{\dagger}_{i} \left[ \partial_\tau + \hat{h}_{i} \right]\psi_{i} \nonumber \\ \mathcal{L}_b &=& |\partial_{\tau} \phi |^2 + \sum_{i=1}^{2}|\partial_{i} \phi |^2 + m^2|\phi|^2 + \lambda |\phi|^4 \nonumber \\ \mathcal{L}_{fb} &=& \sum^{2}_{i = 1}  g_i \phi^{*}\,\psi^{T}_{i} i s_y \psi_{i} + g_i \phi \,\psi^{\dagger}_{i}(-i s_y) \psi^{\dagger T}_{i} \label{YkwEFTN2}
\end{eqnarray}

 where $m$ is the boson mass and $\lambda$ is the four-boson interaction term. Here, we introduced a new term $g_i$ instead of $n_i$ used in Eqn.\ref{frfrmn-HS} because after adding the boson dynamics into the theory, the effective Yukawa theory in Eqn.\ref{YkwEFTN2} is quantitatively different from the one in Eqn.\ref{frfrmn-HS}. As mentioned at the beginning of this subsection, the boson mass term can be considered effectively as that relevant interaction that must be tuned to zero to study the quantum critical properties of the theory.

Recall that all points on the conical phase boundary flow toward the conformal ring manifold. Hence, the effective single-boson Yukawa field theory works well for the entire conical phase boundary, not just the ring manifold.


\section{\label{secN3}
Conformal manifolds and phase boundary in the \texorpdfstring{$\mathcal{N}=3$}{} flavor TI surface}

Here, we study the $\mathcal{N} = 3$ flavor surface, a strong TI phase, subject to strong attractive interactions. We aim to determine all the isolated fixed points and conformal manifolds in the parameter space and then identify the phase boundary of the gapless surface.

We constructed the effective 2D action governing the interaction in Eqns.\ref{Lfree}, \ref{Lfrfmn}. In section \ref{sec:RG}D, we derived the RGE by studying the interaction up to the one-loop level. The interaction matrix $\hat{V}$ in Eqn.\ref{RGeqn} is now a $3\times3$ symmetric matrix, and the interaction parameter space is six-dimensional ($D_p = 6$). Therefore, a brute-force solution would be challenging. Here is where the emergent $SO(3)$ symmetry of the RGE comes to the rescue. It suggests that if they exist, the fixed point matrices $\hat{V}_c$ could be distinguished by their eigenvalues(or traces). Therefore, we shall first look at the RG flow of the eigenvalues of the $\hat{V}$ matrix in section.\ref{secN3}A. 

Once we find the fixed points in terms of the eigenvalues, we use the arguments made in section.\ref{sec:SSB} to identify the geometry of the conformal manifolds, which is nothing but the coset space $G/H$ (see section.\ref{secN3}B).

Since our primary goal is to find a general solution for the conformal manifolds in the parameter space, we use our understanding of the manifold geometry to develop a triad basis for the interaction matrix (section.\ref{secN3}C). We then plug it back into the RGE in Eqn.\ref{RGeqn} to solve for the fixed points/manifolds. Once we solve for the conformal manifolds, we then study their stability properties (section.\ref{secN3}D). We then determine the phase boundary of the gapless surface(section.\ref{secN3}D). Finally, we show that we could study the infrared physics of the phase boundary via an effective Yukawa field theory wherein a single boson strongly interacts with $three$ flavors of surface fermions (\ref{secN3}E).

\subsection{RG flow of the eigenvalues}
Define $\lambda_1$, $\lambda_2$ and $\lambda_3$ as the eigenvalues of the $\hat{V}$ matrix. Plugging the diagonal form of $\hat{V}$ into the RG equation in Eqn.\ref{RGeqn}, we get the following decoupled equations for the eigenvalues,
\begin{eqnarray}
    \frac{d\lambda_i}{dl} &=& - \epsilon\lambda_i + \lambda^2_i \label{RGN3egnvls}
\end{eqnarray}

\begin{table}[b]
\caption{\label{tbleN3egn} Table showing the list of representative fixed points in terms of the eigenvalues of the real symmetric $\hat{V}$ matrix in the $\mathcal{N}=3$ TI surface (Eqn.\ref{RGN3egnvls}). Also shown is the shape of the manifold/coset space formed by each class of the fixed points in the 6-dimensional parameter space. (The fixed points with $'*'$ are representative points on the respective conformal manifolds). $\epsilon = D - 2$, where $D$ stands for spacetime dimensions}
\begin{ruledtabular}
\begin{tabular}{cccc}
  Fixed points ($\lambda^c_1$, $\lambda^c_2$, $\lambda^c_3$) &  Tr$[\hat{V}_c]$  &   subgroup $H$ & coset space
\\
\hline
$(0,0,0)$   &   0  & $SO(3)$ & point  \\
$(0,0,\epsilon)^{*}$ &   $\epsilon$   & $SO(2)$ & 2-sphere  \\ 
$(\epsilon,\epsilon,0)^{*}$  &   $2\epsilon$  & $SO(2)$ & 2-sphere \\  
$(\epsilon,\epsilon,\epsilon)$  &   $3\epsilon$  & $SO(3)$ & point
\end{tabular}
\end{ruledtabular}
\end{table}

It is easier to see that the fixed points are $\lambda^{c}_i = 0, \epsilon$ ($i=1,2,3$). There are a total of eight isolated fixed points. We categorize them into four distinct classes based on their eigenvalues because of the emergent $SO(3)$ symmetry of the RGE. The rotational invariance of the RGE also implies that the fixed point matrices with the same trace could form a conformal manifold, a hypothesis we will soon confirm. Table.\ref{n3table} gives the list of fixed points and their trace.

\subsection{Conformal manifolds as coset space \texorpdfstring{$G/H$}{}}

Having determined the diagonal forms of the fixed points, we will now identify the manifold for each class of fixed points listed in Table \ref{tbleN3egn}. We repeat the procedure used for the $\mathcal{N}=2$ flavor surface. There, we identified the invariant subgroup $H$ of the fixed point matrix (see section.\ref{secN2}C). Then, the conformal manifold is the coset space $G/H$, where $G$ is the RGE's symmetry. This method enables us to understand the geometry of all the conformal manifolds in the parameter space.

\paragraph{The strong and weak coupling fixed points -}
The matrix form of the strong (weak) coupling fixed point is $\hat{V} = \epsilon \hat{I}$ ($\hat{V} = 0 \hat{I}$). In both cases, the invariant subgroup is also $SO(3)$, meaning the coset spaces are isolated fixed points.

\paragraph{Interacting fixed point manifold with $\text{Tr}[\hat{V}_c] = \epsilon$ : } Let us write down the matrix form of $\hat{V}_c$ when $(\lambda_1,\lambda_2,\lambda_3) = (0,0,\epsilon)$ is, 
\begin{eqnarray}
    \hat{V}_c &=& \left(\begin{array}{ccc}
        0 & 0 & 0 \\ 
        0 & 0 & 0 \\
        0 & 0 & \epsilon
    \end{array}\right)
\end{eqnarray}
Since the two diagonal entries are zero, the invariant subgroup is $SO(2)$. Therefore, the coset space is,
\begin{eqnarray*}
    C = \frac{SO(3)}{SO(2)} = S^2
\end{eqnarray*}
Therefore, all the fixed points with trace equal to unity lie on this 2-sphere manifold in the 6-dimensional parameter space. Different points on this manifold can be connected by an $SO(3)$ rotation because they all have the same trace.

\paragraph{Interacting fixed point manifold with $\text{Tr}[\hat{V}_c] = 2\epsilon$ : } The matrix form of $\hat{V}_c$ when $(\lambda_1,\lambda_2,\lambda_3) = (\epsilon,\epsilon,0)$ is, 
\begin{eqnarray}
    \hat{V}_c &=& \left(\begin{array}{ccc}
        \epsilon & 0 & 0 \\ 
        0 & \epsilon & 0 \\
        0 & 0 & 0
    \end{array}\right)
\end{eqnarray}
The group of all rotations that keep the above matrix invariant is $SO(2)$ again. Therefore, the coset space is a 2-$sphere$ in this case, too.

Hence, we have determined the geometry of various classes of fixed point manifolds in the parameter space. In addition to the two isolated fixed points, we identify two unique classes of interacting fixed point manifolds, each forming a 2-sphere within the six-dimensional parameter space.

Understanding the geometry of the fixed point manifold yields crucial insights into its characteristics, particularly the identification of marginal operators ($n_{MAR}$) within the parameter space. With the fixed point manifold being a 2-sphere, the tangent space at any point on the sphere is two-dimensional, indicating two marginal operators at any given point. In other words, $n_{MAR} = 2$ for the two 2-sphere manifolds.

A 2-sphere manifold is a codimension-four ($D_{co} = 4$) manifold in the six-dimensional parameter space. This codimension subspace lies orthogonal to the tangent space defined by the marginal operators at a point on the manifold. All the relevant ($n_R$) and irrelevant ($n_{IRR}$) operators of the conformal manifold must lie in this codimension four subspace. That is,
\begin{eqnarray}
  n_{R} + n_{IRR} = D_{co} = 4  \label{codmnsn}
\end{eqnarray}
 where $n_{IRR}$($n_{R}$) is the number of irrelevant(relevant) operators.

Having explored the concepts of various classes of fixed-point manifolds and their geometries, we are now poised to apply them to derive a general solution to the RG equation. This will enable us to comprehend the details of the stability or instability of the fixed points in these manifolds. 

In the following, we derive an analytical solution for the 2-sphere conformal manifolds. Thereafter, we study the stability of the manifolds by identifying the irrelevant($n_{IRR}$) and relevant($n_{R}$) operators for each. To facilitate this, we develop a triad basis for the interaction matrix $\hat{V}$, discussed in the following subsection.

\begin{table*}
 \caption{Table showing the list of isolated fixed points/fixed point manifolds, the shape of the manifold in the parameter space, and their stability for the $\mathcal{N} = 3$ flavor surface. The coefficients $V^c_i$'s ($i=0,1,2,3$) are defined in Eqn.\ref{Vexp}. $\hat{V}^c$ is the interaction matrix at the fixed point/manifold. $\theta$ and $\phi$ are the polar and azimuthal angles respectively. Also, $\epsilon = D - 2$, where $D$ stands for spacetime dimensions.} 
    \label{n3table}
\begin{ruledtabular}
\begin{tabular}{ccccc}
    Isolated fixed points/Manifolds & Matrix form & Manifold & Stability & Trace[$\hat{V}^c$] \\  \hline
       \makecell{ $V^c_0 = V^c_1 = 0$,\\ $V^c_2 = 0, V^c_3 = 0$} & $\hat{V}^c = 0 \hat{I}$ & point  & Stable weak-coupling fixed point & $0$ \\ \hline \makecell{$V^c_0 = \frac{\epsilon}{3}, V^c_1 = \epsilon$,\\ $V^c_2 = 0, V^c_3 = 0$}  & $\hat{V}^c = \epsilon \, \tilde{e}^{3}_{\theta, \phi} \otimes \tilde{e}^{3}_{\theta, \phi}$ &2-sphere &
          \makecell{2 marginal, 3 irrelevant\\ and 1 relevant operator(s)} & $\epsilon$ \\ \hline \makecell{$V^c_0 = \frac{2\epsilon}{3}, V^c_1 = -\epsilon$,\\ $V^c_2 = 0, V^c_3 = 0$} & $\hat{V}^c = \epsilon \hat{I} - \epsilon \, \tilde{e}^{3}_{\theta, \phi} \otimes \tilde{e}^{3}_{\theta, \phi}$ & 2-sphere & \makecell{2 marginal, 1 irrelevant\\ and 3 relevant operator(s)} & $2\epsilon$\\ \hline \makecell{$V^c_0 = \epsilon, V^c_1 = 0$,\\ $V^c_2 = V^c_3 = 0$} & $\hat{V}^c =  \epsilon \hat{I}$ & point & Unstable strong-coupling fixed point & $ 3\epsilon$ 
    \end{tabular}
    \end{ruledtabular}
\end{table*}

\subsection{General solution: Triad basis}

So far, we have discussed four classes of fixed points or manifolds: two isolated points with co-dimension six ($D_{co}=6$) and two 2-spheres with co-dimension four ($D_{co}=4$). Studying the properties of the first two is relatively simple. So, let us focus on the two 2-sphere conformal manifolds. 

The two conformal manifolds have a co-dimension $D_{co}=4$, and this co-dimension space lies orthogonal to the 2-dimensional tangent space at a point on the 2-sphere. All the irrelevant or relevant operators lie in this co-dimension space. In this respect, to study the stability of the conformal manifold, we need to construct a four-dimensional space on which the interaction matrix $\hat{V}$ could be expanded. This four-dimensional space can be defined at any point on the 2-sphere and is orthogonal to its tangent space.

Let us construct the basis tensors. Note that $\hat{V}$ is a symmetric rank-2 tensor with a non-zero trace in general. The identity matrix $\hat{I}$ will take care of $\text{Tr}[\hat{V}]$. We could use the tensor form of spherical harmonics to define the basis for the remaining traceless part of $\hat{V}$. Generally, a tensor form of the spherical harmonic functions $Y_{jm}$($-j\leq m \leq j$) is suitable to construct a symmetric traceless tensor of rank-$j$. Thus for a rank-2 tensor, we can use the tensor form of spherical harmonic functions with $j=2$.

To realize this, we define three mutually orthonormal vectors $e^1$, $e^2$, and $e^3$ such that $e^{i T} e^j = \delta_{ij}$, called the triad vectors that form a complete set of 3-dimensional space. Generally, the triad vectors should be functions of the three Euler angles, $\theta, \phi$, and $\chi$. 

These triad vectors can be related to an arbitrary point on the surface of a 2-sphere. The azimuthal and polar angles $\phi$ and $\theta$ that identify the triad vectors locate a point on the surface of the two-sphere. This fixes the direction of one of the three axes, let's say $e^{3}$, which should point radially outward from the sphere center. Additionally, there is the freedom to rotate the axes $e^{1}$ and $e^{2}$ around the $e^{3}$ axis by an angle $\chi$. 

Setting $\chi = 0$ for the moment, we define for a given $\theta$ and $\phi$, 
\begin{eqnarray}
    e^{1}_{\theta,\phi} &=& \left(\begin{array}{c}
         \cos \theta \cos \phi \\ \cos \theta \sin \phi \\ -\sin \theta         
    \end{array}\right)\, ,\, e^{2}_{\theta,\phi} = \left(\begin{array}{c}
         -\sin \phi \\ \cos \phi \\ 0        
    \end{array}\right)\nonumber\\ \, e^{3}_{\theta,\phi} &=& \left(\begin{array}{c}
         \sin \theta \cos \phi \\ \sin \theta \sin \phi \\ \cos \theta         
    \end{array}\right) \label{trdbs} 
\end{eqnarray}
 Based on this definition, at the north pole of the 2-sphere, $e^{1}_{0,0} = (1,0,0)^{T}$, $e^{2}_{0,0} = (0,1,0)^{T}$ and $e^{3}_{0,0} = (0,0,1)^{T}$ form the triad basis.
 
  Let us turn to the cases where $\chi \neq 0$ using an orthogonal rotation matrix $R_{\chi}$. The general triad vectors then are defined as,
 \begin{eqnarray}
    \tilde{e}^{1}_{\theta,\phi, \chi} &=& R_{\chi} e^{1}_{\theta,\phi} = e^{1}_{\theta,\phi} \cos \chi + e^{2}_{\theta,\phi} \sin \chi \nonumber\\ \tilde{e}^{2}_{\theta,\phi, \chi} &=& R_{\chi} e^{2}_{\theta,\phi} = - e^{1}_{\theta,\phi} \sin \chi + e^{2}_{\theta,\phi} \cos \chi \nonumber\\ \tilde{e}^{3}_{\theta,\phi} &=& R_{\chi} e^{3}_{\theta,\phi} = e^{3}_{\theta,\phi}
 \end{eqnarray}
 These three vectors, $\tilde{e}^{1}_{\theta,\phi, \chi}$, $\tilde{e}^{2}_{\theta,\phi, \chi}$ and $\tilde{e}^{3}_{\theta,\phi, \chi}$ define the most general form of the triad vectors located at a point on the 2-sphere identified by $\theta$ and $\phi$.

 Hence to study the stability of the fixed point at the point $(\theta,\phi)$ on the 2-sphere, we can construct a four-dimensional space spanned by four rank-2 tensors as basis defined in terms of these triad vectors. Let us make the following definitions for the basis tensors,
\begin{eqnarray}
    \hat{T}_0 &=& \hat{I}\nonumber \\ \hat{T}_1 &=& \tilde{e}^{3}_{\theta, \phi}\otimes \tilde{e}^{3}_{\theta, \phi} - \frac{1}{3} \hat{I}\nonumber \\ \hat{T}_2 &=& \tilde{e}^{1}_{\theta, \phi,\chi} \otimes \tilde{e}^{2}_{\theta, \phi,\chi} + \tilde{e}^{2}_{\theta, \phi} \otimes \tilde{e}^{1}_{\theta, \phi,\chi}\nonumber \\  \hat{T}_3 &=& \tilde{e}^{1}_{\theta, \phi,\chi} \otimes \tilde{e}^{1}_{\theta, \phi, \chi} - \tilde{e}^{2}_{\theta, \phi,\chi} \otimes \tilde{e}^{2}_{\theta, \phi,\chi} 
\end{eqnarray}
As we can see, for given $\theta=\phi=0$, $\hat{T}_1$ is the tensor form of the d-orbital $d_{3z^2 - 1}$, while $\hat{T}_2$ and $\hat{T}_3$ are the tensor representations of $d_{xy}$ and $d_{x^2 - y^2}$ orbitals respectively.

Now let us expand the symmetric matrix $\hat{V}$ in this basis,
\begin{eqnarray}
   \hat{V} = V_0 \hat{T}_0 + V_1 \hat{T}_1 + V_2 \hat{T}_2 + V_3 \hat{T}_3 \label{Vexp}
\end{eqnarray}
where $V_i$($i=0,1,2,3$) are real variables. Given the co-dimension is four, these four components will take care of all the relevant and irrelevant directions.

A keen reader might have noticed a problem here. A $3\times 3$ symmetric matrix should have only six independent parameters, while here we have seven parameters: three Euler angles and the four coefficients $V_i$'s. The fact is that they are not independent. The third Euler angle $\chi$ is redundant here.

To understand why $\chi$ is redundant, let us consider some $SO(3)$ rotations of the interaction matrix $\hat{V}$ defined in Eqn.\ref{Vexp} carefully. If we rotate $\hat{V}$ by varying $\theta$ and $\phi$, we are indeed taken to different points on the manifold. Since the 4-dimensional space we defined in Eqn.\ref{Vexp} is exclusively for a given point on the 2-sphere, as anticipated, these particular rotations effectively take the $\hat{V}$ matrix out of the 4-dimensional space defined for a given point in the conformal manifold.

On the other hand, the rotations by varying $\chi$ do not take the $\hat{V}$ matrix out of the 4-dimensional space.  The $\chi$-rotation only changes the $\hat{T}_2$ and $\hat{T}_3$ tensors, which can be brought back to the original form by simply redefining $V_2$ and $V_3$. Explicitly, one has
\begin{eqnarray}
    R^{T}_{\chi'} \hat{V} R_{\chi'} &=& V_0 \hat{T}_0 + V_1 \hat{T}_1 + V_2 \left(\hat{T}_2 \cos 2\chi' + \hat{T}_3 \sin 2\chi'\right)\nonumber \\ &+& V_3 \left(\hat{T}_3 \cos 2\chi' - \hat{T}_2 \sin 2\chi' \right)\nonumber \\ &=& V_0 \hat{T}_0 + V_1 \hat{T}_1 + V'_2 \hat{T}_2 + V'_3 \hat{T}_3
\end{eqnarray}
where $V'_2 = V_2 \cos 2\chi' - V_3 \sin 2\chi'$ and $V'_3 = V_3 \cos 2\chi' + V_2 \sin 2\chi'$.

In other words, the fixed point solutions are invariant under change in $\chi$, or $SO(2)$ rotations around the third axis $\hat{e}_3$. This can be directly traced to the $SO(2)$ invariance of the fixed point manifold. 
Therefore, we effectively end up with just six independent parameters: the four $V_i$'s and the two Euler angles $\theta$ and $\phi$. 

Plugging this back to the RG equation in Eqn.\ref{RGeqn}, we arrive at the following set of dynamical equations for the four variables,
\begin{eqnarray}
    \frac{d V_0}{dl} &=& - \epsilon V_0 + V^2_0 + \frac{2 V^2_1}{9} + \frac{2}{3} \left( V^2_2 + V^2_3 \right) \nonumber \\ \frac{d V_1}{dl} &=& - \epsilon V_1 + 2V_0 V_1 + \frac{V^2_1}{3} - V^2_2 - V^2_3\nonumber \\ \frac{dV_2}{dl} &=& V_2 \left(- \epsilon + 2 V_0 - \frac{2 V_1}{3}\right)\nonumber \\ \frac{dV_3}{dl} &=& V_3 \left(- \epsilon + 2 V_0 - \frac{2 V_1}{3}\right)\nonumber \\ \frac{d\theta}{dl} &=& 0 \,\, , \,\, \frac{d\phi}{dl} = 0 \label{RGN3}
\end{eqnarray}

We examine the fixed points of the theory. Consistent with the analysis of the RG flow of eigenvalues in section \ref{secN3}A, we identify four distinct classes of fixed points, each characterized by its trace. Note that the coefficient of $\hat{T}_0$ is proportional to the trace. Table. \ref{n3table} lists all the fixed points/manifolds and their properties. Once we solve for $V^{c}_{i}$'s, we then plug them back into the Eqn.\ref{Vexp} to get the general matrix structure of the fixed point manifolds. For the weak and strong coupling fixed points, we find the fixed point interaction matrices are,
\begin{equation}
    \hat{V}_c = 0 \hat{I}, \,\text{and}\, \hat{V}_c = \epsilon \hat{I} 
\end{equation}
  
These are the isolated fixed points in the parameter space. However, for the two other interacting fixed point matrices, the matrix elements are functions of $\theta$ and $\phi$, the polar and azimuthal angles respectively. These two are,  
\begin{equation}
    \hat{V}_c = \epsilon \, \tilde{e}^{3}_{\theta, \phi} \otimes \tilde{e}^{3}_{\theta, \phi},\,\, \text{and}\,\, \hat{V}_c = \epsilon \hat{I} - \epsilon \, \tilde{e}^{3}_{\theta, \phi} \otimes \tilde{e}^{3}_{\theta, \phi}
\end{equation}
 They form manifolds in the shape of a 2-sphere in the six-dimensional parameter space. Thus, we arrived at a general solution for all four classes of fixed points.

\subsection{Properties of the fixed points and manifolds}

Having found the general solutions for the fixed points/manifolds, let us study the stability properties of each of them. This can be studied by linearizing the beta function in Eqn.\ref{RGeqn} around each fixed point. The properties of the weak and the strong coupling fixed points are similar to that of the $\mathcal{N}=2$ theory discussed in section \ref{secN2}. The weak coupling point is stable (all six operators are irrelevant), while the strong coupling fixed point is highly unstable (all six operators are relevant). These two are the isolated fixed points of the theory. In the following, we shall examine the properties of the two 2-sphere conformal manifolds in the parameter space.

\subsubsection{Conformal manifold \texorpdfstring{$\hat{V}^c = \epsilon\, \tilde{e}^{3}_{\theta, \phi} \otimes \tilde{e}^{3}_{\theta, \phi}$}{}}

Let us study the conformal manifold defined by $\hat{V}^c = \epsilon\, \tilde{e}^{3}_{\theta, \phi} \otimes \tilde{e}^{3}_{\theta, \phi}$. This is the one with the trace equal to unity shown in table.\ref{tbleN3egn}. The eigenvalues at any point on the 2-sphere are $(\lambda_1,\lambda_2,\lambda_3)  = (0, 0,\epsilon)$ and independent of both $\theta$ and $\phi$. So the fixed point shown in the table.\ref{tbleN3egn} is a point on this 2-sphere manifold. 

 There exist two marginal operators that span the 2-dimensional tangent space on the conformal manifold, i.e., $n_{MAR} = 2$. Let $\mathcal{O}^{\theta,\phi}_{M,1}(x)$, $\mathcal{O}^{\theta,\phi}_{M,2}(x)$ denote the two marginal operators. At a point $(\theta,\phi)$ on the manifold, we have,
\begin{widetext}
    \begin{eqnarray}
        \mathcal{O}^{\theta,\phi}_{M,1}(x) &=& \sin 2\theta \cos^2 \phi \, \Phi^{\dagger}_1 \Phi_1 + \frac{\sin 2\theta \sin 2\phi}{2} \Phi^{\dagger}_1 \Phi_2 + \cos 2\theta \cos \phi \Phi^{\dagger}_1 \Phi_3 + \sin 2\theta \sin^2 \phi \,\Phi^{\dagger}_2 \Phi_2 \nonumber \\ &+& \cos 2\theta \sin \phi \, \Phi^{\dagger}_2 \Phi_3 - \sin 2\theta \Phi^{\dagger}_3 \Phi_3 + \text{h.c} \nonumber\\ 
        \mathcal{O}^{\theta,\phi}_{M,2}(x) &=& - \sin \theta \sin 2\phi \Phi^{\dagger}_1 \Phi_1 + \sin \theta \cos 2\phi \, \Phi^{\dagger}_1 \Phi_2 - \cos \theta \sin \phi \Phi^{\dagger}_1 \Phi_3 + \sin \theta \sin 2\phi \,\Phi^{\dagger}_2 \Phi_2\nonumber \\ &+& \cos \theta \cos \phi \,\Phi^{\dagger}_2 \Phi_3 + \text{h.c} \label{mrgnlprtsN3}
    \end{eqnarray}
\end{widetext}
where $\Phi_{i}(x) = \psi^{T}_i is_y \psi_i (x)$ is the Cooper pair annihilation operator of flavor $i$($i=1,2,3$). As discussed in the section.\ref{sec:cfnml}, the scaling dimensions of the two marginal operators, $\triangle_{\mathcal{O}_{M,1(2)}} = D$, the space-time dimensions.

Analyzing the beta function in Eqn.\ref{RGeqn} around any point on this manifold reveals three irrelevant operators and one relevant operator. That is
\begin{eqnarray}
    n_{IRR} = 3,\, \text{and}\,\, n_R = 1.\label{stbltyCFM1}
\end{eqnarray}

Therefore, there is a three-dimensional subspace composed entirely of irrelevant operators in the four-dimensional space orthogonal to the tangent space of the manifold. Consequently, if the bare interaction matrix elements lie within this three-dimensional subspace, the theory will flow towards the conformal 2-sphere manifold in the infrared limit.

However, there is one relevant operator in the six-dimensional parameter space. To achieve quantum criticality, we must fine-tune this relevant operator to the fixed point manifold. Given that only one of the six operators requires tuning, we can infer that this conformal manifold is infrared-stable. Appendix \ref{apndxsclngdmnsns} lists the marginal, relevant, and irrelevant operators and their respective scaling dimensions for a few selected points on the conformal manifold for reference. 

\subsubsection{Conformal manifold \texorpdfstring{$\hat{V}^c = \epsilon \hat{I} - \epsilon\, \tilde{e}^{3}_{\theta, \phi} \otimes \tilde{e}^{3}_{\theta, \phi}$}{}}

The conformal manifold defined by $\hat{V}^c = \epsilon \hat{I} - \epsilon\, \tilde{e}^{3}_{\theta, \phi} \otimes \tilde{e}^{3}_{\theta, \phi}$ correspond to the one with trace equal to $2\epsilon$ in the table.\ref{tbleN3egn}. The eigenvalues remain the same throughout the manifold, and they are given by $(\lambda_1,\lambda_2,\lambda_3) = (\epsilon, \epsilon, 0)$.

 Interestingly, we notice here that the functional form of the two marginal operators of this conformal manifold is the same as $\mathcal{O}^{\theta,\phi}_{M,1}(x)$ and $ \mathcal{O}^{\theta,\phi}_{M,2}(x)$ ( Eqn.\ref{mrgnlprtsN3}), the marginal operators of the previous one. However, note that the two manifolds occupy distinct locations within the parameter space. Thus, while the marginal operators appear structurally identical in both cases, they are defined at different points in the parameter space.

Repeating the same process as in the previous case, we find that at any point on the manifold, there are three relevant operators and only one irrelevant operator, all lying in the four-dimensional space normal to the tangent space of two marginal operators. That is
\begin{eqnarray}
    n_{IRR} = 1,\, \text{and}\,\, n_R = 3.\label{stbltyCFM2}
\end{eqnarray}

Overall three operators must be fine-tuned to the fixed point for the system to exhibit quantum critical behavior in the infrared limit. Hence, this conformal manifold is infrared-unstable. See appendix \ref{apndxsclngdmnsns} for the exact form of the relevant and irrelevant operators and their respective scaling dimensions at a few selected points on the conformal manifold. 

\subsection{Phase boundary}

Here, we shall determine the phase boundary of the gapless surface state. Recall that in the $\mathcal{N}=2$ case, the phase boundary is a 2-dimensional surface in the 3-dimensional parameter space, meaning a $D_{p}=1$ manifold. The conformal ring manifold and all its irrelevant operators lie on this phase boundary. Since the parameter space was three-dimensional for the $\mathcal{N}=2$ case, it was straightforward to figure out the exact shape and structure of the phase boundary.

 But things get trickier when it comes to $\mathcal{N}=3$ because the parameter space is six-dimensional. Here, we have two conformal manifolds in the shape of a 2-sphere. However, we shall use the knowledge of the shape of these conformal manifolds and their stability properties to arrive at a structure of the phase boundary.

Recall that one of the conformal manifolds defined by $\hat{V}^c = \epsilon\, \tilde{e}^{3}_{\theta, \phi} \otimes \tilde{e}^{3}_{\theta, \phi}$ is infrared stable. It has only one relevant operator, as shown in Eqn.\ref{stbltyCFM1} (also in Table.\ref{n3table}). Therefore, this conformal manifold is naturally embedded within the phase boundary of the gapless state. Eqn.\ref{stbltyCFM1} suggests there are three irrelevant operators ($n_{IRR} = 3$) at any point on the manifold. Following the spirit of the analysis in the $\mathcal{N}=2$ case, we can illustrate that locally the phase boundary shall be made up of this 3-dimensional subspace of irrelevant flows plus the 2-dimensional 2-sphere conformal manifold. To summarize, we shall have a five-dimensional space as the phase boundary. The sole relevant operator is directed orthogonal to this 5-dimensional phase boundary.

We observe that the infrared physics of most of the 5-dimensional phase boundary is effectively described by the 2-sphere conformal manifold embedded within it, as irrelevant operators flow towards the 2-sphere manifold in the infrared limit.

\subsection{Effective Yukawa field theory for the 5-dimensional phase boundary}

In the previous section, where we studied the quantum criticality of the $\mathcal{N}=2$ phase, we developed an effective single boson Yukawa field theory for the conformal manifold of the shape of a ring in the three-dimensional parameter space (see section \ref{secN2}C). We also found that the theory works well for the phase boundary of the gapless surface state. The two main ingredients that helped us develop this EFT are 1) the interaction matrix at the ring manifold is factorizable, and 2) there is only one relevant direction out of the conformal manifold that could be associated with the boson mass.

We observe that the phase boundary of the $\mathcal{N}=3$ flavor TI surface also satisfies the two ingredients. First, the 2-sphere conformal manifold embedded within the phase boundary is also factorizable. It is evident from its expression, $\hat{V}_c = \epsilon\,\tilde{e}^{3}_{\theta, \phi} \otimes \tilde{e}^{3}_{\theta, \phi}$, already in a tensor product form. Second, Eqn.\ref{stbltyCFM1} suggests that the manifold has only one relevant operator. Therefore, the phase boundary of the $\mathcal{N}=3$ flavor TI surface near the infrared stable conformal manifold can also be expressed by a single boson Yukawa field theory in the infrared limit. 

The effective field theory is similar to the $\mathcal{N}=2$ case but with the three flavors of surface fermions strongly interacting with a single flavor of bosons. For completeness, we shall write down the effective Yukawa field theory below,
\begin{eqnarray}
    \mathcal{Z} &=& \int \prod^{3}_{i=1} D\psi_{i}D\psi^{\dagger}_{i}D\phi^{*}D\phi \,\, e^{-\mathcal{S}[\phi,\psi_i]}\nonumber \\ \mathcal{S}[\phi,\psi_i] &=& \int d^{d}\textbf{r}d\tau \left[\mathcal{L}_{f} + \mathcal{L}_{b} + \mathcal{L}_{fb} \right]\nonumber \\ \mathcal{L}_{f} &=& \sum_{i=1}^{3}\psi^{\dagger}_{i} \left[ \partial_\tau + \hat{h}_{i} \right]\psi_{i} \nonumber \\ \mathcal{L}_b &=& |\partial_{\tau} \phi |^2 + \sum_{i=1}^{2}|\partial_{i} \phi |^2 + m^2|\phi|^2 + \lambda |\phi|^4 \nonumber \\ \mathcal{L}_{fb} &=& \sum^{3}_{i = 1}  g_i \phi^{*}\,\psi^{T}_{i} i s_y \psi_{i} + g_i \phi \,\psi^{\dagger}_{i}(-i s_y) \psi^{\dagger T}_{i} \label{YkwEFTN3}
\end{eqnarray}
where $\phi$ is a complex boson of charge $2e$. $m$ and $\lambda$ are the boson mass and four-boson interaction terms, respectively. The mass parameter is effectively one relevant direction out of the manifold that must be tuned to zero to study the quantum critical dynamics

\begin{figure}[b]
\includegraphics[width=5.5cm, height=5.9cm]{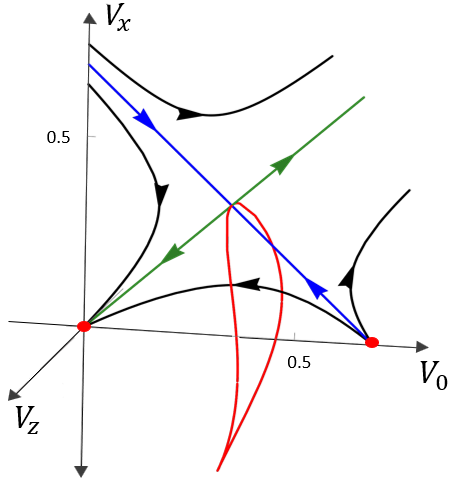}
\caption{RG flows of the parameters $V_0$, $V_z$ and $V_x$ when three-loop effects are taken into account for an $\mathcal{N}=2$ TI surface. The smooth conformal manifold is defined by the Eqn.\ref{qcp3loop}. The perfect ring manifold shown in Fig.\ref{RGflow_N2} is deformed but the closed loop structure remains intact. The definitions of other RG flow lines are the same as in fig.\ref{RGflow_N2}.}
\label{RGflow_N23loop}
\end{figure}

\section{\label{sec:hghloop}Effects of \texorpdfstring{$N_L$}{}-loops, \texorpdfstring{$N_L=2,3,4,…$}{}}
Further inclusion of effects of multiple-loop diagrams in the renormalization group equations can lead to the breaking of the emergent $SO(\mathcal{N})$ symmetry in the case of $\mathcal{N}=2,3$ studied in Section \ref{secN2},\ref{secN3}. The number of loops $N_L$ is related to the number of four fermion vertices $N_v$. $N_v=N_L+1$.
In the context of the $\epsilon$-expansion, these contributions with multiple loops are also represented as higher-order effects so that symmetry is only weakly broken. For instance, the $N_L$-loop effects are proportional to $\epsilon^{L+1}$, $L\geq 1$. For two-loop effects, the contributions are of the order of $\epsilon^3$, while for three-loop effects, the contributions are of the order of $\epsilon^4$.

The main issue is whether the manifolds remain intact when higher loop effects or higher-order contributions are present.
To understand these more subtle multiple-loop or higher-order effects, we factorized the $\beta$-function in the RGE into two pieces
\begin{eqnarray}
\frac{d \hat{V}}{dl}=\beta (\hat{V}) +\beta_{M}(\hat{V}) \nonumber\\
\end{eqnarray}
where $\beta(\hat{V})$ represents all the one-loop effects that already have been studied while $\beta_M(\hat{V})$ includes other multiple-loop effects that were not included in the studies in section \ref{secN2}.

The stability of the conformal manifolds turns out to be dependent on the structure of $\beta_M$. Below we discuss two different classes of high-loop effects.
Type A is related to  higher-loop one-particle irreducible processes that diagrammatically appear in a form of vertex corrections. Type B is related to the fermion field renormalization effects. 
The smooth conformal manifolds that we have seen in the one-loop RGE appear to  be stable with respect to Type A class effects as they only induce perturbative distortions.
However, we also find that these manifolds are unstable with respect to the fermion field renormalization effects in Type B. Note Type B effects also result in breakdowns of $SU(\mathcal{N})$ symmetry in Green's functions that play very important role
in the discussion of emergent $SO(\mathcal{N})$ in RGEs. 

\subsection{A three-loop effect of Type A: Distorted Conformal Manifolds}

If one-particle irreducible loop effects are taken into account in RGEs, we can express these subset higher-loop effects in term of $\beta_M$ in the following form, 

\begin{eqnarray}
\beta^A_M &=&\hat{V} \Gamma_z \hat{V}, \Gamma_z=A+B\sigma_z+C\sigma_x.
\end{eqnarray}

For instance, 
if three loop effects in Fig.\ref{fnmndgrms} (b) are considered, $A$ and $B$ themselves are quadratic functions of $V$ but we can show that $C$ in this limit vanishes, i.e,
\begin{eqnarray}
A =\frac{V^2_{11} + V^2_{22}}{2}, B=\frac{V^2_{11} - V^2_{22}}{2}, C=0.
\end{eqnarray}
One can then study the effect of $\beta_M$ that breaks the $SO(\mathcal{N})$ symmetry. As $\beta(\hat{V})$ is $SO(\mathcal{N})$  symmetric leading to various manifolds of different co-dimensions, we can study the effect on the flows within the conformal manifolds as well as the flow along all other directions orthogonal to the tangent space at a particular point in the manifolds. This can be carried out straightforwardly.
For $\mathcal{N}=2$ and $D_p=3$, the conformal manifold has co-dimension $D_{co}=2$ and the manifold is a one-dimensional ring while there is only one direction orthogonal to the manifold embedded in the phase boundary. In the $\epsilon$-expansion approach for $N_L=3$-loop effects, one finds that the overall effect of $\beta_M(\hat{V})$ is perturbative on the RGE flows near the conformal manifold, the manifold of strong coupling fixed points. The main effect is to slightly distort the shape of the ring rather than reducing the dimension of the manifold. For instance, for $\mathcal{N}=2$ and $D_p=3$, there still exists a one-dimensional ring-like manifold of strong coupling fixed points that represent the universality of quantum criticality. Details are presented below.

Let us see explicitly if the one-dimensional conformal manifold exists when the three-loop effects are included in the RG equation. Let us express the interaction matrix $\hat{V}$ in terms of Pauli matrices as shown in Eqn.\ref{N2dcmp1}. The RG equation is then (See fig.\ref{fnmndgrms} for the Feynman diagram),
\begin{subequations}
\begin{eqnarray}
    \frac{dV_0}{dl} &=& -\epsilon V_0 + V^2_0 + V^2_x + V^2_z + V^4_0 + V^4_z\nonumber\\ &+& 6 V^2_0 V^2_z + V^2_x \left(V^2_0 + V^2_z\right) \\ \frac{dV_x}{dl} &=& V_x \left( -\epsilon + 2 V_0 + 2 V_0 \left( V^2_0 + 3 V^2_z \right) \right) \\ \frac{dV_z}{dl} &=& V_z \left( -\epsilon + 2 V_0 + 4 V_0 \left( V^2_0 + V^2_z - \frac{1}{2} V^2_x \right) \right)
\end{eqnarray}    
\end{subequations}
Solving for the fixed points of the theory by setting the LHS to zero, we find that a one-dimensional conformal manifold still exists in the parameter space. The equation of the closed loop is found to be,
\begin{eqnarray}
    V^{2}_{x,c} + V^{2}_{z,c} &=& V^{2}_{0,c} \nonumber \\ V^2_{0,c} + 3 V^2_{z,c} + 1 &=& \frac{\epsilon}{2 V_{0,c}}\label{qcp3loop}
\end{eqnarray}

  In the former case when only one-loop effect was taken into account, $V_0$ was fixed at $\epsilon/2$ (see fig.\ref{RGflow_N2}). On the other hand, here we see that $V_0$ is a function of $V_z$. So we still have a closed loop but it is rather in a distorted shape as compared to the perfect ring manifold in the one-loop limit. The RG flow lines and the conformal manifold (red loop) are shown in fig.\ref{RGflow_N23loop}. Consequently, the phase boundary would also be a distorted 2-dimensional surface instead of a perfect cone. Essentially, the three-loop correction modifies the one-loop result only slightly. The conformal manifold and phase boundary are altered but remain intact in the three-loop approximation.
  
We also want to remark that the conclusion above can be generally true in the presence of higher loops, and the manifold shall be quite robust against this class of symmetry-breaking effects. In fact, one can show that there shall still exist a co-dimension two manifold even when other higher loop effects of similar structures with $N_L=4,5,6…$ beyond the simple three-loop one are included in the $\epsilon$ expansion. The generality of this conclusion about the conformal manifold beyond the $\epsilon$-expansion approach employed here needs to be further investigated and explored in the future, likely involving some numerical simulations.

\begin{table*}
\caption{\label{tableN2hghloop}Table showing the list of fixed points and their stabilities for the $\mathcal{N} = 2$ flavor surface when two-loop effects are included. Also, $\epsilon = D - 2$ where D is the spacetime dimensions.}
\begin{ruledtabular}
\begin{tabular}{cccc}
 Isolated fixed points/Manifolds  & Stability & Invariant subgroup($H$) & Coset space\\
 \hline
  $\hat{V}_c = 0 \hat{I}$   & Three irrelevant operators. Stable fixed point &$Z_2 \otimes Z_2$ & $I$  \\
 $\hat{V}_c = \left(\frac{\epsilon}{2} +  \frac{a}{2}  \epsilon^2  \right)\left(\hat{I} \pm \sigma_x \right)$  & One relevant, two irrelevant operators.
 &$Z_2$ & $Z_2$  \\ 
 $\hat{V}_c = \left(\frac{\epsilon}{2} + a \epsilon^2 \right)\left(\hat{I} \pm \sigma_z \right)$  & Two relevant, one irrelevant operators.
 &$Z_2$ & $Z_2$  \\
 $\hat{V}_c =  \left(\epsilon + 2 a \epsilon^2 \right)\hat{I} $ & Three relevant operators. Unstable fixed point. &$Z_2\otimes Z_2$ & $I$
 \end{tabular}
\end{ruledtabular}
\end{table*}

\subsection{A two-loop effect of type B: Inducing flow in the Manifold}

Now we are going to tackle more delicate effects of the field renormalization. Including the two-loop effect in Fig.\ref{fnmndgrms} (c), we show that

\begin{eqnarray}
\beta^B_M &=& (\eta_i+\eta_j) V_{ij},\,\,\, \eta_i=\alpha \sum_k V_{ik} V_{ki}
\end{eqnarray}
where $\eta_i$ is directly induced by the field renormalization of the $i$th flavor $\psi_i$. 

When such a term is present in the RGEs, the dynamic symmetry is drastically reduced. In the case of a topological surface with $\mathcal{N}=2$ Dirac cones, the emergent $SO(2)$ symmetry appearing in the one-loop RGEs is reduced to
a discrete symmetry group of $\mathcal{Z}_2 \otimes \mathcal{Z}_2$. The two $\mathcal{Z}_2$ transformations are

\begin{eqnarray}
\mbox{I)} V_z \rightarrow -V_z; \mbox{II)} V_x \rightarrow  -V_x.
\label{Z2T}
\end{eqnarray}
One can easily verify that the resulting RGEs with $\beta^B_M$ are invariant under either of these two transformations.

As we will see below, the term of $\beta^B_M$ can induce a flow within the conformal manifold discussed in the previous section, although the RGE flow is very slow in the $\epsilon$-expansion where the flow velocity is proportional to 
$\epsilon^3$. Nevertheless, such a flow breaks the conformal manifold into two pairs of discrete fixed points. Within each pair, two fixed points are related to each other by one of the $Z_2$ transformations introduced before.
In addition, there is one non-interacting fixed point and one strong coupling fixed point, both of which are singlet under the $\mathcal{Z}_2 \otimes \mathcal{Z}_2$ symmetry group. 

In this limit, we identify a total of six fixed points. Among them,
one is non-interacting fermion fixed points. Two are strong coupling fixed points which are infrared stable in the phase boundary manifold. The other three are infrared unstable along one or both directions in the phase boundary manifold and can be 
associated with multi-criticality. Details of them are discussed below.

\begin{figure}[b]
\includegraphics[width=5.5cm, height=7.1cm]{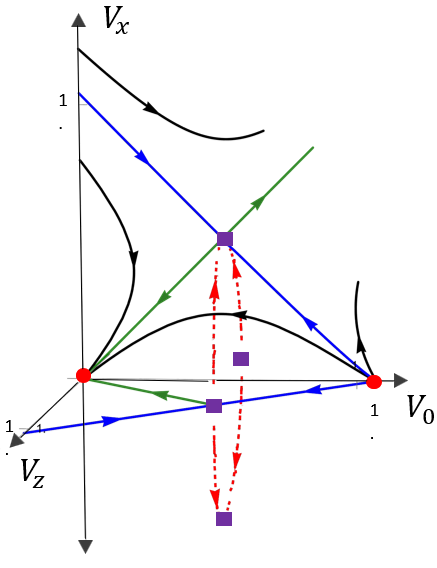}
\caption{RG flows of the $\mathcal{N}=2$ flavor interacting TI surface when two-loop field renormalization effects are included. The smooth ring manifold in Fig.\ref{RGflow_N2} breaks into four isolated fixed points (violet squares) (see Table.\ref{tableN2hghloop}). The dashed lines represent the relevant or irrelevant flows, which were marginal at the one-loop level. The other RG flow lines have the same definition as in Fig.\ref{RGflow_N2}} 
\label{RGflow_N22loop}
\end{figure}

Upon including the field renormalization effects, the RGE expressed in terms of $V_0,V_x,V_z$(see Eqn.\ref{N2dcmp1}) takes the form,
\begin{eqnarray}
    \frac{dV_0}{dl} &=& - \epsilon V_0 + V^2_0 + V^2_z + V^2_x\nonumber \\  &-& 2a V_0 \left[ V^2_0 + 3V^2_z + V^2_x\right]\nonumber \\ \frac{dV_z}{dl} &=& V_z \left[ -\epsilon +  2V_0 - 2a \left(3V^2_0 + V^2_z + V^2_x\right) \right]\nonumber \\ \frac{dV_x}{dl} &=& V_x \left[ -\epsilon +  2V_0 - 2a \left(V^2_0 + V^2_z + V^2_x\right) \right]\label{RGEN22loop}
\end{eqnarray}
where $a=\frac{1}{32}$ and $\epsilon = D - 2$.
\subsubsection{The fate of the conformal manifolds}
The sign of $a$ determines how the RGE flow can be induced by the symmetry-breaking field renormalization effect in the conformal manifold (or around the ring in this case) and how the smooth manifold can break down into a few discrete, strong coupling fixed points.

To visualize this, we introduce a dynamic polar angle $\theta$, which locates a point around the ring by
\begin{eqnarray}
    \tan \theta = \frac{V_z}{V_x}
\end{eqnarray}

Taking into account the last two equations in Eqn.\ref{RGEN22loop}, one can show that a finite flow is induced around the ring given by,
\begin{eqnarray}
    \frac{d\theta}{dl} = -2a V^2_0 \sin 2\theta
\end{eqnarray}

If we treat the effect perturbatively and assume $V_0$ still remains constant in the manifold, the induced flow is periodical around the ring with a period equal to $\pi$ rather than $2\pi$. This leads to the formation of IR stable fixed points $\theta = 0, \pi$ and infrared unstable fixed points at $\theta = \frac{\pi}{2}, \frac{3\pi}{2}$ when $a$ is positive. 

So in the ring manifold, two fixed points with $V_z = 0$ become stable in the presence of the symmetry-breaking effect, and two other fixed points with $V_x = 0$ become infrared unstable (see below Fig.\ref{RGflow_N22loop}). Note that because of a finite flow around the ring manifold, no marginal operators exist associated with these four fixed points. 

Table.\ref{tableN2hghloop} lists the fixed points of the RGE in Eqn.\ref{RGEN22loop}
. Fig.\ref{RGflow_N22loop} shows the RG flow lines of the solutions to Eqn.\ref{RGEN22loop}.  Appendix.\ref{apndxsclngdm2loop} lists the relevant and irrelevant operators and their scaling dimensions.

\begin{figure}[b]
\includegraphics[width=7cm, height=7cm]{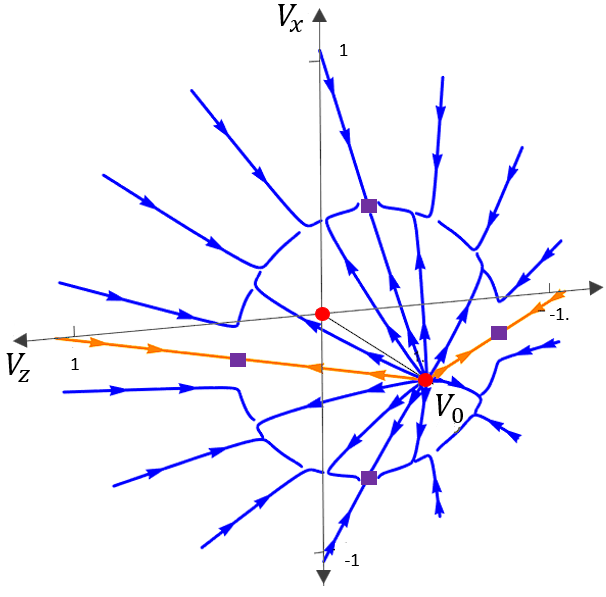}
\caption{RG flow lines that lie on the phase boundary in the interaction parameter space. The set of all blue lines flow towards the fixed points $\hat{V}_c = \left(\frac{\epsilon}{2} + \frac{a}{2} \epsilon^2 \right)\left(\hat{I} \pm \sigma_x \right)$, while the orange lines are the only ones flowing towards the fixed points $\hat{V}_c = \left(\frac{\epsilon}{2} + a \epsilon^2 \right)\left(\hat{I} \pm \sigma_z \right)$. The violet squares denote the isolated fixed points, part of the ring manifold in one-loop order (see Fig.\ref{RGflow_N2}).}
\label{phsbndryN22loop}
\end{figure}

\subsubsection{Phase boundary}
Let us study the effects of fermion field renormalization on the phase boundary of the gapless surface. At the one-loop level, it was conical in shape with the universality class determined by the
conformal ring manifold. Once we include the field renormalization effect, we see that the phase boundary is still two-dimensional but slightly distorted. The universality class of the phase boundary is now determined by the isolated fixed points $\hat{V}_c = \left(\frac{\epsilon}{2} + \frac{a}{2} \epsilon^2 \right)\left(\hat{I} \pm \sigma_x \right)$, depending on the sign of $V_x$ in the parameter space (see Fig.\ref{phsbndryN22loop}). The $Z_2$ invariance of the RGE implies both the fixed points are in the same universality class.

One notices a cut at $V_x = 0$ separating the phase boundary on either side (dashed orange lines in Fig.\ref{phsbndryN22loop}). This cut is the sole irrelevant flow line to the multi-critical fixed points $\hat{V}_c = \left(\frac{\epsilon}{2} + a \epsilon^2 \right)\left(\hat{I} \pm \sigma_z \right)$. Any point on this irrelevant line flows towards one of these two fixed points depending on the sign of $V_z$ in the infrared limit. Hence, we find a one-dimensional curve with a universality class distinct from that at the phase boundary.

\section{\label{sec:cnclsn} Conclusion}

In this article, we have explored various emergent conformal-field-theory states on surfaces of a 3D topological insulator (cubic crystal) in the presence of strong attractive interactions. Specifically, we considered the cases of interacting TI surfaces with two flavors of 2-component Dirac fermions (weak TI) ($\mathcal{N}=2$) and three flavors of 2-component Dirac fermions (strong TI) ($\mathcal{N}=3$). We itemize the main results below.

 A) {\bf Phase boundary of Gapless topological surfaces} We have constructed effective field theories for interacting surface topological states and identified the phase boundaries of weakly interacting gapless surface fermions via analyzing the renormalization group flow. The dimension of interaction parameter space $D_p$  where our studies are carried out is defined in terms of the number of surface fermion flavors $\mathcal{N}$ as $D_p=\frac{\mathcal{N}(\mathcal{N}+1)}{2}$. The cases of $\mathcal{N}=2,3$ have been the focus of the current studies.

In the $\mathcal{N}= 2$ weak TI phase with $D_p=3$, we illustrate a two-dimensional conical phase boundary in the three-dimensional parameter space spanned by the interaction matrix elements. The phase boundary separates the weakly interacting gapless surface fermions from other gapped superconducting phases. At the phase boundary, the universality of generic surface topological quantum critical points is determined by a ring-like co-dimension two conformal manifold or a one-sphere manifold formed by strongly interacting fixed points.

When it comes to an $\mathcal{N} = 3$ surface with $D_p=6$, the phase boundary becomes a five-dimensional manifold embedded in the six-dimensional interaction parameter space, separating the gapless fermions and gapped superconducting phases. Generic quantum critical behaviors at the phase boundary are now governed by a co-dimension four conformal manifold, i.e., a two-sphere embedded in the six-dimensional space.

Generally, for a TI surface with attractive interactions, the phase boundary of weakly interacting gapless fermion states is a $D_p-1$-dimensional manifold in the $D_p$-dimensional parameter space.  The smooth conformal manifolds where all the renormalization group flows within the phase boundary manifold are channeled into are infrared stable within the $D_p-1$ dimension phase boundary and hence dictate all the universal properties of generic surface topological critical points.

B) {\bf The infrared stable conformal manifolds} In $D_p-1$ -dimension phase boundaries, there can be multiple conformal manifolds, as well as isolated strong coupling fixed points with very different infrared stabilities.
The infrared stable conformal manifolds mentioned in A), which play a paramount role in our studies of surface topological quantum critical points, are always uniquely characterized by a finite number of marginal operators $n_{MAR}$. The dimension of these smooth manifolds itself is precisely defined by $n_{MAR}$, the number of marginal operators associated with the manifold. 

 One important property of a marginal operator $\mathcal{O}_{M}(x)$ is that its scaling dimension $\triangle_{\mathcal{O}_M} = D$, the spacetime dimensions of the theory. Hence, the correlation function of the marginal operators at their respective conformal manifold obeys the power law $<\mathcal{O}_{M}(x)\mathcal{O}(0)> \sim \frac{1}{|x|^{2D}}$. Verifying this scaling behavior experimentally could provide direct evidence for the existence of conformal manifolds. 

In addition, there are always a finite number of infrared irrelevant operators, $n_{IRR}$. However, there shall be only one relevant operator representing the direction perpendicular to the phase boundary so that $n_{IRR} +n_{MAR}=D_p-1$ for these infrared stable manifolds.

For $\mathcal{N} =2$ weak TI phase with a one-dimensional conformal manifold, we find $n_{IRR}=n_{MAR}=1$ and $D_p=3$. 
For $\mathcal{N}=3$ strong TI phase, the conformal manifold with $\text{Tr} [\hat{V}_c]=\epsilon$ has exactly the desired property.  It has one single relevant operator and multiple irrelevant and marginal operators: $n_{IRR}=3$, $n_{MAR}=2$ and $D_p=6$ (See Table. \ref{tableN2} , \ref{n3table}.).

Unlike a more standard quantum critical point that can be often associated with a single isolated Wilson-Fisher fixed point, universal properties of generic surface topological phase transitions or quantum critical points appear to be dictated by a lower dimensional conformal manifolds that are embedded in the $D_p-1$ dimension phase boundaries.

It is worth mentioning that there appears to be an intimate connection between the geometry of the conformal manifolds in the parameter space and the emergent dynamic symmetries of the RG equation. Assume the symmetry group of the RG equation is $G$ and the invariant subgroup of a fixed point is H, we have found that the corresponding conformal manifold is precisely the coset space of G induced by the subgroup H, i.e. $G/H$ (see Section.\ref{sec:SSB}). However, this emergent dynamic symmetry group G is sufficient but not necessary for the appearance of the conformal manifolds. In our studies, we also show that conformal manifolds are robust to the effects of certain class of G-symmetry breaking terms, although there can be various distortions.

C) {\bf Infrared unstable manifolds and strong coupling fixed points}: Apart from the infrared stable manifolds summarized in B), there can be other infrared unstable manifolds as well as isolated strong coupling fixed points (See Table.\ref{tableN2} and Table.\ref{n3table}).

The isolated strong coupling fixed points or other manifolds appearing in the phase boundaries turn out to be highly unstable in the limit of infrared, even when RG flows are confined within the $D_p-1$-dimension phase boundaries.  In fact, there are multiple relevant operators associated with them, and for that reason, they do not represent generic surface topological quantum critical points.
In both the $\mathcal{N}=2,3$ case, around the isolated strong coupling fixed points, there are no irrelevant or marginal operators. That is $n_{IRR}=n_{MAR}=0$ and all the operators become relevant. Therefore, the number of relevant operators in these cases is maximal, equal to $D_p$, the dimension of the interaction parameter space.

In the case of $\mathcal{N}=3$, there is also another two-sphere manifold ($\text{Tr} [\hat{V}_c]=2\epsilon$) with three relevant operators instead of a single one. And $n_{IRR}=1$, $n_{MAR}=2$ (See Table. \ref{n3table}). This manifold is also infrared unstable with respect to the flows within the phase boundary. It has three relevant operators, two more than the infrared stable manifold discussed in B). This manifold, together with other strong coupling fixed points, can be identified as a candidate for surface topological multiple critical points rather than generic surface topological quantum critical points.

D) We have also studied the stability of these conformal manifolds in the cases of $\mathcal{N}=2,3$ against higher loop $SO(\mathcal{N})$-symmetry breaking effects.
While $SO(\mathcal{N})$-symmetry breaking three-loop one-particle irreducible effects only result in distortions (see Fig.\ref{RGflow_N23loop}), two-loop $SO(\mathcal{N})$-symmetry breaking field renormalization effects can further break the conformal manifolds into isolated fixed points (see Table.\ref{tableN2hghloop} or Fig.\ref{RGflow_N22loop}). 
In the latter case, the manifold can severely constrain the RGE flows between discrete, strong coupling fixed points living in the manifold, leading to an envelope of the flows.

The overall stability of these conformal manifolds against $SO(\mathcal{N})$-symmetry breaking effects further depends on how the symmetry is broken. This emergent symmetry is a sufficient but not necessary condition for the conformal manifold.

E) {\bf Emergent boson fields at conformal manifolds} At the infrared stable conformal manifolds, we show that the four-fermion interactions are factorizable in the $\mathcal{N}$-dimension flavour space. Therefore, these conformal manifolds at the phase boundaries can be more conveniently studied via an emergent boson field that interacts with gapless surface fermions (see Eqns.\ref{YkwEFTN2},\ref{YkwEFTN3}). 

In such an effective theory, $\mathcal{N}=2,3$ flavors of surface fermions strongly interact with a single flavor of complex boson, resulting in conformal fields in the limit of massless bosons. 
The boson mass is directly related to that sole relevant operator associated with the conformal manifolds, orthogonal to the phase boundary in the $D_p$-dimension parameter space.

F) {\bf Stability of Fermion degrees of freedom} Although the focus of this research project is on interacting gapless fermions, we have also studied the stability of having multiple numbers of Dirac fermions on the surface of non-interacting TI in a cubic crystal. For instance, a strong TI with a given bulk topological invariant can have either $\mathcal{N} =1$ or $\mathcal{N} =3$ TI surfaces (see Fig.\ref{phsdgrm}). 

Questions can be raised about the stability of the respective surface states. We found that the two distinct surfaces can be deformable without closing the bulk gap under the time-reversal symmetry (see Fig.\ref{dfrmtn_TI}). However, this requires lowering the crystal symmetry of the TI bulk. One has to break translation and inversion symmetries in addition to introducing infinite-range hopping terms to the lattice Hamiltonian. 

The same set of arguments can also be applied to the weak TI phase. This suggests that the surface with arbitrary $\mathcal{N}$-flavors of 2-component Dirac fermions is stable provided we do not break the crystal's symmetry.

The important overall takeaway from our study is that for an interacting TI surface, the number $\mathcal{N}$ of the surface fermions plays a paramount role in setting the phase boundary manifold of the gapless topological surface fermions as well as possible families of emergent conformal fields. The dynamic properties of conformal manifolds emerging in the $D_p-1$ dimension phase boundary are in fact fully determined by $\mathcal{N}$.

Only the infrared stable conformal manifolds in the phase boundary can be applied to 
identify all the possible universality classes of generic surface topological quantum critical points. For instance, we have demonstrated that the $\mathcal{N} = 1$  and $\mathcal{N} = 3$ TI surfaces, although characterized by the same bulk topological invariants, result in 
completely different classes of conformal fields at surface topological critical points when strong attractive interactions are present. 

One of the future directions is perhaps to further understand the consequences of the 
conformal manifolds that we have exclusively associated with the universalities of surface topological quantum critical points. One very unique aspect of conformal manifolds is the existence of multiple marginal operators which define the dimension of the manifolds. It is fascinating to further explore the dynamic implications of these manifolds in the context of surface transport dynamics or hydrodynamics phenomena.
Specially at which energy scales can the emergent $SO(\mathcal{N})$ symmetry and resultant manifolds be probed and how does the breakdown of conformal manifolds into the discrete fixed points manifest in infrared physics ?

We acknowledge the support of an NSERC(Canada) Discovery Grant under contract No RGPIN-2020-07070 and useful discussions with Julio Parra Martinez and Mark Van Raamsdonk on RGEs and CFTs.

\appendix

\section{Derivation of the effective 2D interacting theory from the 3D lattice model}
\label{EFT_deriv}
\subsection{Domain wall model}
In this section, we develop an effective low-energy continuum model for the surface states. We show the derivation explicitly for a specific case of an isotropic cubic lattice with a single Dirac cone surface and then argue that the results can be extended to more general cases of TI surfaces with $\mathcal{N}$ flavors of surface fermions.

Consider an isotropic case of the bulk Hamiltonian given in Eqn.\ref{anisotropicHamiltonian}, which is when $t_x = t_y = t_z = t$. In this case, when $1/3 < t < 1$, the TI is in a non-trivial phase and the surface normal to $z$-axis hosts a Dirac cone at the TRI point $\textbf{Q}_1 = (0,0)$. In the bulk BZ, the mass gap is negative at the TRI point $\textbf{K} = (0,0,0)$, while it is positive at all the remaining seven bulk TRI points. In this context, to derive the surface states, we shall expand the bulk Hamiltonian to first order in $\textbf{k}$ about $\textbf{K}$ to get,
\begin{eqnarray}
    \mathcal{H}_0 (\textbf{k}) &=& \vec{\alpha}.\vec{k} +  m \beta + \mathcal{O}(k^2) \nonumber
\end{eqnarray}
where $m_1 = 1 - 3t$ is the bulk energy gap. 
Now consider the TI surface normal to the $xy$-plane at $z=0$. Since we broke the translation symmetry in $z-$direction, we transform $k_z \rightarrow -i\partial_z$. Also, we write the mass gap as a function of $z$ such that $m(z) = 1 - 3 t (z) = - m$ when $z \rightarrow -\infty$ and $m(z) = + m$ if $z \rightarrow +\infty$. Given that there exists a surface state at $(0,0)$ point in the surface BZ, it means a zero-energy solution exists for the Schrodinger equation at $k_x = k_y = 0$. After some rearrangement, we arrive at the following equation for the surface state,
\begin{eqnarray}
    \left( \frac{\partial}{\partial z} + \Gamma_s m_1 \right) \varphi_1 (0, z) = 0 \label{srfcsteqnappndx}
\end{eqnarray}
where we define $\Gamma_s = \beta \alpha_z$ and $\varphi_1 (k_x, k_y, z)$ is a 4-component spinor wavefunction. This implies that the surface state, if it exists at $z=0$, must be an eigenstate of the operator $\Gamma_s$ with eigenvalue $+1$. That is,
$\Gamma_s \varphi_1 = + \varphi_1$. Now, projecting the Hamiltonian to the two-dimensional subspace formed by the surface states having positive eigenvalue to the $\Gamma_s$ operator, we arrive at the following effective surface Hamiltonian,
\begin{eqnarray}
    \mathcal{H}^{(1)}_{\text{surf}} (\textbf{k}) &=& s_y k_x - s_x k_y \label{rfcstapndx}
\end{eqnarray}
The eigenstates will be 2-component spinors.
This is exclusively for the low-energy surface fermions in and around the surface TRI point $\textbf{Q}_1 = (0,0)$. The effective Hamiltonians of the surface fermions at the other three TRI points can be derived similarly. The corresponding Lagrangians are given in Eqn.\ref{Lfree} in the main text.

Now suppose the bulk topological phase is such that the mass gap is negative at the bulk TRI point $\textbf{K} = (0,0,\pi)$ while positive everywhere else. That means we have to expand the lattice Hamiltonian around the $k_z = \pi$ point. In this case, the Schrodinger equation for the surface states takes the form,
\begin{eqnarray}
    \left( \frac{\partial}{\partial z} - \Gamma_s m_1 \right) \varphi_1 (0, z) = 0
\end{eqnarray}
On comparing this with Eqn.\ref{srfcsteqnappndx}, we see that the surface state, if it exists, must be an eigenstate of the $\Gamma_s$ operator but with eigenvalue $-1$. The effective surface Hamiltonian will still have the same form as in Eqn.\ref{srfcsteqnappndx}, but the difference here is that the pseudo-chirality of the surface state is different.

\subsection{Effective 2D interacting theory}
In this section, we use the dimension reduction method to derive the surface Hamiltonian starting with a bulk interaction theory. We follow closely the path outlined in Ref.\cite{saran(2022)}. 
Let us begin with a general interaction Hamiltonian for the 3D TI,
\begin{eqnarray}
    \mathcal{H}_{\text{I}} &=& -V\sum_{\textbf{k},\textbf{q}, \textbf{p}} \sum^{N}_{j=1}  c^{\dagger}_{\textbf{k},z_j} s_y\otimes \hat{I} c^{\dagger}_{\textbf{q}-\textbf{k},z_j}c_{-\textbf{p},z_j} s_y\otimes \hat{I} c_{\textbf{q} + \textbf{p},z_j}\nonumber\\ \label{HBCS}
\end{eqnarray}
Here the interaction is modelled as on-site in the z-direction. The matrix $s_y \otimes \hat{I}$ is to emphasize that the pairing is between the time-reversed partners. The attractive interaction strength $V$ is such that only electronic states within an energy window $[-\Lambda, \Lambda]$ measured from the Dirac point are interacting. Here $\Lambda$ is the UV-cut-off. If phonons mediate the interaction, then $\Lambda$ becomes the Debye frequency.

In the limit of $\Lambda \ll m_i$ for all $i=1,2,3,4$, the bulk fermions is essentially non-interacting. The only particles that lies within this small energy window are the surface fermions if they exist. These surface fermions are defined in terms of the flavor index $i$ which is essentially the location of the surface TRI point $\textbf{Q}_i$. We have developed a domain wall model for the low-energy surface fermions. 
To write down the effective low-energy model for the interacting Hamiltonian, 
let us replace the annhilation(creation) operator $c_{\textbf{k},z_j} (c^{\dagger}_{\textbf{k},z_j})$ in the interaction Hamiltonian $\mathcal{H}_{I}$ in Eqn.\ref{HBCS} with the field operator corresponding to domain wall fermions. Let $\Psi_i$ be the field operator defined in 3D for the domain wall fermion of flavor $i$. The exact definition is $\bra{\textbf{k}}\Psi^{\dagger}_i(\textbf{r},z)\ket{0} = \varphi_{\textbf{k},i} (\textbf{r},z)$. Since we are in the continuoum limit, the sum over the lattice sites in $z$-direction turns out into an integral over $z$ divided by the size $L$ of the material. Let us set that to unity since it is of no consequence to our effective theory. So we finally have,
\begin{eqnarray}
    \mathcal{H}^{\text{eff}}_{\text{I}} &=& - V\sum_{\textbf{k},\textbf{p}} \sum^{\mathcal{N}}_{i,j = 1} \int dz \,\, \\ && \Psi^{\dagger}_{\textbf{k},i}(z) s_y\otimes \hat{I} \Psi^{\dagger}_{\textbf{q}-\textbf{k},i}( z)\Psi_{\textbf{p},j}(z) s_y\otimes \hat{I} \Psi_{\textbf{p} + \textbf{q},j}(z)\nonumber
\end{eqnarray}
Here $\textbf{k},\textbf{q}$ and $\textbf{p}$ are defined up to the limit of respective mass gaps. That is, $|\textbf{k}| \leq m_i$ and $|\textbf{p}| \leq m_j$.Here $i,j$ runs over the surface Dirac cones. Now one can separate the $z$-dependence out of the spinor field using the relation given in Eqn.\ref{srfcsteqnappndx}. Therefore, this would allow us to write down the interaction Hamiltonian in terms of the 2-component spinor field $\psi_{i}(\textbf{r})$ as,
\begin{eqnarray}
    \mathcal{H}^{\text{eff}}_{\text{I}} &=& -V\sum_{\textbf{k},\textbf{q},\textbf{p}}  \sum^{\mathcal{N}}_{i,j = 1}  \\ && \psi^{\dagger}_{\textbf{k},i} s_y \psi^{\dagger}_{\textbf{q}-\textbf{k},i} \psi_{-\textbf{p},j}s_y \psi_{\textbf{q}+\textbf{p},j}  \int dz \,\, f^{2}_{i}(z) f^{2}_{j}(z)\nonumber
\end{eqnarray}
where $f_{i}(z) = \sqrt{m_i} e^{- m_i |z|}$. This is obtained by solving the Schrodinger equation given in Eqn.\ref{srfcsteqnappndx} for the surface states. 
Integrating out $z$, we arrive at an effective 2-dimensional theory of attractively interacting surface fermions as,
\begin{eqnarray}
     \mathcal{H}^{\text{eff}}_{\text{I}} &=& -\sum_{\textbf{k},\textbf{q}}  \sum^{\mathcal{N}}_{i,j = 1} V_{ij} \psi^{\dagger}_{\textbf{k},i} s_y \psi^{\dagger}_{\textbf{q}-\textbf{k},i} \psi_{-\textbf{p},j}s_y \psi_{\textbf{q}+\textbf{p},j} \nonumber \\ \label{Hint} 
\end{eqnarray}
where,
\begin{eqnarray}
    V_{ij} = V \frac{m_i m_j}{m_i + m_j}\nonumber
\end{eqnarray}
where $m_{i}$ and $m_j$ are the bulk mass gaps of the Dirac cones between which the scattering of the singlet pair of fermions occurs due to effective attractive interaction.

\begin{figure}[b]
 \includegraphics[width=8.6cm, height=5.7cm]{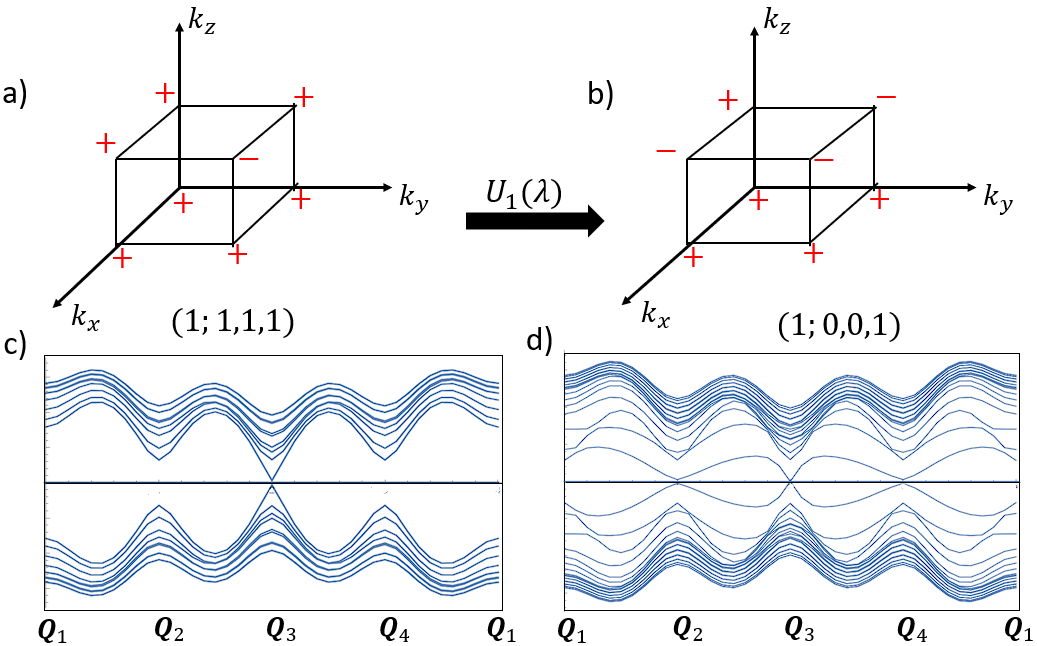}
 \caption{ The figure illustrates how the unitary operator $U_1(k_z,\lambda)$ (Eqn.\ref{U1_ltce}) changes the bulk topology. a) Mass signs ($\delta_{\Gamma_i}$) at the eight TRI points in the bulk BZ before the unitary operation. The topological phase is $(1;1,1,1)$. b) Bulk mass signs after the unitary operation. The bulk topology changes to $(1;0,0,1)$. c) and d) show the surface spectrum found by solving the Hamiltonian in Eqn.\ref{untry1H} before ($(\lambda = 0)$) and after ($\lambda = 1$) the unitary evolution, respectively. $\textbf{Q}_{i}$'s are surface TRI points (See Fig.\ref{phsdgrm} for exact definition). } 
\label{dfrmtn_TIprt1}
\end{figure} 

\section{  Action of non-local unitary operator $U_{1}(k_z,\lambda)$ on the TI bulk \label{appndxuntry1}}

 Here, we study the action of the unitary operator $U_{1}(k_z,\lambda)$ (Eqn.\ref{U1_ltce}) on the bulk Hamiltonian in Eqn.\ref{anisotropicHamiltonian}. In the main text, we argued that the unitary operation changes the bulk topology in addition to deforming the surface states. We show this explicitly here by using a practical example (See Fig.\ref{dfrmtn_TIprt1}).

Consider the isotropic case of the Hamiltonian in Eqn.\ref{anisotropicHamiltonian}, just as in the main text. That is, we set $t_{i} = t \, \forall i=x,y,z$. Under the unitary operation, the Hamiltonian  evolves as,
\begin{widetext}
 \begin{eqnarray} \hat{\mathcal{H}}_{\textbf{K}}(\textbf{k}, \lambda) &=& U_1^{-1}(k_z, \lambda) \hat{\mathcal{H}}_{\textbf{K}}(\textbf{k}) U_1(k_z, \lambda)\nonumber  \\ &=& \sigma_z \left(\alpha_x \sin k_x + \alpha_y \sin k_y \right)\left( \cos \lambda \pi \phi_{\textbf{k}} + i \sigma_y \otimes \gamma \sin \lambda \pi \phi_{\textbf{k}} \right) + \alpha_z \sin k_z - \sigma_z \beta (t\cos k_x + t\cos k_y) \nonumber \\ &+& \beta (1 - t\cos k_z) \left( \cos \lambda \pi \phi_{\textbf{k}} + i \sigma_y \otimes \gamma \sin \lambda \pi \phi_{\textbf{k}} \right) \label{untry1H}
 \end{eqnarray}
\end{widetext}
We observe that the unitary operation is non-local and breaks both translation and inversion symmetries for intermediate values of $\lambda$. While both the symmetries are restored when $\lambda$ approaches unity, the Hamiltonian still has infinite-range hopping terms. In short, the unitary operation possesses the same set of properties as $U(k_z,\lambda)$ (Eqn.\ref{U_efl_tce}), the one discussed in the main text (see Section.\ref{sec:TIdfm}C). 

Fig.\ref{dfrmtn_TIprt1} illustrates the action of the unitary operator on the TI bulk characterized by the topological phase $(1;1,1,1)$. We see that the mass signs ($\delta_{\Gamma_i}$'s) (Eqn.\ref{TI_inv1}) at the $k_z = \pi$ plane in the bulk BZ flips their sign. However, it is simultaneously followed by an interchange of the mass points separated by $\textbf{K} = (\pi,\pi,0)$, resulting in the change of bulk topological invariants. The change of the topological invariants is not surprising as the three weak invariants $(v_1v_2v_3)$ are robust only if the crystal translation symmetry is respected. In this case, we break the crystal translation symmetry by introducing momentum coupling, and the unitary operation is non-local.

Figs.\ref{dfrmtn_TIprt1}c,\ref{dfrmtn_TIprt1}d show the surface spectrum before and after the unitary evolution, respectively. The unitary operation deforms the TI surface from a single Dirac cone phase to three Dirac cones phase. However, unlike the full unitary operation discussed in the main text, the location of the Dirac cones is not complementary to each other on the surface BZ.

\section{Renormalization group study of the interacting TI surface in $D=2+\epsilon$ dimensions up to two-loop level\label{appndxRGE}}
Here, we show the derivation of the RG equation for a TI surface with generic $\mathcal{N}$ flavors of attractively interacting surface fermions. First, we identify the Lagrangian shown in Eqns.\ref{Lfree},\ref{Lfrfmn} as the bare Lagrangian with bare fields $\psi_{i,0}$'s and bare couplings $V^0_{ij}$. Here we write down the renormalized Lagrangian as follows,
\begin{eqnarray}
    \mathcal{L} &=& \sum_{i=1}^{\mathcal{N}} Z_{\psi_i} \psi^{\dagger}_{i,R} \left[ \partial_\tau + \hat{h}_{1} \right]\psi_{i,R} \nonumber\\ &-& \mu ^{-\epsilon}\sum^{\mathcal{N}}_{i,j = 1} \sum_{k=1}\tilde{V}^{R}_{ik} Z^V_{kj} \,\,\psi^{\dagger}_{i,R}s_y \psi^{\dagger T}_{i,R} \psi^{T}_{j,R}s_y \psi_{j,R}  
\end{eqnarray}
where $Z_{\psi_i} $ is the field renormalization factor of the $i$the flavor of fermion. $\psi_{i,R}$ is the renormalized fermion field. $Z^V_{kj}$ is the renormalization factor of the interaction matrix element. Here $\tilde{V}^{R}_{ik}$ is the renormalized dimensionless interaction matrix. $\mu$ is the renormalization scale. 

The renormalized action is obtained from the bare action by a simple field rescaling as,
$\sqrt{Z_{\psi_i}} \psi_{i,R} = \psi_{i,0}$. With this rescaling, we see that the bare and the renormalized interaction strengths are related by
\begin{eqnarray}
    V^0_{ij} &=& \frac{\mu^{-\epsilon} \sum_{k}\tilde{V}^{R}_{ik} Z^V_{kj}}{Z_{\psi_i} Z_{\psi_j}} \label{RGcond}
\end{eqnarray}
Note that the LHS is independent of the renormalization scale $\mu$. By taking the derivative with respect to $\mu$ on both sides of the above equation, we can derive the RGE of the system. 

To facilitate discussions on the RGE, we make a few definitions here. From Eqn.\ref{RGcond}, we can see that the $bare$ dimensionless interaction matrix element has the following form,
\begin{eqnarray}
    \tilde{V}^0_{ij} = \frac{ \sum_{k}\tilde{V}^{R}_{ik} Z^V_{kj}}{Z_{\psi_i} Z_{\psi_j}}\label{bereintrc}
\end{eqnarray}
Plugging this back to the RGE in Eqn.\ref{RGcond} taking derivative w.r.t $\ln \mu$, we have
\begin{eqnarray}
    - \epsilon \tilde{V}^0_{ij} + \frac{\partial \tilde{V}^0_{ij}}{\partial \ln \mu} = 0
\end{eqnarray}
Now we use the chain rule to write,
\begin{eqnarray}
    \frac{\partial}{\partial \ln \mu} &=& \sum_{l,m} \frac{\partial \tilde{V}^R_{lm}}{\partial \ln \mu} \,\,\frac{\partial}{\partial \tilde{V}^{R}_{ml}} = \sum_{l,m} \beta_{lm} \frac{\partial}{\partial \tilde{V}^{R}_{ml}}  
\end{eqnarray}
Plugging this back, we get,
\begin{eqnarray}
    \sum_{l,m} \beta_{lm} \frac{\partial \tilde{V}^0_{ij}}{\partial \tilde{V}^{R}_{ml}} &=& \epsilon \tilde{V}^0_{ij}\label{btafunc}
\end{eqnarray}
From this equation, one can extract the beta function very easily.
In a similar fashion, the anomalous dimension becomes,
\begin{eqnarray}
    \eta_{\psi_i} = \frac{\partial \ln Z_{\psi_i}}{\partial \ln \mu} = \sum_{l,m} \beta_{lm} \frac{\partial \ln Z_{\psi_i}}{\partial \tilde{V}^{R}_{ml}}\label{anmlsdef}
\end{eqnarray}

We calculate the renormalization factors using the dimensional regularization and the minimal subtraction (MS) scheme. 
\subsection{Vertex renormalization}
The one-loop diagram shown in Fig.\ref{fnmndgrms}(a) contributes to the vertex renormalization factor $Z^V_{ij}$. The bare four-point scattering amplitude at one-loop order reads,
\begin{eqnarray}
    &&\Gamma_{ij}^{(4)}(\textbf{q},\tilde{V}^0_{ij}) =  -\tilde{V}^0_{ij}\mu^{-\epsilon}\\&& -8\mu^{-2\epsilon}\sum_{l} \tilde{V}^0_{il}\tilde{V}^0_{lj} \int \frac{d^{D}\textbf{p}}{(2\pi)^D} \text{Tr}\left[G_{l}(\textbf{p})s_y G^T_{l} (\textbf{q}-\textbf{p}) s_y \right]\nonumber
\end{eqnarray}
Here $\textbf{q} = \textbf{k}_1 + \textbf{k}_2 = \textbf{k}_3 + \textbf{k}_4$, where $\textbf{k}_1, \textbf{k}_2$ and $\textbf{k}_1, \textbf{k}_4$ are the incoming and outgoing momenta respectively in the $2+1D$ spacetime. We evaluate the integral using the dimensional regularization technique at arbitrary $D$-spacetime dimensions. Then, we write $D = 2 + \epsilon$ and expand the function to zeroth order in $\epsilon$. We get,

\begin{eqnarray}
     &&\Gamma_{ij}^{(4)}(\textbf{q},\tilde{V}^0_{ij}) = \\ && -\mu^{-\epsilon}\tilde{V}^0_{ij} -\frac{4}{\pi}\mu^{-2\epsilon} \sum_{l} \tilde{V}^0_{il} \tilde{V}^0_{lj} \left[ -\frac{2}{\epsilon} + \ln \frac{1}{\textbf{q}^2} -\gamma + 1 + \mathcal{O}(\epsilon) \right]\nonumber
\end{eqnarray}
The renormalized four-point function has the definition,
\begin{eqnarray}
    \Gamma_{ij, R}^{(4)} (\textbf{q}, \tilde{V}^R_{ij}) = Z_{\psi_i} Z_{\psi_j} \Gamma_{ij}^{(4)}(\textbf{q}, \tilde{V}^0_{ij})\label{rnrmforpnt}
\end{eqnarray}
Now we use the relation between the bare and the renormalized coupling constants derived in Eqn.\ref{bereintrc} to write down the renormalized four-point function in terms of the renormalized interaction strength.
Demanding the renormalized four-point function to be finite up to second order in the interaction strength, the vertex renormalization factor $Z^{V}_{ij}$ becomes,
\begin{eqnarray}
    Z^V_{ij} &=& \delta_{ij} + \frac{16}{2\pi\epsilon} \tilde{V}^{R}_{ij}
\end{eqnarray}.

\subsection{Field strength renormalization}
The field strength renormalization effects start appearing at the two-loop level (see Fig.\ref{fnmndgrms}(c)). The bare 2-point correlation function reads,
\begin{eqnarray}
    \Gamma^{(2)}_i(\textbf{k}, \tilde{V}^0_{ij}) &=& -i k_0 + \vec{s}.\vec{k} + \Sigma_i (\textbf{k}) \label{bre2pntcrltr}
\end{eqnarray}
where $\Sigma_i (\textbf{k})$ is the self-energy of the surface fermion of flavor $i$. It reads,
\begin{widetext}
\begin{eqnarray}
     \Sigma_i(\vec{k},k_0) &=& - 8 \mu^{-2\epsilon}\sum_{j} \tilde{V}^0_{ij} \tilde{V}^0_{ji} \int \frac{d^Dq}{\left(2\pi\right)^D} \frac{d^D p}{\left(2\pi\right)^D} \,\,  s_y G^T_i(\textbf{q}-\textbf{k}) s_y \text{Tr}\left[s_yG^T_j(\textbf{q}-\textbf{p})s_y G_j(\textbf{p}) \right] \label{slfenrgy}
 \end{eqnarray}    
\end{widetext}
In $D = 2 + \epsilon$ spacetime dimensions and upto zeroth order in $\epsilon$, the electron self-energy reads,
\begin{eqnarray}
    &&\Sigma_i(\vec{k},k_0) = \\ &&\frac{16}{\left(4\pi\right)^2}\left( -i k_0 + \vec{s}.\vec{k}\right)\sum_{j} \tilde{V}^0_{ij} \tilde{V}^0_{ji}\left( -\frac{1}{\epsilon} + \ln \frac{1}{\textbf{k}^2} + C\right) \nonumber 
\end{eqnarray}
where $C$ is a numerical constant.
Plugging this back to Eqn.\ref{bre2pntcrltr}, we get,
\begin{eqnarray}
   && \Gamma^{(2)}_i(\textbf{k}, \tilde{V}^0_{ij}) = \\ && \left(- i k_0 + \vec{s}.\vec{k} \right)\left(1 - \frac{16}{\left( 4\pi\right)^2\epsilon}  \sum_{j} \tilde{V}^0_{ij} \tilde{V}^0_{ji} + O(\epsilon^{0}) \right)\nonumber
\end{eqnarray}
The bare and the renormalized 2-point functions are related by,
 \begin{eqnarray}
     \Gamma^{(2)}_{i,R}(\textbf{k}, \tilde{V}^R_{ij}) = Z_{\psi_i} \Gamma^{(2)}_i(\textbf{k}, \tilde{V}^0_{ij})
 \end{eqnarray}
 As before, we calculate the field strength renormalization factor $Z_{\psi_i}$ by demanding that the renormalized two-point function is finite upto second order in the renormalized interaction strength. We get,
\begin{eqnarray}
     Z_{\psi_i} &=& 1 +  \frac{16}{\left(4\pi\right)^2\epsilon} \sum_{j} \tilde{V}^R_{ij} \tilde{V}^R_{ji} 
 \end{eqnarray}
\subsection{Renormalization group equation} 
Plugging these results back to Eqn.\ref{bereintrc}, we obtain the bare coupling constant $\tilde{V}^0_{ij}$ as a function of the renormalized coupling constant $\tilde{V}^R_{ij}$. $\tilde{V}^0_{ij}$ is then expanded in powers of $\tilde{V}^R_{ij}$ to the third order. Then, we take the derivative with respect to the $\tilde{V}^R_{ij}$ and plug the results into Eqn.\ref{btafunc} to obtain the beta function as,
\begin{eqnarray}
    \beta_{ij} &=& \tilde{V}^R_{ij} \left(\epsilon + \frac{32}{\left(4\pi\right)^2} \sum_{n} \left(\tilde{V}^R_{in} \tilde{V}^R_{ni} +  \tilde{V}^R_{jn} \tilde{V}^R_{nj}\right) \right)\nonumber\\ &-&  \frac{16}{2\pi}\sum_{k} \tilde{V}^{R}_{ik}\tilde{V}^{R}_{kj}\label{RGEappnd}
\end{eqnarray}
Having derived the RGE, we remark on the definitions/conventions used in the main text for the Eqn.\ref{RGEappnd} to simplify the discussions. In the main text, we use $V_{ij}$ to represent the dimensionless renormalized coupling instead of $\tilde{V}^R_{ij}$. We rescaled the coupling matrix as $\frac{16}{2\pi} V_{ij} \rightarrow V_{ij}$. Also, the flow parameter $\mu$ is redefined as $\mu = e^{-l}$. With these definitions, the RGE of Eqn.\ref{RGEappnd} takes the form,
\begin{eqnarray}
    \frac{dV_{ij}}{dl} &=& -\left(\epsilon + \frac{1}{32} \sum_{n} \left(V_{in} V_{ni} +  V_{jn} V_{nj}\right) \right)V_{ij}\nonumber\\ &+& \sum_{k} V_{ik}V_{kj}
\end{eqnarray}
In the $\mathcal{N}=2$ case, the RGE is the same as shown in Eqn.\ref{RGEN22loop} 
So we find that the parameter $a$ defined in Eqn.\ref{RGEN22loop} has the value $a = 1/32$. If the two-loop field renormalization effects are ignored, then the above RGE is reduced to Eqn.\ref{RGeqn}.   

Using Eqn,\ref{anmlsdef}, we calculate the anomalous dimension of the surface fermion of flavor $i$ as,
\begin{eqnarray}
    \eta_{i} = \frac{1}{32} \sum_{j} V_{ij} V_{ji}\label{anmlsdmnsn}
\end{eqnarray}

\section{Table of scaling dimensions of the operators at the selected points in the conformal manifolds}
\label{apndxsclngdmnsns}

Below, we shall list the set of relevant, irrelevant, and marginal operators and their respective scaling dimensions at some selected points on the conformal manifold. For brevity, we denote the Cooper pair operator by the form $$\Phi_i = \psi^{T}_{i}i s_y \psi_i.$$
\subsection{\texorpdfstring{$\mathcal{N}=2$}{} TI surface}

\begin{table}[H]
 \caption{Irrelevant, marginal and relevant operators at a point $(V^c_0, V^c_z, V^c_x) = (\frac{\epsilon}{2}, \frac{\epsilon}{2}\cos \phi, \frac{\epsilon}{2} \sin \phi)$ on the ring manifold. Here, $D$ is the spacetime dimensions.} 
    \label{scldmnsnN2}
    \renewcommand{\arraystretch}{1.5}
\begin{tabular}{|c|c|c|}
\hline \hline\,\,\, & Operator &\makecell{Scaling\\ dimension}  \\ \hline
\makecell{Irrelevant\\ operators} & \makecell{$\left( \cos \phi - 1\right) \Phi^{\dagger}_1 \Phi_1 + \sin \phi \, \Phi^{\dagger}_1 \Phi_2 $\\ $-\left( \cos \phi + 1\right) \Phi^{\dagger}_2 \Phi_2$ + h.c} & $2D-2$ \\

\hline \makecell{Marginal\\ operators} & 

\makecell{$\cos \phi \Phi^{\dagger}_1 \Phi_1 - \sin \phi \Phi^{\dagger}_1 \Phi_2$ \\ $ -  \cos \phi \Phi^{\dagger}_2\Phi_2$ + h.c} & $D$\\

\hline \makecell{Relevant\\ operators} &

\makecell{$\left(1 + \cos \phi\right) \Phi^{\dagger}_1 \Phi_1 +  \sin \phi \Phi^{\dagger}_1 \Phi_2 $\\ $\left(1-\cos \phi\right) \Phi^{\dagger}_2 \Phi_2$ + h.c} & $2$ \\ \hline \hline
 \end{tabular}
\end{table}

In the $\mathcal{N}=2$ surface, the conformal manifold is a ring. The exact form is given by $\hat{V}_c = \frac{\epsilon}{2} \hat{I} + \frac{\epsilon}{2} \cos \phi \, \sigma_z + \frac{\epsilon}{2} \sin \phi \, \sigma_z$ We found that the manifold has one relevant and one irrelevant direction in addition to the marginal direction which is tangential to the ring (See table \ref{tableN2}). By linearizing the beta function around a general point on the manifold, we found those three operators and their respective scaling dimensions at any point on the manifold. We list them in Table.\ref{scldmnsnN2}.

\subsection{\texorpdfstring{$\mathcal{N}=3$}{} TI surface}
For the $\mathcal{N}=3$ case, we have two conformal manifolds as shown in table.\ref{n3table}. The eigenstates of the linearized beta function for arbitrary values of $\theta$ and $\phi$ are complex expressions. Therefore, we will select two random points on the conformal manifolds and determine the irrelevant, relevant, and marginal operators at those specific points on the manifold. Tables.\ref{scalndmnsnN300},\ref{sclndmnsnN3200} lists the operators and their respective scaling dimensions at the point $(\theta, \phi) = (0,0)$ on the respective conformal manifolds, while Tables.\ref{sclndmnsnN3pi2pi4},\ref{sclndmnsnN32pi2pi4} list them at the points $(\theta, \phi) = (\frac{\pi}{2},\frac{\pi}{4})$ on the two manifolds.
   

\begin{table}[H]
 \caption{Relevant, irrelevant, marginal operators and their respective scaling dimensions at $\hat{V}_c = \left(\begin{array}{ccc}
        0 & 0 & 0 \\ 
        0 & 0 & 0 \\
        0 & 0 & \epsilon
    \end{array}\right)$($\theta = \phi = 0$ point on the manifold ).  Here, $D$ is the spacetime dimensions.}
    \label{scalndmnsnN300}
    \renewcommand{\arraystretch}{1.5}
\begin{tabular}{|c|c|c|}
\hline \hline 
\,\,\, & Operator & Scaling dimension \\

\hline

\multirow{2}{*}{Irrelevant operators}
 & $ \Phi^{\dagger}_{2} \Phi_2$

 &  $2D-2$ \\ \cline{2-3} 
&
 $ \Phi^{\dagger}_{1} \Phi_2 + \text{h.c}$   &
 
 $2D-2$ \\ \cline{2-3}
 &
 
 $ \Phi^{\dagger}_{1} \Phi_1$  & 
 $2D-2$ 
 
 \\ 
 \hline
 
 \multirow{2}{*}{Marginal operators} &
 
 $ \Phi^{\dagger}_{2} \Phi_3 + \text{h.c}$  &
 
 $D$ \\ \cline{2-3} &

  $ \Phi^{\dagger}_{1} \Phi_3 + \text{h.c}$ &
  
 $D$ \\
  
 \hline Relevant operators &
 
 $ \Phi^{\dagger}_{3} \Phi_3$ & 
 
 $2$ 
 
 \\ \hline \hline
 \end{tabular}
\end{table}

\begin{table}[H]
 \caption{Relevant, irrelevant, marginal operators and their respective scaling dimensions at $\hat{V}_c = \frac{\epsilon}{2}\left(\begin{array}{ccc}
        1 & 1 & 0 \\ 
        1 & 1 & 0 \\
        0 & 0 & 0
    \end{array}\right)$ (
 $\theta = \frac{\pi}{2}, \phi = \frac{\pi}{4}$ point on the manifold).  Here, $D$ is the spacetime dimensions.} \label{sclndmnsnN3pi2pi4}
    \renewcommand{\arraystretch}{1.5}
\begin{tabular}{|c|c|c|}
\hline \hline \,\,\, & Operator & \makecell{Scaling \\dimen-\\sion} \\ \hline
\multirow{2}{*}{\makecell{Irrelevant \\ operators}} &

$ \Phi^{\dagger}_{3} \Phi_3$ &  

$2D-2$ \\ \cline{2-3} &

$- \Phi^{\dagger}_{3} \Phi_1 +  \Phi^{\dagger}_{2} \Phi_3 + \text{h.c}$ & 

$2D-2$ \\ \cline{2-3} &

 $ \Phi^{\dagger}_{1} \Phi_1$  $-\Phi^{\dagger}_{1} \Phi_2$ + $ \Phi^{\dagger}_{2} \Phi_2$ + \text{h.c} &
 
 $2D-2$ \\ \hline
 
 \multirow{2}{*}{\makecell{Marginal \\ operators}}  & 
 
 $\Phi^{\dagger}_{1} \Phi_3 + \Phi^{\dagger}_{2} \Phi_3 + \text{h.c}$  &

 $D$  \\ \cline{2-3} &
  
  $- \Phi^{\dagger}_{1} \Phi_1 +  \Phi^{\dagger}_{2} \Phi_2 + \text{h.c}$ &
  
  $D$ \\ \hline 
  
  \makecell{Relevant \\ operators} & 
  
  $ \Phi^{\dagger}_{1} \Phi_1 +  \Phi^{\dagger}_{1} \Phi_2 +  \Phi^{\dagger}_{2} \Phi_2 + \text{h.c}$ &
  
  $2$ \\ \hline \hline
 \end{tabular}
\end{table}


\begin{table}[H]
 \caption{Relevant, irrelevant, marginal operators and their respective scaling dimensions at $\hat{V}_c = \left(\begin{array}{ccc}
        \epsilon & 0 & 0 \\ 
        0 & \epsilon & 0 \\
        0 & 0 & 0
    \end{array}\right)$ ( $\theta = \phi = 0$ point on the manifold).  Here, $D$ is the spacetime dimensions.
   } \label{sclndmnsnN3200} 
   \renewcommand{\arraystretch}{1.5}
\begin{tabular}{|c|c|c|}
\hline \hline \,\,\, & Operator & Scaling dimension \\ \hline
 Irrelevant operators &
 
 $ \Phi^{\dagger}_{3} \Phi_3$ & $2D-2$
 
 \\ \hline
 \multirow{2}{*}{Marginal operators}  &
 
 $ \Phi^{\dagger}_{2} \Phi_3 + \text{h.c}$ & $D$ \\ \cline{2-3} &
 
 $ \Phi^{\dagger}_{1} \Phi_3 + \text{h.c}$ & $D$ 
 \\  \hline \multirow{2}{*}{Relevant operators}  & 
 
 $ \Phi^{\dagger}_{1} \Phi_1$ & 
 
 $2$ \\ \cline{2-3} & 
 
  $ \Phi^{\dagger}_{1} \Phi_2 + \text{h.c}$
  
  & $2$ \\ \cline{2-3} & 
  
   $ \Phi^{\dagger}_{2} \Phi_2$ &
   $2$
   \\ \hline \hline
 \end{tabular}
\end{table}

\begin{table}[H]
 \caption{Relevant, irrelevant, marginal operators and their respective scaling dimensions at $\hat{V}_c = \frac{\epsilon}{2}\left(\begin{array}{ccc}
       1 & -1 & 0 \\ 
       -1 & 1 & 0 \\
       0 & 0 & 2
   \end{array}\right)$ ( $\theta = \frac{\pi}{2}, \phi = \frac{\pi}{4}$ point on the manifold).  Here, $D$ is the spacetime dimensions.
  }  \label{sclndmnsnN32pi2pi4} 
  \renewcommand{\arraystretch}{1.5}
\begin{tabular}{|c|c|c|}
\hline \hline \,\,\, & Operator & \makecell{Scaling \\dimen-\\sion}  \\  \hline
\makecell{Irrelevant \\ operators} &

$\Phi^{\dagger}_{1} \Phi_1 +  \Phi^{\dagger}_{1} \Phi_2 + \Phi^{\dagger}_{2} \Phi_2 + \text{h.c}$ & $2D-2$

\\ \hline 

\multirow{2}{*}{\makecell{Marginal \\operators}}  & 

$ \Phi^{\dagger}_{1} \Phi_3 +  \Phi^{\dagger}_{2} \Phi_3 + \text{h.c}$ &  $D$ \\ 

\cline{2-3} & 

 $-  \Phi^{\dagger}_{1} \Phi_1 +  \Phi^{\dagger}_{2} \Phi_2 $ & $D$ \\

 \hline \multirow{2}{*}{\makecell{Relevant \\operators}} 
 
 & $ \Phi^{\dagger}_{3} \Phi_3$ & $2$ 

 \\ \cline{2-3} &
 
  $- \Phi^{\dagger}_{1} \Phi_3 +  \Phi^{\dagger}_{2} \Phi_3 + \text{h.c}$ & $2$ 
 
 \\ \cline{2-3} &

  $ \Phi^{\dagger}_{1} \Phi_1 -  \Phi^{\dagger}_{1} \Phi_2 +  \Phi^{\dagger}_{2} \Phi_2$ &  $2$ 
   \\ \hline \hline 
 \end{tabular}
\end{table}

\section{Table of scaling dimensions at the isolated fixed points after symmetry-breaking two-loop effects are included \label{apndxsclngdm2loop}}

In section \ref{sec:hghloop}B, we found that the ring conformal manifold breaks down into isolated fixed points when two-loop fermion field renormalization effects are included. In Tables.\ref{tblen22loop1},\ref{tblen22loop2}, we list the scaling dimensions of the relevant and irrelevant operator(s) at two of these four fixed points. Notice the presence of an operator at the fixed point with its scaling dimension proportional to $a\epsilon^2$, second order in $\epsilon$. It suggests that the operator was marginal at the one-loop level. 

\begin{table}[hbt!]
 \caption{Scaling dimensions of the relevant and irrelevant operators at the isolated fixed point $\hat{V}_c = \left(\frac{\epsilon}{2} + a \epsilon^2 \right)\left(\hat{I} + \sigma_z \right)$.  Here, $D$ is the spacetime dimensions. The numerical constant $a = \frac{1}{32}$.} 
    \label{tblen22loop1}
\renewcommand{\arraystretch}{1.5}
\begin{tabular}{|c|c|c|}
\hline \hline \,\,\, & Operator &\makecell{Scaling dimension}  \\ \hline

\makecell{Irrelevant\\ operators} & 

\makecell{$ - 2\Phi^{\dagger}_2 \Phi_2$} & 

\makecell{$2D-2$} \\
\hline  

\multirow{2}{*}{\makecell{Relevant\\ operators}}  &

$2 \Phi^{\dagger}_1 \Phi_1$ &

$2 + 2a\epsilon^2$ \\ \cline{2-3} &

 $\Phi^{\dagger}_1 \Phi_2$+ h.c &  $D-a\epsilon^2$ \\ \hline \hline
 \end{tabular}
\end{table}

\begin{table}[H]
 \caption{Scaling dimensions of the relevant and irrelevant operators at the isolated fixed point $\hat{V}_c = \left(\frac{\epsilon}{2} + \frac{a}{2} \epsilon^2 \right)\left(\hat{I} + \sigma_x \right)$.  Here, $D$ is the spacetime dimension. The numerical constant $a = \frac{1}{32}$.} 
    \label{tblen22loop2}
\renewcommand{\arraystretch}{1.5}
\begin{tabular}{|c|c|c|}
\hline \hline \,\,\, & Operator &\makecell{Scaling\\ dimension}  \\ \hline

 \multirow{2}{*}{\makecell{Irrelevant\\ operators}} & 

$- \left( \Phi^{\dagger}_1 \Phi_1 + \Phi^{\dagger}_2 \Phi_2 \right) +  \Phi^{\dagger}_1 \Phi_2$ + h.c &

$2D-2 + a\epsilon^2$ \\ \cline{2-3} &

 $\Phi^{\dagger}_1 \Phi_1 - \Phi^{\dagger}_2 \Phi_2 $ &

 $D + a\epsilon^2$ 
 
 \\ \hline 
 \makecell{Relevant\\ operators} & \makecell{$  \Phi^{\dagger}_1 \Phi_1 + \Phi^{\dagger}_2 \Phi_2  +  \Phi^{\dagger}_1 \Phi_2$ + h.c } & \makecell{$2 + a\epsilon^2$ } \\ \hline \hline
 \end{tabular}
\end{table}

\bibliography{intrcTIbib}

\end{document}